\documentclass[graybox,natbib]{svmult}

\usepackage{mathptmx}       
\usepackage{helvet}         
\usepackage{courier}        
\usepackage{type1cm}        
\usepackage{makeidx}         
\usepackage{graphicx}        
\usepackage{multicol}        
\usepackage[bottom]{footmisc}

\usepackage{amsmath}  
\newcommand{\aap}{A\&A}
\newcommand{\aapr}{A\&AR}
\newcommand{\aaps}{A\&AS}
\newcommand{\araa}{ARAA}
\newcommand{\apj}{ApJ}
\newcommand{\aj}{AJ}
\newcommand{\apjl}{ApJL}
\newcommand{\apjs}{ApJS}
\newcommand{\aplett}{ApL}
\newcommand{\bain}{BAN}
\newcommand{\mnras}{MNRAS}
\newcommand{\nat}{Nature}
\newcommand{\pasa}{PASA}

\newcommand{\farcs}{\mbox{\ensuremath{.\!\!^{\prime\prime}}}}

\makeindex             

\begin{document}

\title*{HI in the Outskirts of Nearby Galaxies}
\titlerunning{HI in Galaxy Outskirts}
\author{A. Bosma}
\institute{A. Bosma \at Aix Marseille Univ, CNRS, LAM, Laboratoire d'Astrophysique de Marseille, Marseille, France, \email{bosma@lam.fr}}
%
%
%
\maketitle

\abstract*{The H{\sc i} in disk galaxies frequently extends beyond the optical image, and can
trace the dark matter there. I briefly highlight the history of high spatial
resolution H{\sc i} imaging, the contribution it made to the dark matter
problem, and the current tension between several
dynamical methods to break the disk-halo degeneracy. I then turn to the flaring problem, which 
could in principle probe the shape of the dark halo. Instead, however, a lot of attention is now
devoted to understanding 
the role of gas accretion via galactic fountains. The current $\rm \Lambda$\,cold dark matter theory
has problems on galactic scales, such as the core-cusp problem, which can be addressed with
H{\sc i} observations of dwarf galaxies. For a similar range in rotation velocities, galaxies of type Sd 
have thin disks, while those of type Im are much thicker. After a few comments on modified Newtonian dynamics and on irregular galaxies, I close with statistics on the H{\sc i} extent of galaxies.}

\abstract{The H{\sc i} in disk galaxies frequently extends beyond the optical image, and can
trace the dark matter there. I briefly highlight the history of high spatial
resolution H{\sc i} imaging, the contribution it made to the dark matter
problem, and the current tension between several
dynamical methods to break the disk-halo degeneracy. I then turn to the flaring problem, which 
could in principle probe the shape of the dark halo. Instead, however, a lot of attention is now
devoted to understanding
the role of gas accretion via galactic fountains. The current $\rm \Lambda$\,cold dark matter theory
has problems on galactic scales, such as the core-cusp problem, which can be addressed with
H{\sc i} observations of dwarf galaxies. For a similar range in rotation velocities, galaxies of type Sd 
have thin disks, while those of type Im are much thicker. After a few comments on modified Newtonian dynamics and on irregular galaxies, I close with statistics on the H{\sc i} extent of galaxies.}

\section{Introduction}
\label{sec:intro}
In this review, I will discuss the development of H{\sc i} imaging in nearby galaxies, with emphasis on the galaxy outskirts, and take stock of the subject just before the start of the new surveys using novel instrumentation enabled by the developments in the framework of the Square Kilometer Array (SKA), which was originally partly inspired by H{\sc i} imaging (\citealt{Wilk91}). I  refer to other reviews on more specific subjects when appropriate. Issues related to star formation are dealt with by Elmegreen and Hunter (this volume),
and Koda and Watson (this volume).

In the late 1950s, it became clear that there was more to a galaxy than just its optical image. This was principally due 
to the prediction of the 21\,cm H{\sc i} hyperfine structure line by Van de Hulst in 1944, and its detection in the Milky Way (\citealt{EP51, MO51}). A first rotation curve of the Milky Way was determined by \citet{KMW54}. The first observations of M31 were done using the Dwingeloo 25\,m telescope (\citealt{Hulst57}), and a mass model for this galaxy considered by \citet{Schmi57}. A possible increase in the mass-to-light  ($M/L$) ratio in the outermost parts of M31 prompted the photoelectric study of its light distribution by \citet{Vauc58}, who confirmed that the
local $M/L$ ratio at the last point measured  by Van de Hulst et al. exceeds the 
expectation from a model with constant $M/L$ ratio throughout this galaxy's disk, in contrast with an 
earlier model by \citet{Schwa54}. The outer H{\sc i} layer of the Milky Way was found to be warped (\citealt{Bur57, Kerr57}), and flaring (\citealt{Gum60}). Dark matter in the Local Group was inferred by \citet{KW59}, who attributed the high mass of the Local group to intergalactic gas, and explained the warp as due 
to our Milky Way moving through it. 

\section{HI in Galaxies and the Dark Matter Problem: Early Work}
\label{sec:early-darkm}

Early H{\sc i} work on external galaxies was done with single-dish telescopes, often
prone to side lobe effects (cf. discussion after \citealt{Sal78}). Progress
was slow, and hampered by low spatial resolution---a small ratio of H{\sc i} radius to beam size 
(\citealt{Bos78}, his Chapter 3.4)---of the observations. 
With the Dwingeloo telescope, work on M31 was followed by major axis measurements of M33 and M101 (\citealt{Vol59}).  The Parkes telescope was used to image the LMC (\citealt{McGee66}), 
the SMC (\citealt{Hin67}), NGC~300 (\citealt{Shob67}), and M83 (\citealt{Lew68}). 
In the northern hemisphere, data were reported for several galaxies by \citet[][for M31]{Rob66}, \citet{Rob72} and \citet{Dav74}. In an Appendix, \citet{kcf70} discusses rotation
data for the LMC, SMC, NGC~300 and M33, and, for the latter two galaxies, found that the turnover velocity indicated by the low-resolution
H{\sc i} data is larger than the one indicated by fitting an exponential disk to the optical surface photometry, 
suggesting the presence of matter in the outer parts with a different distribution than that in the inner parts.

\begin{figure*}[htb]
\sidecaption
\includegraphics[scale=0.365, angle=0.0]{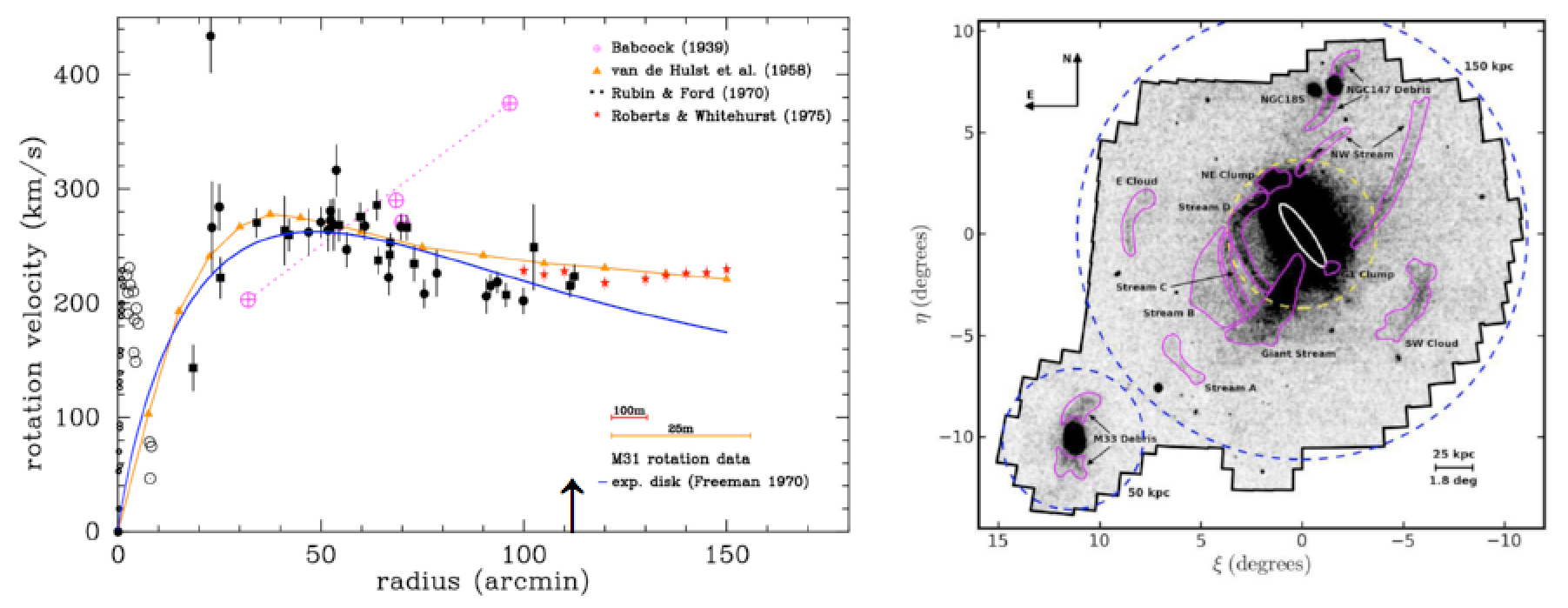}
\caption{{\it Left}  panel: rotation curves of M31, as determined by \citet[][purple points]{Bab39}, 
\citet[][orange points]{Hulst57}, \citet[][black points]{Rub70}, and \citet[][red points]{Rob75}. 
The blue line indicates the expected maximum disk rotation curve based on an exponential disk with the 
scale length given in \citet{kcf70} based on the study of \citet{Vauc58}. The arrow indicates the optical radius.
{\it Right} panel: modern picture of the outskirts of M31, as determined from the PAndAS survey (reproduced with
permission from \citealt{Fer16}), 
where the outline of the optical image is in white. Note the change in scale}
\index{M31}
\label{fig:m31}
\end{figure*}

The quest for higher spatial resolution was first achieved  routinely with the Caltech interferometer,
(e.g., for M101, \citealt{Rog71a, Rog71b}),  and results for five Scd galaxies were summarized in \citet{Rog72},
who found high $M/L$ material in the outer parts of these galaxies. Work began on M31 with
the Cambridge half-mile telescope with a first report (\citealt{EB73}) showing ``normal" $M/L$ values. Their data were in disagreement with data by Roberts on the same galaxy, as
debated in a meeting in Besan\c con (\citealt{Rob74, Bal74b}). This problem was settled by new data reported
by \citet{New77}, who by and large confirmed the data of \citet{Rob75}.
Meanwhile, H{\sc i} work with the Westerbork Synthesis Radio Telescope (WSRT) had started, with as first target M81. 
The resulting rotation data, extending
beyond the optical image, were discussed in \citet{Rob73}, with curves for M31 (300\,ft data), M81 (WSRT and
300\,ft data), M101 (from Rogstad's Caltech data) and the Milky Way. These data, in particular for M31, also suggested that there could be more than meets the eye in the outer 
parts of galaxies, i.e., material with high $M/L$ ratio (cf. Fig.~\ref{fig:m31}, left panel).
\index{H{\sc i} rotation curves}
\index{dark matter} 

A second line of argument for material with high $M/L$ ratio in the outer parts of galaxies came from theory.
Early numerical experiments, e.g., by \citet{Hohl71}, showed that simulating 
a galaxy disk with $N$-body particles and letting
it evolve generated the formation of a bar-like structure with relatively high velocity dispersion. Since strong 
bars are present only in about 30\% of the disk galaxies,
\index{bar instability in disk galaxies}
\citet{OP73} proposed to stabilize the disk by immersing it in a halo of ``dark matter", in such a way that the
gravitational forces of the halo
material acting on the disk could prohibit the bar to form. This requires that out to the radius of the ``optical" disk,
the mass of the halo is $1 - 2.5$ times the disk mass. Once hypothesized, it followed that the masses
of galaxies exterior to this radius could be extremely large. This was subsequently investigated by 
\citet{EKS74}, as well as \citet{OPY74}.
While the former considered five galaxies, amongst which
IC~342, and data on binary galaxies, the latter put forward at least half a dozen
probes (the rotation curve data discussed above, the Local Group timing argument already considered in \citealt{KW59}, binary galaxy samples, etc.). Both papers concluded that there must be additional, dark, matter beyond the optical radius of a galaxy, since the mass increases almost linearly with radius, and the light converges asymptotically. 

\citet{Atha02, Atha03}, using much-improved $N$-body simulations with live haloes (i.e., haloes responding
to gravitational forces), confirmed the above results for the initial phases of the evolution. She found,
however, that at later times, when the bar evolves and increases in strength, the halo material at
resonance with the bar will actually help the bar grow and become stronger.
Hence the 
theoretical picture in \citet{OP73} is now superseded. Likewise, \citet{Atha08} showed that the Efstathiou-Lake-Negroponte global
stability criterion (\citealt{ELN82}), popular in semi-analytic galaxy formation models of galaxies, is not really valid.

Although some cosmologists developed the theory of galaxy formation, and came up with a two-stage model 
(e.g., \citealt{WR78}), the debate about the validity of the data, and even the notion of dark matter itself,
continued for several years, witness papers disputing the idea (e.g., \citealt{Bur75, MT76}).
New data on this topic came from efforts improving the statistics of binary galaxies and clusters of
galaxies, but the most convincing evidence came from fresh data on the rotation curves of spiral galaxies, discussed in Sect.~\ref{sec:darkm-ab}.
 
\section{Warps}
\label{sec:warps}

One observational problem hindering the acceptance of the existence of the dark matter
indicated by the H{\sc i} data concerned the presence of large-scale non-circular motions in the outer parts, as
frequently emphasized by sceptical observers (e.g., \citealt{Bal74a, Bal74b}). \citet{EKS74} simply rejected this 
hypothesis---without stating a reason---even though the
one galaxy they show data for, IC~342, was found to be clearly warped (\citealt{New80}).

\index{warps}
\begin{figure*}[htb]
\centering
\includegraphics[scale=0.384, angle=0.0]{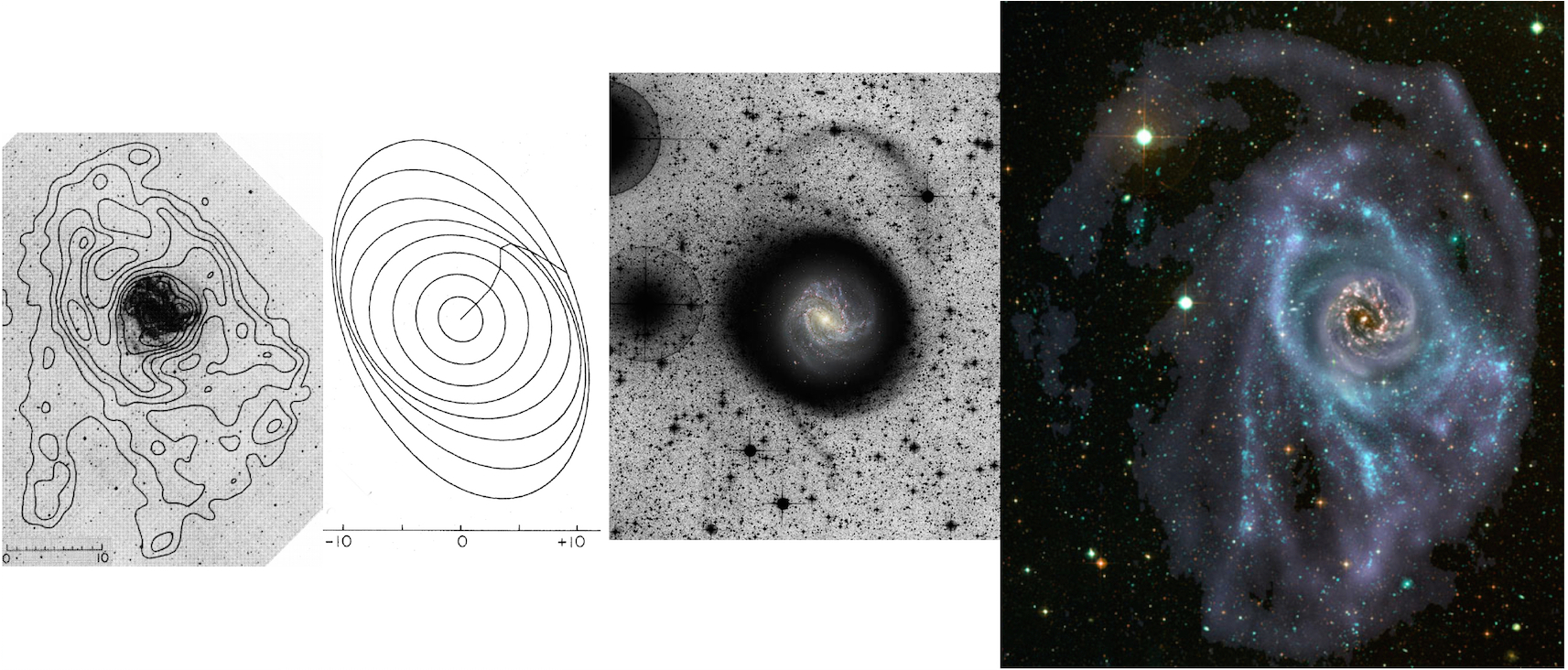}
\caption{{\it Left} two panels: H{\sc i} distribution and tilted ring model of M83 (reproduced with permission
from \citealt{Rog74}); next to it a deep optical image obtained by \citet{Mal97} with a shallower colour image superimposed (credit: D. Malin). The {\it rightmost} image was
constructed from {\it GALEX} UV data, optical images, and the H{\sc i} distribution obtained in the Local Volume HI Survey conducted with CSIRO's Australia Telescope Compact Array (ATCA; credit: B.~Koribalski and A.~R.~L\'opez-S\'anchez). All images are on the same scale}
\index{M83}
\label{fig:m83}
\end{figure*}

\citet{Rog74} came up with a novel approach to model the velocity field of
M83, using new data obtained with the two-element Owens Valley interferometer. They introduced the notion
of what later was called the ``tilted ring model'', by modelling a galaxy with a set of concentric annuli each having
a different spatial orientation. Although they were forced to use a model rotation curve on account of M83's
low inclination, later on rotation curves were derived by determining the rotation velocity for each annulus separately (\citealt{Bos78, Bos81a}). It is of interest to display the results of \citet{Rog74}
side by side with more modern results, as has been arranged in Fig.~\ref{fig:m83}. Since the outer contour in the 
1974 H{\sc i} image is $1.37 \times 10^{20}$\,cm$^{-2}$, the presence of extended H{\sc i} disks in M83-like systems is detectable 
with the upcoming large H{\sc i} surveys, even for those with shorter integration times such as planned for 
the APERTIF-SNS and WALLABY surveys.

\begin{figure*}[ht]
\centering
\includegraphics[scale=0.28, angle=0.0]{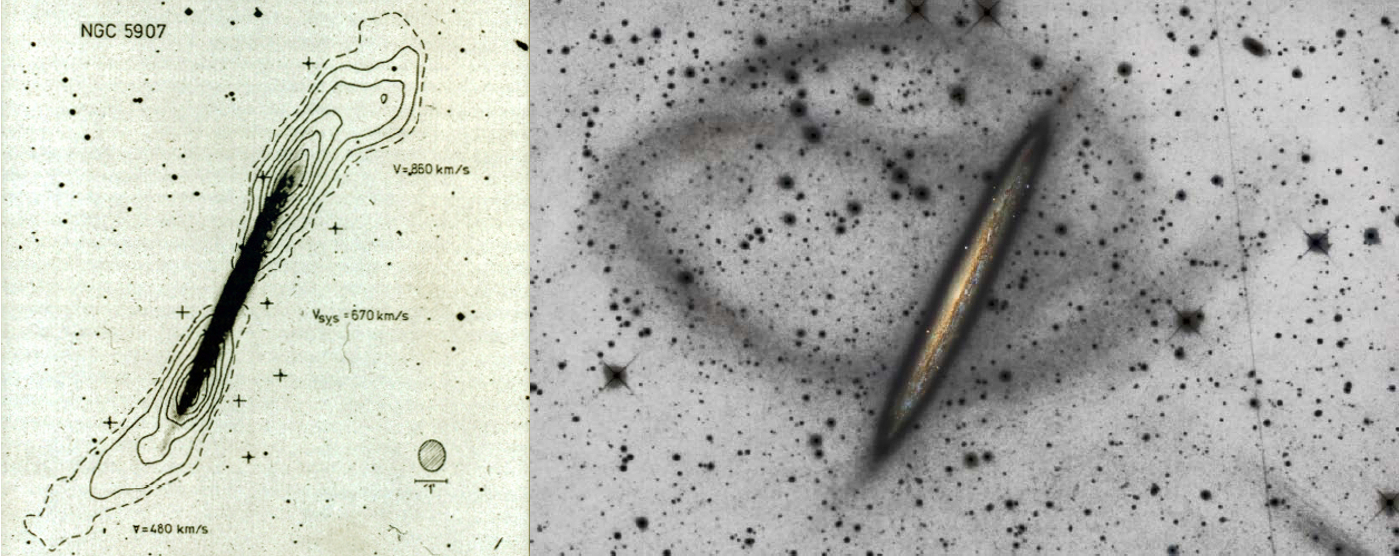}
\caption{{\it Left}: two H{\sc i} channel maps at either side of the systemic velocity of NGC~5907, superimposed on
an optical image of the galaxy, indicate clearly the 
presence of a warp in the H{\sc i} distribution (reproduced with permission from \citealt{San76}). {\it Right}: at the same scale, a deep optical image 
showing the presence of a diffuse tidal stellar stream around this galaxy (reproduced with permission from \citealt{Mar08})}
\index{NGC~5907}
\label{fig:n5907}
\end{figure*}

The first study of H{\sc i} warps in spiral galaxies seen edge-on was done by \citet{San76}, whose clearest case was 
NGC~5907. This galaxy has subsequently been studied extensively in the optical regime, in order to find the presence of extraplanar light. A ``faint glow" was found by \citet{Sac94a}, whose report made the cover of Nature, but later work 
by \citet{Zhe99} indicated the presence of an arc. A deeper picture was published by 
\citet{Mar08}, and is reproduced in Fig.~\ref{fig:n5907}. Further modelling of the warp
has been done in \citet{All15}.
The surmise in \citet{San76} that warps are occurring in galaxies seemingly free of signs of
interactions thus turns  out to be not justified, but the stellar mass involved in the stream appears minor compared to that in the main galaxy. As is the case for M83, the streamer indicates a past interaction or merging event, a process which could have caused the warp as well.

\citet{Rog76, Rog79} found warps in M33 and NGC~300, and I found several more during my thesis work, as
discussed below. \citet{Bri90} outlined a set of ``rules of behaviour" for warps by studying H{\sc i} images 
of a number of warped galaxies studied by the end of the 1980s.
They start in the region beyond the optical radius ($r_{\rm opt}$), and become prominent at
the Holmberg radius ($r_{\rm Ho}$), while the line-of-nodes turns in such a manner that the direction in which it is turning is leading with respect to the spiral arms. This behaviour is due to differential precession, as already noted in \citet{KW59}  and was privately discussed in 1976 between Rogstad and Bosma for five cases then known. \citet{New80} shows clearly that the data of IC~342 do not support this picture, but his work was not
considered by \citet{Bri90}. Note that for the giant low surface brightness disk galaxy Malin~1, the H{\sc i} velocity field observed by \citet{Lel10} shows clearly that this galaxy also violates Briggs's rule no. 3, and that the faint
giant low surface brightness disk imaged recently by \citet{Gal15} is in the warped part of the galaxy.

\section{Further Data on HI in Galaxies and the Dark Matter Problem}
\label{sec:darkm-ab}

\index{H{\sc i} rotation curves} 
\index{dark matter}
In my thesis work (\citealt{Bos78, Bos81a, Bos81b}) I collected extensive H{\sc i} data on a number of galaxies using the 
WSRT and its 80-channel receiver. I produced warp models using tilted ring models, and derived mass models from rotation curves ignoring possible
vertical motions associated with the warps. For the photometry, new observations were obtained (cf. \citealt{PCK79}), 
and the resulting plot of the local $M/L$ ratios was reported in \citet{Bos78} and \citet{Bos79}.
I augmented the sample with 
literature data so as to arrive at a comprehensive figure of 25 rotation curves split over six panels of different morphological types. This Figure was reproduced in an influential review article by \citet{Fab79}, and gained 
widespread attention. 
The local $M/L$ ratios in the outer parts of galaxies
were found to be very large in quite a number of cases, establishing the presence of dark
matter.

At present, such data are considered as primary evidence for the presence of dark matter in spiral galaxies. 
Contemporary data
by \citet{Rub78}, followed by a systematic study of galaxies of type Sc (\citealt{Rub80}), Sb (\citealt{Rub82}) and Sa 
(\citealt{Rub85}) show also flat and rising  rotation curves, but it has been convincingly shown, first by
\citet{Kal83} and later by \citet{Kent86, Kent87, Kent88}, as well as by \citet{ABP87}, that for most of the
galaxies in Rubin's survey the
data do not go out far enough in radius to unambiguously
demonstrate the need for dark matter (see also \citealt{Ber16}). 

\begin{figure*}[t]
\centering
\includegraphics[scale=0.507, angle=0.0]{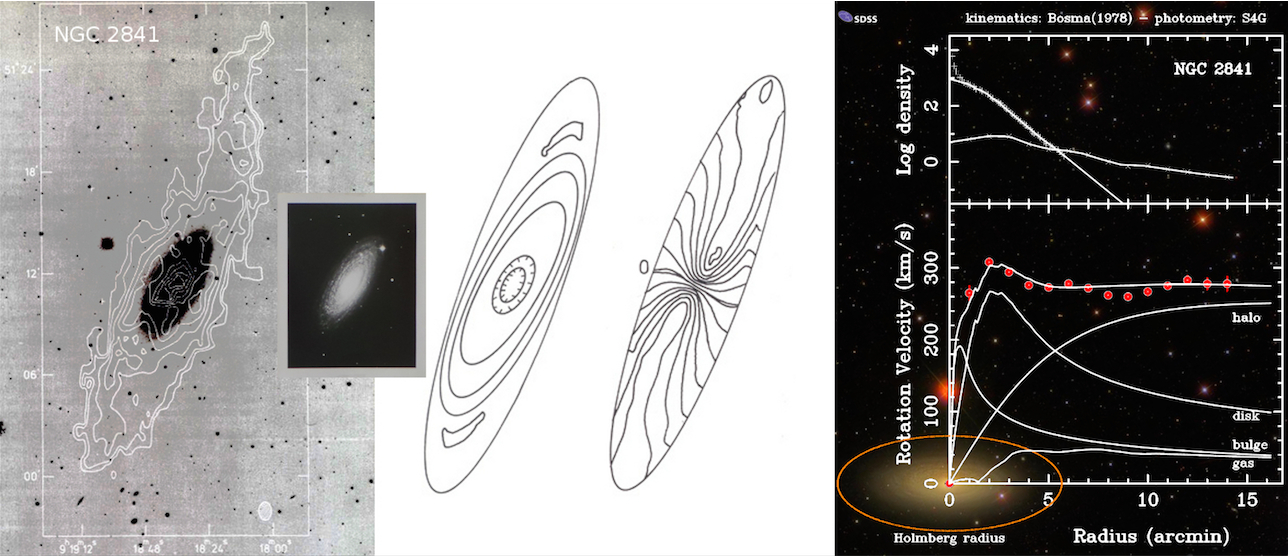}
\caption{At {\it left}, the H{\sc i} distribution in the galaxy NGC~2841 observed by \citet{Bos78}, overlaid on 
a deep IIIaJ image provided by H.C. Arp; the inset shows the Hubble Atlas image (\citealt{San61}).
The {\it middle} panels show the warp model. At {\it right} a mass model
of the galaxy adjusted to the H{\sc i} rotation data in \citet{Bos78},
calculated as described in \citet{ABP87}, using surface photometry data from the
 {\it Spitzer} Survey of Stellar Structure in Galaxies (S$^4$G; \citealt{Mun15}), and overlaid on scale on a Sloan Digital Sky Survey  (SDSS) colour image. The orange ellipse 
around the galaxy outlines the \citet{Hol58} dimensions. Montage as in \citet{Rob88}, 
suggested in this form by Bernard Jones (private communication)}
\index{NGC~2841}
\index{warps}
\label{fig:n2841}
\end{figure*}

\begin{figure*}[t]
\centering
\includegraphics[scale=0.549, angle=0.0]{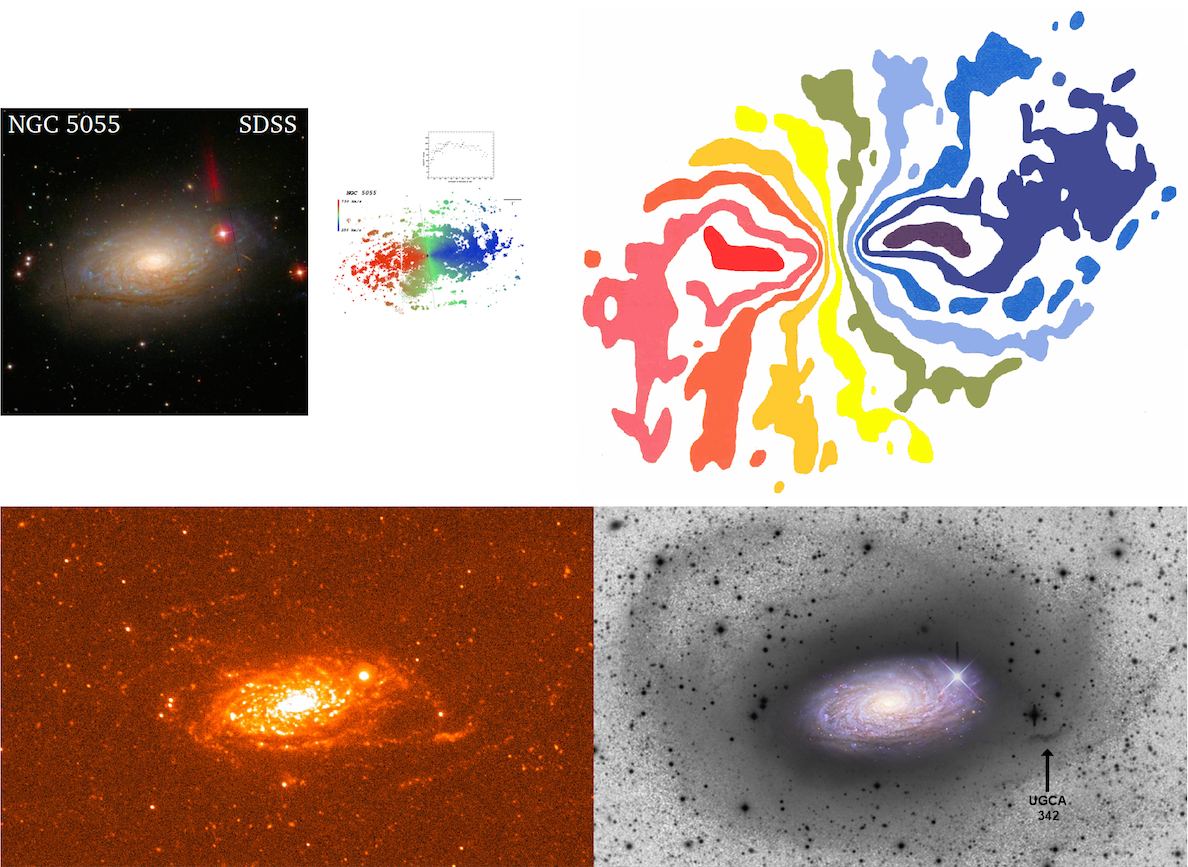}
\caption{{\it Top left}, 
a SDSS colour image of NGC~5055, with next to it a representation of the H$\alpha$ velocity field
from the GHASP survey (reproduced with 
permission from \citealt{Blais04}), and just above it the H$\alpha$ rotation curve of \citet{Bur60}. 
The {\it top right} panel shows the velocity field of NGC~5055 shown
on the front cover of \citet{Bos78}. The colour lookup table used there is such that the boundaries of each coloured
patch correspond to equispaced isovelocity contours. {\it Bottom left} is the {\it GALEX} near-UV image
(data taken from the NASA/IPAC Extragalactic Database, NED) and {\it bottom right} a deep optical image from \citet{Mar10} (credit: D. Martinez-Delgado 
and T. Chonis). All images 
are on the same scale. The star-forming patch west of the centre of 
NGC~5055, UGCA~342, can already be seen on the deep IIIaJ image (\citealt{PCK79}) also
shown as Fig.~4.3.1b in \citet{Bos78}}
\index{NGC~5055}
\index{warps}
\label{fig:n5055}
\end{figure*}

For this review, I will concentrate on the most extended H{\sc i} disks found in \citet{Bos78, Bos81a, Bos81b}, and show as example the galaxy NGC~2841 (Fig.~\ref{fig:n2841}). The H{\sc i} extends out to $\sim$2.5 times the Holmberg radius, i.e.,  $\sim3.5 \times r_{\rm opt}$. The tilted ring model describing the warp leaves out the northernmost feature, an asymmetry further emphasized in Baldwin et al. (1980), although of much smaller amplitude 
than that in the H{\sc i} distribution in M101. 
A similar extended H{\sc i} disk was found for NGC~5055 (Fig.~\ref{fig:n5055}), which became later a prototype of a Type I extended UV disk
(\citealt{Thi07}), and for which \citet{Mar10} found extensive streamers, some of which are not associated with
the H{\sc i} gas or the young stars. In \citet{Bos78, Bos81b} I summarize data on the extent of the H{\sc i} disks in my sample, and found a wide variety of
values of the ratio $r_{\rm HI}$/$r_{\rm opt}$, where $r_{\rm HI}$ is defined as the isophote where the H{\sc i} column
density is $1.82 \times 10^{20}$\,cm$^{-2}$, with a mean of 2.2 $\pm$ 1.1. 

\section{The Disk-Halo Degeneracy in the Dark Matter Problem}
\label{sec:darkm-dh}

\citet{vAS86} pointed out that mass modelling based on the assumption that the $M/L$ ratio
is constant as function of radius in the disk contains a degeneracy: it is a priori not clear whether a maximum
disk, i.e., a disk so massive that its rotation curve fits the inner parts of the observed one 
without overshooting it, is the correct answer. Already it is not entirely justified to assume that 
the $M/L$ ratio is constant throughout the disk, even though this is customary, since if there is a colour gradient in bright spirals, it is usually in the sense that the outer parts are bluer. Yet another problem is that the assumption of maximum disk generally leads to haloes which are
cored, and thus do not follow a Navarro-Frenk-White (NFW) model (e.g., \citealt{Nav98}).

\begin{figure*}[htb]
\centering
\includegraphics[scale=0.223, angle=0.0]{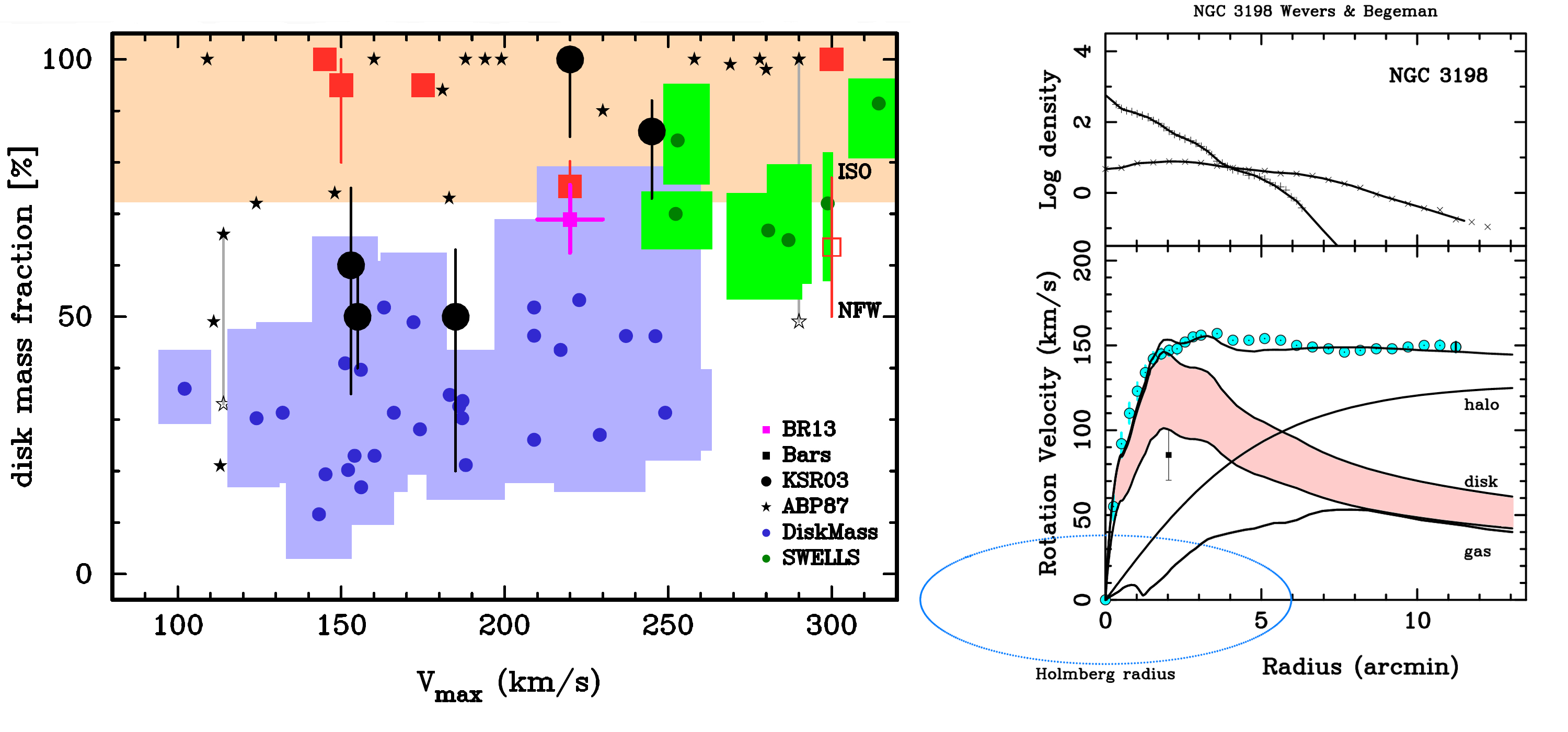}
\includegraphics[scale=0.280, angle=0.0]{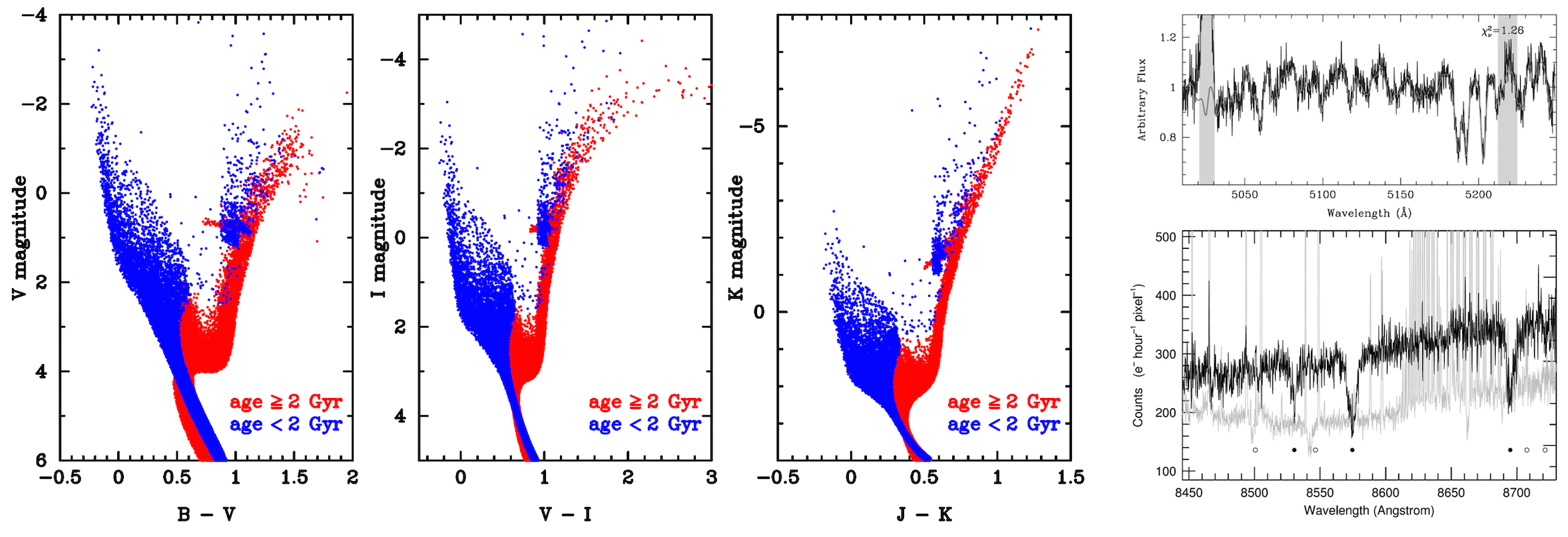}
\caption{{\it Top left}: disk mass fraction as function of the maximum velocity of the rotation curve, determined
with several methods. ``BR13" (\citealt{Bov13}) indicates the value for our Galaxy.  ``Bars" concern the determination using gas flow models in barred spirals (see text). ``KSR03" concern five galaxies studied
by \citet{KSR03}. Black filled stars concern the results from \citet{ABP87} for their 
``maximum disk with no $m=1$" models, except for two vertical lines at $V_{\rm max} = 114.0$ and 280.0\,km/s
which indicate also the ``no $m=2$" models. For the DiskMass project, the results are taken as in \citet{Cou15},
but the error bars are replaced by the area spanned by them. A similar representation has been done
for the SWELLS survey (\citealt{Bar12, Dut13}). {\it Top right}:
mass models for NGC~3198 according to the method described in \citet{ABP87}, and shown in detail in \citet{Bos99}; {\it Bottom left}: colour-magnitude diagrams calculated with IAC-STAR (\citealt{Apa04}) in a manner similar to that used in \citet{Ani16}. {\it Bottom
right}: spectra obtained by \citet{Wes11} in the Mg{\sc i} region, and by \citet{Ber05} in the Ca{\sc ii} region 
(reproduced by permission)}
\index{NGC~3198}
\label{fig:veldisp}
\end{figure*}

\index{dark matter}
\index{disk-halo degeneracy}
This disk-halo degeneracy is a serious problem which is even now under debate. Relating a value of the $M/L$ ratio
to a disk colour, and working out whether
``reasonable" stellar populations can then be assumed, does not appear
entirely satisfactory, due to possible variations in the initial mass function (IMF). 
Various groups have thus tried to marshall dynamical arguments to break the degeneracy, 
but the answers are mixed. Mechanisms
of spiral structure generation, in particular the swing amplifier mechanism (\citealt{Toom81}), depend on the
ratio of disk mass to halo mass. \citet{ABP87} used this in detail to set a range of allowed values
of the disk/halo mass ratio, by remarking that most spirals are dominated by an $m = 2$ component, and thus
requiring that an $m = 2$  spiral be allowed to amplify, while at the upper end suppressing an $m = 1$ component. 
This is further discussed in \citet{Bos99} and illustrated in Fig.~\ref{fig:veldisp} (top right panel).

Fuller dynamical modelling has been done for a number of galaxies.  \citet{KSR03} calculated spiral structure
models based on potentials derived from $K'$-band photometry and compared the gas flow in these with
the observed velocity fields for a sample of five galaxies. Similarly, for bars, the gas flow can be calculated in a potential derived from imaging in the near- or mid-infrared. Seeking to fit the amplitude of the jump in radial velocity across a dust lane, which outlines the location of a shock in the flow, will constrain the $M/L$ ratio of the disk (\citealt{LLA96, Wei01, Wei04, Zan08}). For the face-on barred spiral NGC~1291, \citet{Frag16}
calculated a range of models trying to fit the shape of the dust lane. Most of the barred spiral models require
close to maximum disk, while for the spiral models a range of values depending on the maximum rotational
velocity has been found, as shown in Fig.~\ref{fig:veldisp} (top left panel). 

The study of the lensing galaxy associated with the quad-lens Q2237+0305 done by \citet{Tro10} shows that at least the central part of that galaxy needs a maximum $M/L$ value. Note that that galaxy is barred, and its
bulge thus presumably of the boxy/peanut type. \citet{Bar12} and \citet{Dut13} also find maximum 
bulges in the SWELLS survey. \citet{Dut13} suggest that the IMF in bulges is more like the Salpeter one,
and in disks is closer to the Chabrier one.

The major method favouring non-maximum disks is based on the analysis of stellar velocity dispersions. This is not straightforward, since one has to assume either a disk thickness when the galaxy is face-on, or a ratio of radial to vertical velocity dispersion when the galaxy is edge-on. The results of  \citet{Bot93, Bot97}, 
\citet{Kre05} and the more recent
DiskMass project (\citealt{Ber11, Mar13}, and references therein) all point to sub-maximum disks. Yet doubts have been expressed, as in \citet{Bos99}, from which we quote \begin{quotation} ... as argued by Kormendy (private communication, see also Fuchs's contribution---\citealt{Fuc99}) , the influence of younger stellar populations could result in lower measured velocity dispersions.\end{quotation}
\index{stellar velocity dispersions}

\citet{Ani16} recently quantified this concern, and I show in Fig.~\ref{fig:veldisp} (lower left panel) a variation on their principal result. As argued in Aniyan et al., the stellar populations in the blue-visible region, where the Mg{\sc i} velocity dispersion is measured, are heavily contaminated by the presence of young stars, which might have lower velocity
dispersions than the old stars. On the other hand, in the infrared (e.g., $K$-band), the light of the red giant branch stars
dominate. Hence there is a mismatch between the stellar populations used to measure the velocity dispersion and those used to estimate the disk scale height, with as a result that the $M/L$ ratio of the disk is underestimated. Calculations by \citet{Ani16} show that this can roughly explain the difference between the results of the
DiskMass project (sub-maximum disks), and those from several other dynamical estimators (maximum disks).

Although the authors of the DiskMass project were aware of a stellar population effect, witness
the extensive discussions of this in, e.g., \citet{Ber05, Ber10a, Ber10b} and \citet{Wes11}, they argue clearly
for the use of the Mg{\sc i} region as the preferred region for doing velocity dispersion work, since the Ca{\sc ii} region has
1) larger intrinsic line widths, 2) a higher background, and 3) more scattered light (cf. \citealt{Ber05}, and the 
spectra shown in Fig.~\ref{fig:veldisp}, lower right panel). Most galaxies they discuss have been observed only in
the Mg{\sc i} region, and, in the few cases where data at both wavelength regions were available, no
difference was noticed.

It should be possible, however, to investigate a possible systematic effect of stellar population differences on
the velocity dispersions further, since even in the $i$-band the older stellar populations are 
substantially more prominent than in the $g$-band (see Fig.~\ref{fig:veldisp}, lower row, second panel from left). 
There are now several spectrographs being built which will have a setup allowing
the simultaneous measurement of the velocity dispersions of the same galaxy in both the Mg{\sc i} region and the Ca{\sc ii}
region. In particular, the WEAVE spectrograph  (\citealt{Dal16}) has high
spectral resolution, and can thus suitably be used to test whether Ca{\sc ii}-derived velocity dispersion measurements
are systematically larger than Mg{\sc i}-derived ones in face-on spiral galaxies. 

\section{Flaring of the Outer HI Layer: Probing the Shape of the Dark Matter Halo}
\label{sec:flare}

As mentioned in Sect.~\ref{sec:intro}, a flaring H{\sc i} disk was found in the Milky Way (Gum et al. 1960).
An early study of the flaring H{\sc i} disk in the edge-on galaxy NGC~891 was made by \citet{pck81}, who 
found that the H{\sc i} thickness as function of radius observed by \citet{San79} could best be
modelled by assuming that the disk potential was furnished mainly by the old stellar disk. 
For the face-on galaxies NGC~3938, NGC~628 and NGC~1058 a constant gas velocity dispersion 
as function of radius was found, of order 10\,km/s (\citealt{pck82, Shos84, pck84}).
It was thus thought that flaring H{\sc i} disks could be used to probe the shape of the dark halo.

\subsection{Early Work on Case Studies}
\label{subsec:cases}

With the advent of the high-sensitivity, high-resolution capabilities of the Very Large Array (VLA), NGC~4565 and NGC~891 were re-observed by \citet{Rup91}. These observations indicated clearly the warps in 
these systems, as well as asymmetries in the H{\sc i} distribution.
A further effort was made by \citet{Oll96a, Oll96b} for the small
galaxy NGC~4244, who found that the best fit was a very flattened halo, with an axial ratio $c/a$ 
of $\sim$0.2. \citet{Bec97} found a similar result for NGC~891, but if the vertical velocity dispersion
was smaller than the radial one, a value of $\sim$0.4-0.5 could be found.
Applying the same technique to observations of our Galaxy, \citet{Oll00} found a
relatively round halo ($c/a \sim$0.8), but needed to adopt rather 
low values for some Galactic constants. Using a completely different method (lensing and optical
spectroscopy) for the massive galaxy SDSS J2141-0001, \citet{Bar12}
recently found an oblate dark halo, with $c/a = 0.75 \pm ^{0.27}_{0.16}$.

Also in the 1980s-90s, efforts were made to use another probe of the shape of the dark matter halo, i.e.,
polar ring galaxies. Here, there was a remarkable shift of the ``most likely" halo shape: the early studies of the
polar ring galaxy A0136-0801 by \citet{Schwe83}, and a few others (\citealt{Whitm87}) indicated a nearly round halo, while a later study of \citet{Sac94b} showed the halo around NGC~4650A as flat as E6. However, \citet{Arna97} showed that the polar ring in that galaxy is very massive, and has spiral structure.
\citet{Khop14} worked on a larger sample, and showed that a variety of halo shapes describe the data.
\index{polar ring galaxies}

\begin{figure*}[htb]
\centering
\includegraphics[scale=0.197, angle=0.0]{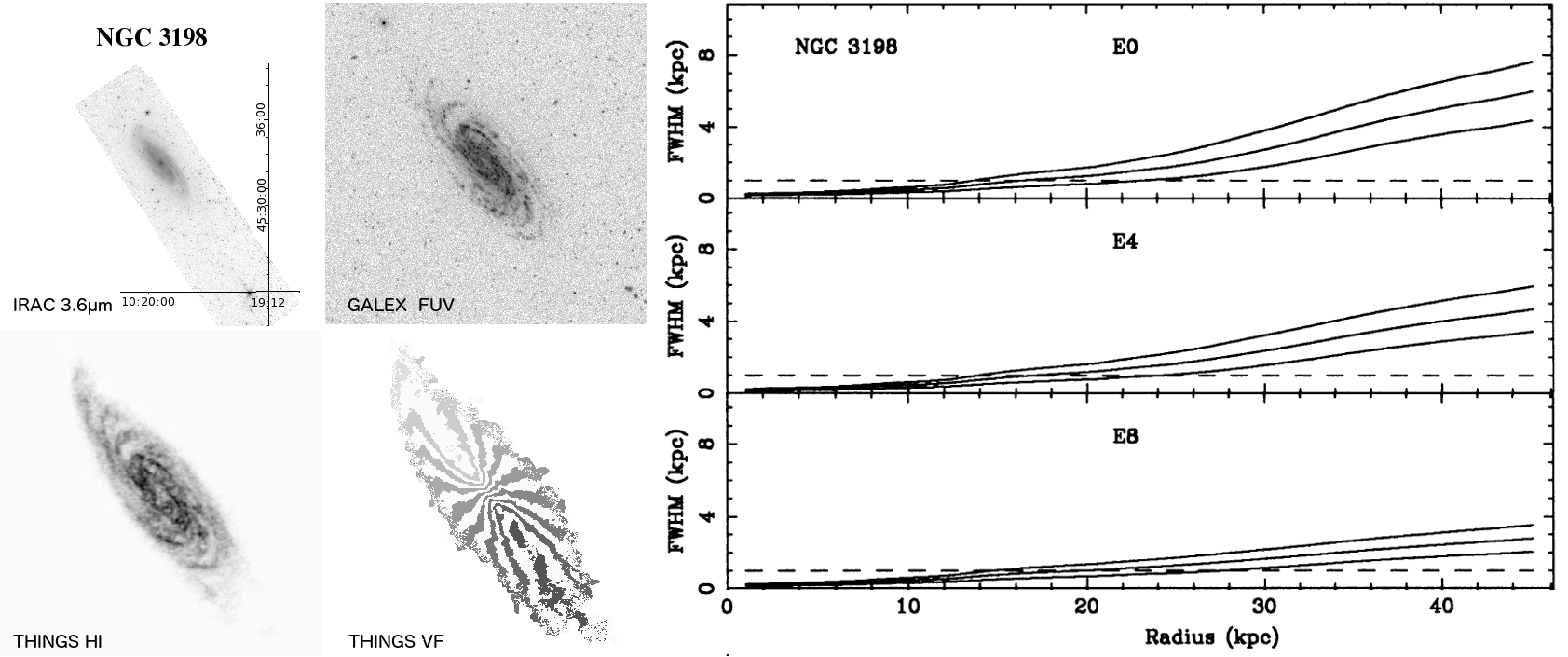}
\caption{{\it Spitzer} 3.6\,$\mu$m, {\it GALEX} far-UV, and H{\sc i} data from the THINGS survey for NGC~3198 (data taken from NED). At {\it right} is the flaring calculation from Bosma 1994, using the mass model in the top right panel of Fig.~\ref{fig:veldisp}.
The three curves are for constant velocity dispersions of 6, 8 and 10\,km/s, and the halo flattening is expressed
as for elliptical galaxies}
\index{NGC~3198}
\label{fig:flaring}
\end{figure*}

The expected H{\sc i} disk flaring was calculated for NGC~3198 and DDO~154 by \citet{Bos94},
and the result for NGC~3198, based on the maximum disk model in Fig.~\ref{fig:veldisp} (top right panel),
is shown in Fig.~\ref{fig:flaring}. This calculation assumes equilibrium, and shows that the flaring depends
on the velocity dispersion of the H{\sc i} disk, as well as the axial ratio of the dark halo. For a dwarf
galaxy, assumed also to be flat, the flaring starts well inside the optical radius,
 since the dark matter dominates also
in the inner parts of the disk. Thus the study of small, thin edge-on galaxies seems ideal for this problem.
\index{flaring H{\sc i} disks}

\subsection{Recent Results for Small, Flat Galaxies}
\label{subset:flatgal}

``Super thin" galaxies were studied initially by \citet{Goa81}, and later investigated by, e.g.,
\citet{Uson03}, \citet{Mat03}, \citet{Mat08} and in thesis work by 
\citet{OBri10c, OBri10b, OBri10d, OBri10a} and 
\citet{Pet13, Pet16c, Pet16a, Pet16b, Pet16d}. A lot of work hides behind these results, since H{\sc i} disks are frequently warped, and candidate galaxies have to be observed first before judging whether they are suitable for further study for the flaring problem. Even if the obviously warped galaxies are excluded from further study in this respect, subtle variations can influence the result. \citet{Mat03} thus considered for their best case, UGC~7321, several models:
a smooth distribution, a warped one, a flaring one, and a model with an H{\sc i} halo, which either corotates
or lags, and combinations of these. None of them give a very satisfactory fit if attention is paid to detail.
\citet{OBri10a} found $c/a = 1.0 \pm 0.1$ for their best case, again UGC~7321.
\citet{Pet13} extended the database of O'Brien et al, but \citet{Pet16d} found in the end
only good flaring models for two galaxies,  ESO~274-G001 and UGC~7321. For ESO~274-G001, they 
found an oblate halo with shape $c/a = 0.7 \pm 0.1$, while for UGC~7321 they found a prolate halo,
with $c/a = 1.9 \pm ^{0.1}_{0.3}$ in case the H{\sc i} is treated as optically thin, and $2.0 \pm 0.1$ in case
the optical thickness of the H{\sc i} is taken into account. They point out that \citet{OBri10a} did not consider 
prolate halo shapes, hence the difference.
\index{super thin galaxies}

Most of these late-type Sd galaxies have small maximum rotation velocities, and low star formation activity.  Systematic studies by \citet{Kar99} have resulted in an extensive catalogue of these flat galaxies. Such data can be explored further in the era of large galaxy surveys. 

\citet{Pet16c} emphasize that it cannot be assumed that the H{\sc i} in galaxies is optically thin,
and constructed a modelling procedure. \citet{Bra97} already emphasized this point twenty years ago
in a study of seven nearby spiral galaxies at high spatial and spectral resolution with the VLA, and 
argued that there is significant opacity in the high surface brightness H{\sc i} gas.
 \citet{Bra09} show this in
their M31 data with specific examples, and estimate that the total H{\sc i} gas mass in that galaxy is about 
30\% higher than the value inferred using the assumption of optically thin gas.

\subsection{Large Galaxies with a High Star Formation Rate: Accretion}
\label{subsec:accrete}

\begin{figure*}[htb]
\centering
\includegraphics[scale=0.46, angle=0.0]{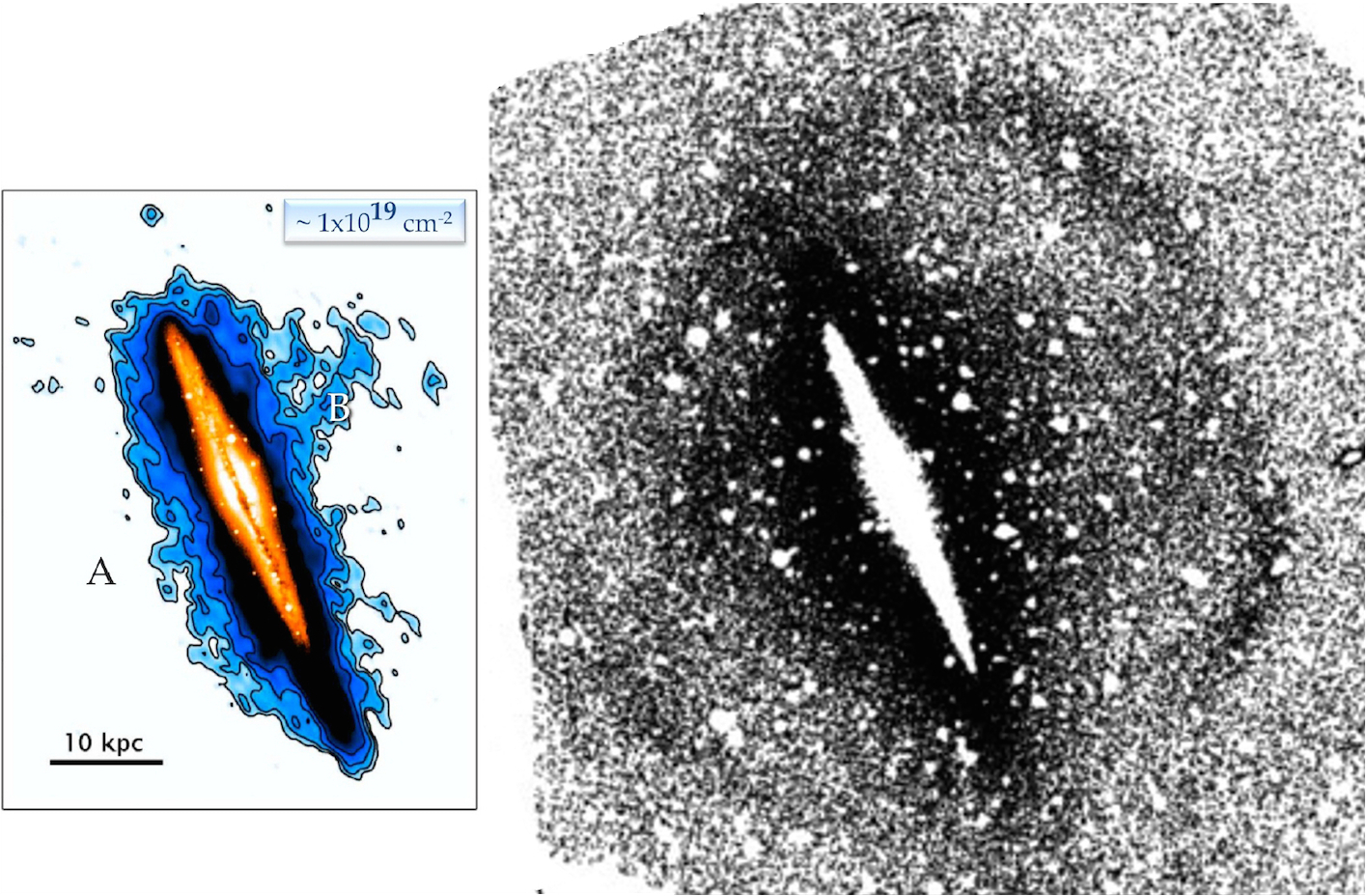}
\caption{At {\it left} the H{\sc i} image of NGC~891 (reproduced with permission from \citealt{Oos07}), with an outer contour column density of
$1.0 \times10^{19}$\,cm$^{-2}$, and at 
{\it right}, on the same scale, the image of the RGB stars (reproduced with permission from \citealt{Mou10})}
\index{NGC~891}
\label{fig:n891-comp}
\end{figure*}

Deep imaging (for that time) of M101 (\citealt{Hul88}) showed that there are
numerous holes in the H{\sc i} distribution in the main disk. In addition, H{\sc i} gas was detected
at velocities not corresponding to the disk rotation, indicating the possible
presence of the equivalence with the Galactic high velocity clouds (HVCs), either 
due to accretion of fresh gas, or a collision with remnant gas clouds related to a tidal interaction
event between this galaxy and a smaller neighbour. Deeper H{\sc i} images of edge-on galaxies also
became available, in particular for the galaxy NGC~891 (cf. Fig.~\ref{fig:n891-comp}).
The observations of this very actively star forming galaxy show a thick H{\sc i} disk (\citealt{Swa97}), and 
yet more sensitive data show a more extensive thick H{\sc i} disk and a streamer (\citealt{Oos07}). 

The main result for the thick H{\sc i} disk in NGC~891 is that its rotation rate is lagging with respect to  the thin 
disk, as already shown by \citet{Swa97}, and more extensively in \citet{Fra05}. 
In \citet{Oos07} more modelling of NGC~891 was done, with a catalogue of possibilities: 
a thin disk, a strong warp along the line of sight, a flaring disk, a disk and corotating halo, a 
lagging halo with constant gradient, a lagging halo with high velocity
dispersion, a lagging halo with a radial inflow motion, and a lagging halo with velocity gradient 
increasing in the inner parts.

\begin{figure*}[htb]
\centering
\includegraphics[scale=0.40, angle=0.0]{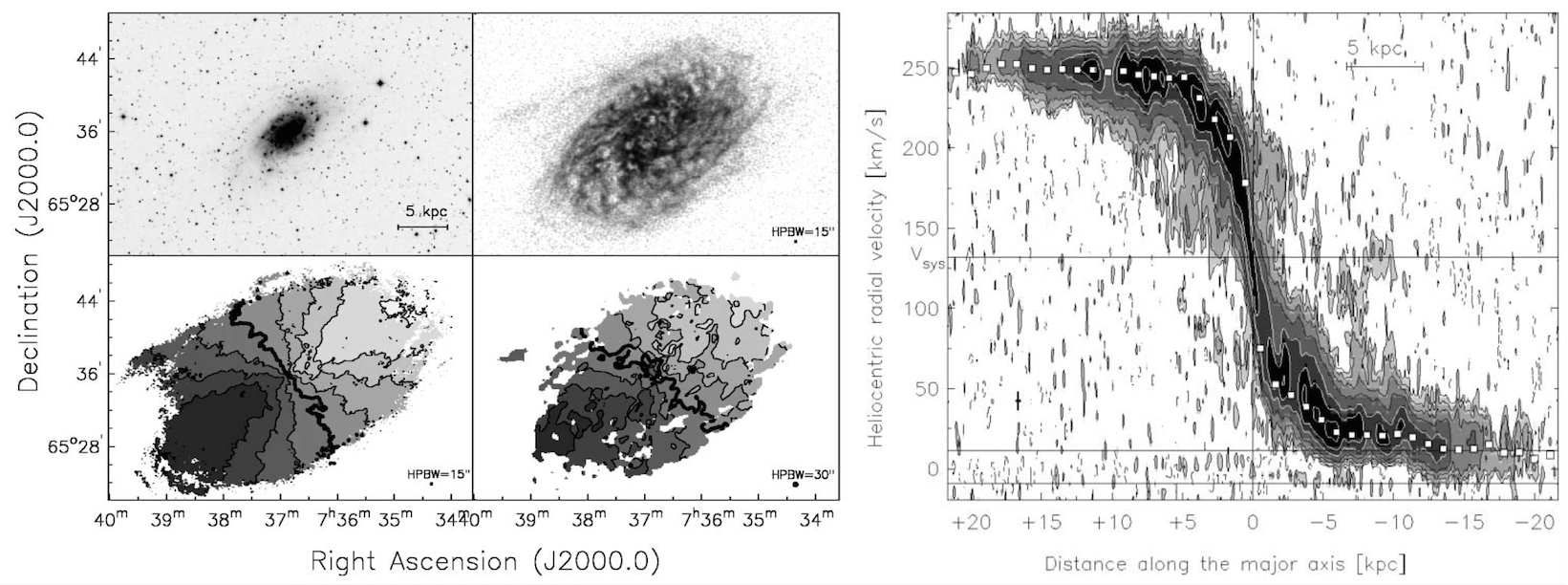}
\caption{VLA observations of the galaxy NGC~2403 (reproduced with permission from \citealt{Fra01}). 
{\it Left} panel of four: {\it top}: Optical image
from the Digital Sky Survey ({\it left}) and H{\sc i} distribution ({\it right}) of the main disk component; {\it bottom}: 
velocity field of the main disk component ({\it left}) and of the anomalous gas component ({\it right}).
{\it Right} panel: position-velocity diagram along the major axis, clearly showing the anomalous 
velocities (the ``beard"). The derived rotation curve of the main disk component is superimposed,
and the lines at the bottom of the plot show the range of velocities where contamination 
by galactic emission was filtered out}
\index{NGC~2403}
\label{fig:n2403}
\end{figure*}

It was realized that there could also be an effect of lagging H{\sc i} halo gas seen in the observations of more
face-on galaxies. Indeed, a position-velocity diagram for the galaxy NGC~2403 
(\citealt{Fra01, Fra02}; Fig.~\ref{fig:n2403})
shows that there is an anomalous component with $\sim$10\% of the total H{\sc i} mass.
The rotation velocity of the anomalous gas is $25-50$\,km/s lower than that of the disk. 
Its velocity field has non-orthogonal major and minor axes implying an overall inflow motion of 
$10-20$\,km/s toward the  centre of the galaxy.
A similar phenomenon is also seen in the THINGS
observations of the galaxy NGC~3198 shown in Fig.~\ref{fig:gasdisp} (upper panel), and is better brought out by the
deeper imaging of this galaxy by \citet{Gen13}, who find that $\sim$15\% of the total H{\sc i} gas mass is 
in a thick disk with a scale height of about $3 \pm 1$\,kpc.

\citet{Fra06} explored an initial model, where galactic fountain activity leads to clouds being sent up in the halo, 
and then falling back to the disk again, but this did not explain the lagging of the rotation of the extraplanar 
gas, nor the inflow towards the disk. \citet{San08}, in a  review, discuss the signatures of H{\sc i} gas accretion around nearby galaxies, and found an average visible H{\sc i} accretion rate of $\sim0.2\,M_{\odot}$\,year$^{-1}$. This is an order of magnitude too 
low to sustain the current star formation rates in some of the galaxies studied. 
\citet{Fra08} improved their galactic fountain model by considering the interaction of the fountain
clouds with the
hot coronal gas in the halo, the presence of which is expected theoretically, and ought to
be observable in X-ray observations (see \citealt{Hod13} for
a detection of that gas in NGC~891). A higher accretion rate can then be obtained, which is more
of order of the star formation rate, so that the latter could be sustained from infalling gas over a 
longer timescale (e.g., \citealt[][and references therein]{Fra14}).
For a recent in-depth review of this accretion model, see \citet{Fra16}.
\index{gas accretion}

To follow up on this, the HALOGAS project was undertaken with the WSRT, but, while first results have been published, a conclusive paper has still to be completed. From presentations at conferences, 
it appears that NGC~891 is a relatively rare case, given its active star formation and extensive extraplanar gas distribution, which has been detected at other wavelengths as well (e.g., radio continuum: \citealt{All78},
H$\alpha$: \citealt{Rand90}, X-ray: \citealt{Hod13}). 
\citet{Dah06} argue on the basis of several examples that the presence of radio continuum thick disks is correlated with active star formation.

\begin{figure*}[htb]
\centering
\includegraphics[scale=0.1375, angle=0.0]{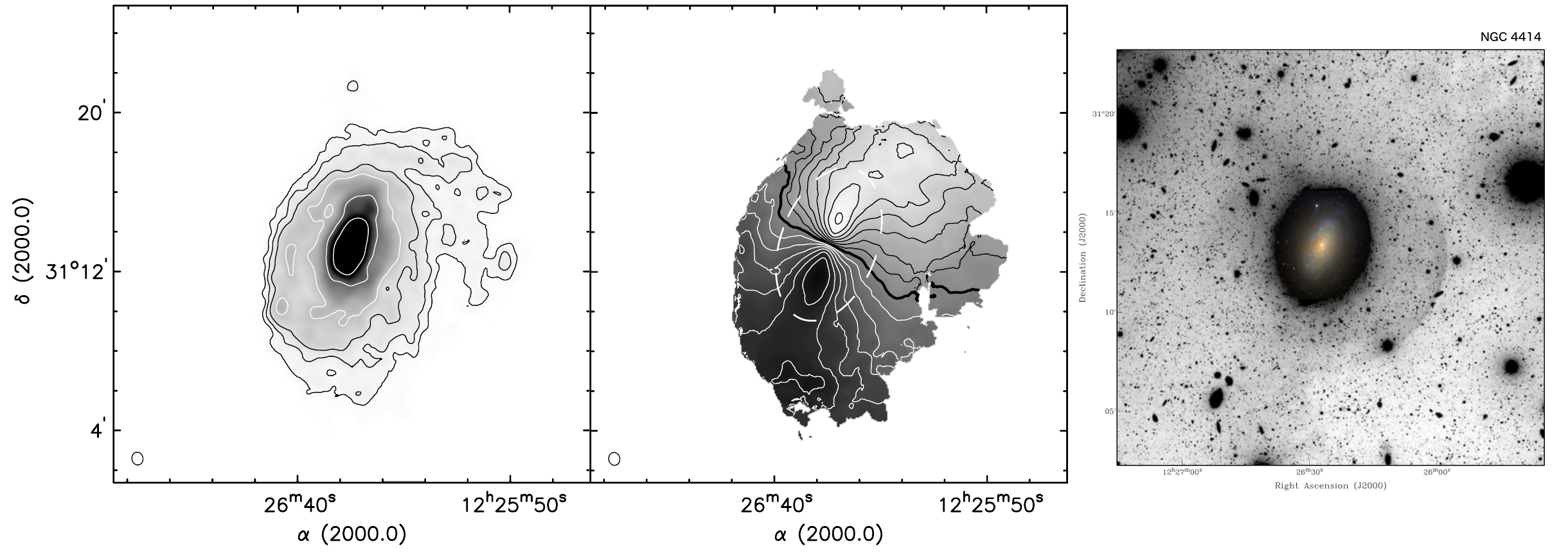}
\caption{Deep H{\sc i} imaging of the galaxy NGC~4414 (reproduced with permission from \citealt{dBlo14}). From {\it left} to {\it right}, the H{\sc i} image, with the outer column density contour being $2\times 10^{19}$\,cm$^{-2}$,  the velocity field indicating a strong warp in the northwestern 
part of the galaxy, and a deep optical image showing an outer shell feature, with a colour SDSS image 
overlaid in the inner parts. The outer isophotes of the SDSS image have a lower axial ratio than those 
of the outer isophotes of the main galaxy body in the deeper image}
\index{NGC~4414}
\label{fig:n4414-deep}
\end{figure*}

Several HALOGAS case studies have been published. New data on NGC~4414 (\citealt{dBlo14}) 
show that the outer warped  H{\sc i} layer is dynamically somewhat offset from the central
parts, while on a deep optical image a shell structure is seen in the western part. Both the warp
and the thickening of the main galaxy (the inclination is higher in the central parts of the galaxy than 
the outline of the main galaxy body seen in Fig.~\ref{fig:n4414-deep}) could be due to an interaction 
which caused also the shell structure itself.

Concurrently, the HALOGAS galaxies were observed with the Green Bank Telescope (GBT), to look for further
H{\sc i} gas at lower column densities. \citet{dBloP14} found a cloud near NGC~2403 in the GBT data, 
which has an H{\sc i} mass of $4 \times 10^6$\,$M_{\odot}$ which is 0.15\% of the total H{\sc i} mass of the galaxy.
It seems to link up with a $\sim$10$^7$\,$M_{\odot}$ filament in the anomalous H{\sc i} gas described in
\citet{Fra02}. Likewise, \citet{Hea15} report about a few HVC like clouds with a total H{\sc i} mass of $4\times 10^6\,M_{\odot}$ around NGC~1003. The very low column density H{\sc i} filament between M31 and 
M33 discovered by \citet{Bra04} was reobserved by \citet{Wol13}, and shown to be composed for $\sim$50\%
of discrete clouds, embedded in more diffuse gas. To conclude, it can be said that there is H{\sc i} around disk galaxies which could be accreted, but the amount of mass is not enough to sustain the star formation rate 
in the main disk (cf. \citealt{Hea15}).

This gas accretion problem has almost completely superseded the attention given to the flaring of
the H{\sc i} gas layer.  \citet{All15} have taken the latter problem up again, and find that the H{\sc i} disk in 
NGC~5907 is flaring, 
while they do not exclude a moderate flaring for other edge-on galaxies. \citet{Kal08} show a
significant flaring for the gas disk of the Milky Way, in agreement with earlier studies. Note that although
a lot of modelling of the stellar disk of edge-on galaxies has been done assuming no flaring, the
work by \citet{Saha09} and \citet{Str16} shows evidence for a mild flaring of the older
stellar disk in a number of edge-on galaxies.

\subsection{Velocity Dispersions in the Outer HI Layers of Spiral Galaxies}
\label{subsec:disp-spiral}
\index{gas velocity dispersions}

\begin{figure*}[htb]
\centering
\includegraphics[scale=0.3365, angle=0.0]{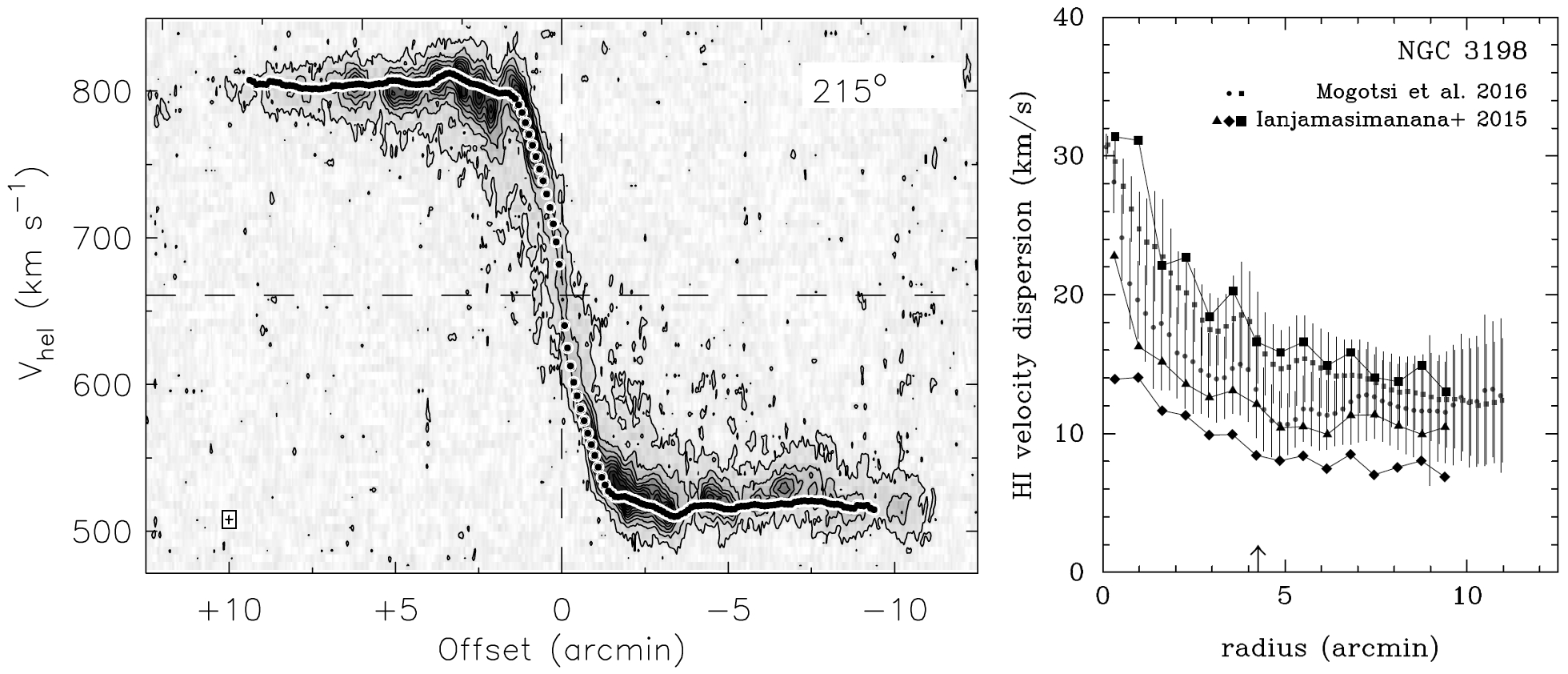}
\includegraphics[scale=0.252, angle=0.0]{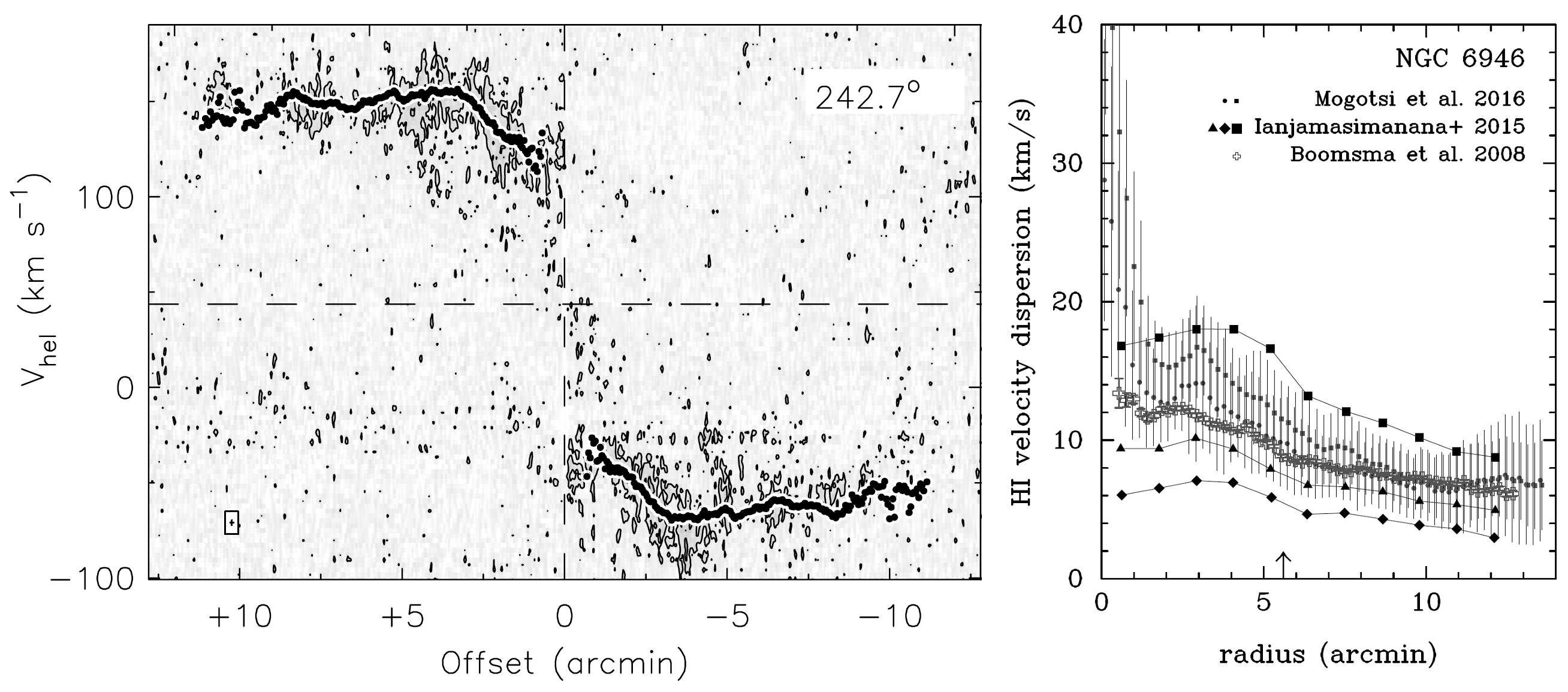}
\includegraphics[scale=0.3175, angle=0.0]{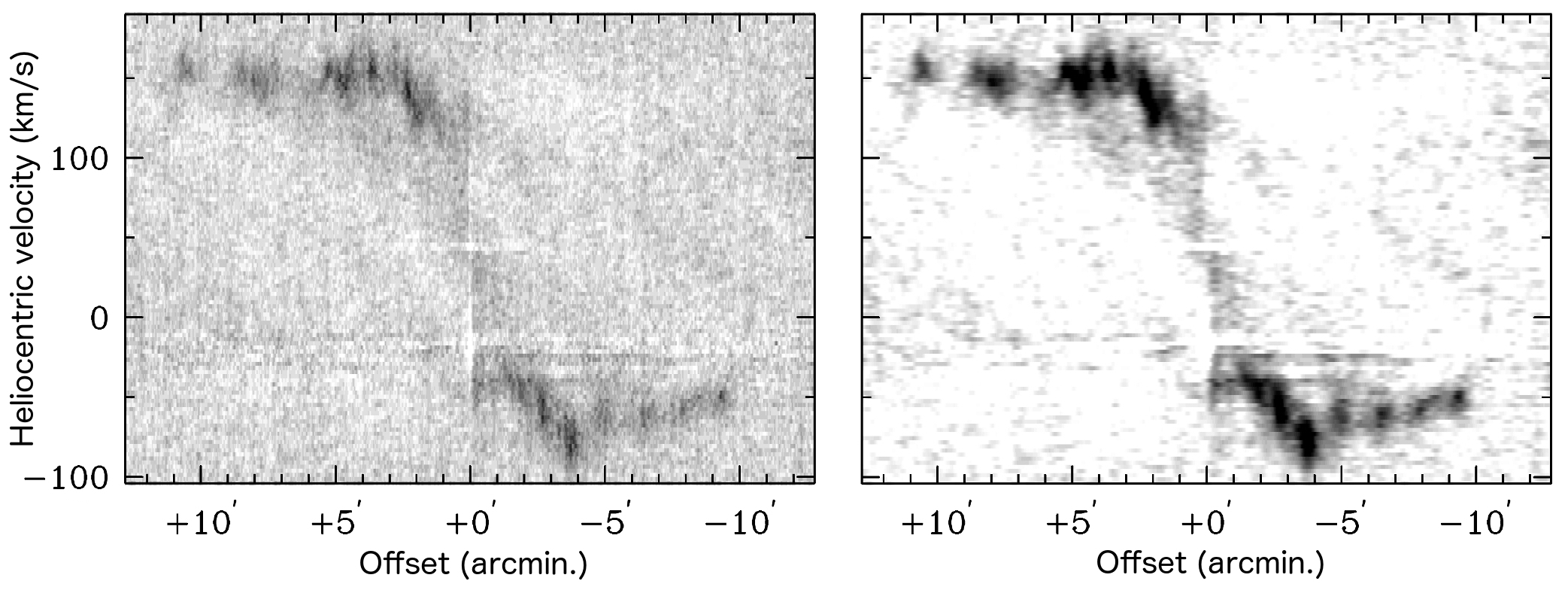}
\caption{{\it Top}: Position-velocity diagram along the major axis of NGC~3198 (reproduced with permission
from \citealt[][{\it left}]{dBlo08}), 
and various radial profiles of the H{\sc i} gas velocity dispersion ({\it right}). {\it Middle}: idem for NGC~6946.
{\it Bottom}: Position-velocity diagrams along the major axis of NGC~6946, using
the available data cube from NED: at left the data with beam size of $6\farcs{0} \times 5\farcs{6}$, 
and at right the $18\farcs{0} \times 18\farcs{0}$ data. The ``beard" shows up rather well in
the smoothed data. The horizontal stripes are due to incomplete subtraction of galactic foreground emission}
\index{NGC~6946}
\label{fig:gasdisp}
\end{figure*}

As discussed at the start of this Section, the flaring problem depends in part on the determination of 
the vertical velocity dispersion of the H{\sc i} gas, and early work with the WSRT
indicated a value around 10\,km/s, with a slight enhancement in star forming regions as found
for NGC~628 by \citet{Shos84}.  \citet{Dick90} determined that for NGC~1058 the velocity dispersion in the gas outside the optical image is remarkably constant at 6\,km/s. Further work on this was done by \citet{Kam93a} and \citet{Kam93b}, in particular for the relatively face-on galaxy NGC~6946, where the velocity dispersion was determined after derotating the data cube by resetting the intensity-weighted mean velocities of the individual profiles to the same central velocity before adding them, as suggested already by \citet{Bou92}. It was shown more clearly that the higher velocity dispersions are related to star formation activity.

In the framework of the THINGS project, \citet{Tam09} determine H{\sc i} gas velocity dispersions
for 11 galaxies, and found again evidence for lower velocity dispersions in the extended H{\sc i} envelopes beyond
the optical image and high velocity dispersions in the inner parts of actively star forming galaxies. If correct,
the model calculation in Fig.~\ref{fig:flaring} (right panels) suggests that a gradient in the velocity dispersion from, e.g., 
10 to 6\,km/s, as seen in several cases, can lead to the absence of significant flaring.

There are various ways to go about the determination of the H{\sc i} gas velocity dispersion, and \citet{Tam09}
took the simplest way, i.e., considering the second moment map derived from the profiles in the data cube.
However, various geometric effects need to be taken into account. One is that the H{\sc i} might not be a single
layer, but in a combination of a low velocity dispersion thin disk and a higher velocity dispersion thick disk.
This has been taken up recently for a number of galaxies 
in the THINGS sample (\citealt{Ian12, Ian15, Mog16}). The results for NGC~3198 are shown 
in Fig.~\ref{fig:gasdisp}, 
side by side with the position-velocity diagram along the major axis. This galaxy does not have the most 
favourable orientation for this work (perhaps too edge-on), but the results are illustrative of the
difficulties involved. \citet{Ian15} use ``super-profiles", i.e., profiles corrected
for the effect of rotation, and stacked in annuli $0.2\,R_{25}$ wide. They fit both a double Gaussian
to low- and high-velocity components, as well as a single Gauss fit
which allows comparison with older data in the literature. The low-velocity component is thought to
represent the cold neutral medium in the thin H{\sc i} disk, and the higher-velocity component the warm
neutral medium in a thicker disk. \citet{Mog16} present in an Appendix the results from fitting a
single Gaussian to the profiles as well as taking the second moment. The comparison
in Fig.~\ref{fig:gasdisp} shows that the results  differ widely. In the very central parts, the data are
influenced by the broadening of the profiles inside a spatial resolution element (i.e., beam smearing). 
Outside, there is a large difference between the second moment and a Gaussian fit, on account 
of the skewness of the profiles. As discussed already in Sect.~\ref{subsec:accrete}, for NGC~3198,
this skewness is due to the presence of a thick disk with lagging rotation (\citealt{Gen13}). 

For the more face-on galaxy NGC~6946, the results from both studies also show again the presence of
a low- and high-velocity dispersion component. Older WSRT data by \citet{Boom08} show an H{\sc i} velocity
dispersion profile similar to the one found in \citet{Kam93a}, and the profile determined by Gaussian fitting of \citet{Mog16} agrees reasonably well with this. However, the single Gauss 
fit by \citet{Ian15} is definitely lower, presumably due to the selection of only profiles with signal-to-noise ratio of three or more to be included into the stacked profiles.
What is most striking, however, is that the velocity dispersions in NGC~6946 are lower than those
in NGC~3198, despite the fact that NGC~6946 is forming stars more actively. The 
difference is perhaps due to the presence of more extraplanar gas in NGC~3198. 
Moreover, NGC~3198 is seen relatively edge-on, so there are line of sight
integration effects. However, analysis of the publicly available data cube of NGC~6946 suggests another
cause: if I smooth the cube from $6\farcs{0} \times 5\farcs{6}$ to $18\farcs{0} \times 18\farcs{0}$, the ``beard" shows up 
in the smoothed version of the position velocity diagram (cf. Fig.~\ref{fig:gasdisp}, lower panels). Thus considerations
of spatial resolution and sensitivity, as well as the spatial orientation of the galaxy with respect to  the line of sight all
play a role in the outcome of the studies of this problem. And, of course, if the question is asked: ``what 
about the magnetic field?", it is strong in NGC~6946, and much weaker in NGC~3198. \citet{Beck07} shows
that in the outer parts of NGC~6946 the magnetic field energy density is higher than the energy in turbulent motions, so this cannot be neglected in the calculations of the thickness of the gas layer, at least for that galaxy.

\subsection{Star Formation in Warped HI Layers}
\label{subsec:warped-hi}
\index{warps}

Occasionally, star formation is seen in the warped H{\sc i} layers, as in, e.g., NGC~3642 (\citealt{Ver02}),
where the outer disk of low surface brightness has a different spatial orientation than the inner parts. This galaxy
was later taken as a prototype of a Type III radial surface brightness profile in \citet{Lai14}, i.e., having an
exponential disk with in the outer parts an upbending profile. This situation is noted 
also for Malin 1 (see also Sect.~\ref{sec:warps}). \citet{pck07} notes that for the survey of warps in edge-on
galaxies by \citet{Gar02}, the onset of the warp is rather abrupt, and occurs, for most of the galaxies
considered, just outside the truncation radius if there is one. However,  faint star forming regions can be 
found in a number of XUV disks, as shown in Fig.~\ref{fig:n5055} (lower left) for NGC~5055 and
Fig.~\ref{fig:flaring} (top, second panel from left) for NGC~3198. \citet{Kor09} present H{\sc i} observations of the  NGC~1510/1512 system,
and state that the faint disk of NGC~1512, which contains strong spiral features, is warped with respect to  the inner disk.
\citet{Rad14} find that in the warped H{\sc i} disk of NGC~4565, the young stellar populations participate in
the warp, but not the older ones. Detection of molecular gas in some of these galaxies is discussed in Koda and Watson (this volume).

\section{The Core-Cusp Problem}
\label{sec:core-cusp}

The {$\rm \Lambda$}CDM theory of galaxy formation and evolution predicts that in dark
matter-dominated galaxies the tracers of the potential should indicate a cuspy NFW profile (\citealt{NFW96, NFW97}). However, this has not been observed, as already remarked by \citet{Moo94} and \citet{Flo94}.
Observations of late-type low surface brightness disk galaxies were done by a number of groups, first
in H{\sc i}, and later augmented with long-slit H$\alpha$ observations in the inner parts 
(e.g., \citealt{dBlo01, dBlo02, dBlo03}). These
confirmed the presence of cores in such galaxies, or, at the most, mildly cusp 
slopes with $\alpha = -0.2\pm0.2$, not
compatible with the NFW profiles. These results were (e.g., \citealt{Swa03}) and still are the subject 
of intense debate.  CO and three-dimensional  H$\alpha$ observations (\citealt{Sim03, Sim05, Blais04, Kuz06, Kuz08}) 
were also brought to bear on this problem. A review has been given in \citet{dBlo10}.

\begin{figure*}[ht]
\centering
\includegraphics[scale=0.156, angle=0.0]{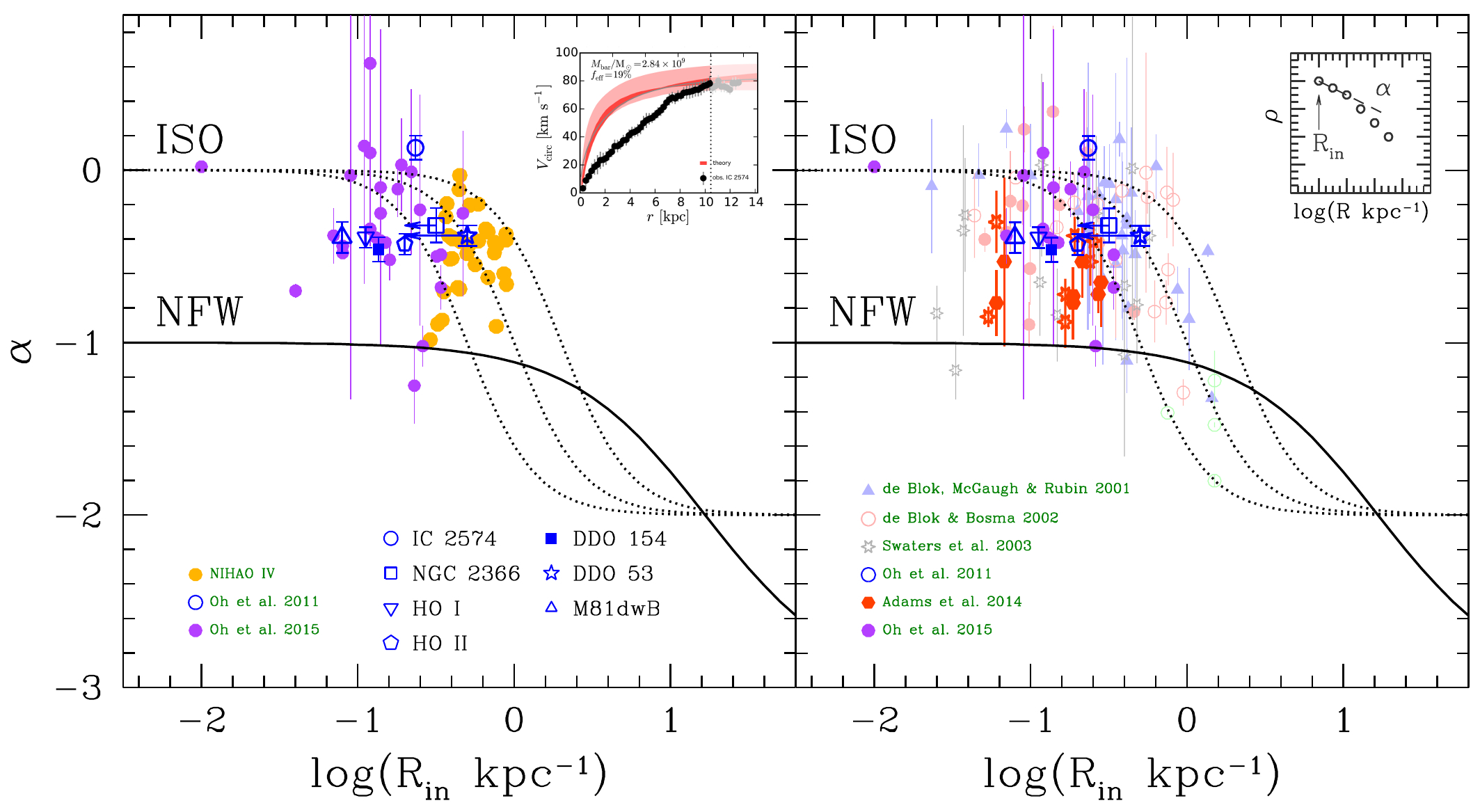}
\caption{{\it Left}: results from \citet{Oh11, Oh15} for the slope of the dark matter density profile as function of
the radius of the innermost point on the rotation curve. The orange points are from hydrodynamical
zoom simulations from the NIHAO project (\citealt[][their Fig. 13]{Tol16}). The inset shows the current
data for IC~2574 and the APOSTLE simulation result for it (reproduced with permission from \citealt{Oman16}).
{\it Right}: a similar plot which collects older data 
(\citealt{dBlo01, dBlo02, Swa03}), the data from \citet{Oh11}, data based on stellar and gas velocity
dispersions (\citealt{Ada14}), and the selected data for 15 galaxies from \citet{Oh15} as described in the text.
The dotted lines converging on $\alpha$ = 0 represent the theoretical variations in slope for ISO haloes with $R_{\rm C} = 0.5$, 1 and 2\,kpc. The line converging on $\alpha = -1$ shows the variation in slope for 
$c/V_{\rm 200} =  8.0/100$}
\index{core-cusp problem}
\label{fig:corecusp}
\end{figure*}

Projects such as THINGS, and LITTLE THINGS, using H{\sc i} observations at higher resolution and sensitivity, 
keep finding cores (\citealt{Oh08, Oh11, Oh15}). These observations can now be compared with modified predictions
of the {$\rm \Lambda$}CDM theory of galaxy formation and evolution, where star formation and feedback have
been added to the ingredients (``subgrid physics") of the numerical simulations. The results are given in
Fig.~\ref{fig:corecusp}, together with the results from cosmological zoom simulations of the NIHAO project
({\citealt{Tol16}). The addition of baryonic physics changes the prediction for the
dark matter profile towards a slope shallower than the NFW profile. However, the current resolution achieved
in the NIHAO project does not yet probe the full range in inner radii comparable to those of the
observations. The debate is ongoing about whether the
{$\rm \Lambda$}CDM theory should be further revised to accommodate the new observational results. I 
select here a couple of issues which can be addressed with H{\sc i} data in the context of galaxy outskirts.

\citet{Oman15, Oman16} emphasize that there is a large variety in the data of late-type low surface brightness dwarfs, and find that the observations of the rotation properties of some dwarfs are closer to their models than others. They suggest that observers somehow do not have the systematics of galaxy inclinations under control. \citet{Read16} claim good agreement with {$\rm \Lambda$}CDM by discarding half their sample of four galaxies: IC 1613 is thought to be in disequilibrium due to starburst activity, and for DDO~101 the distance is too uncertain. 

\begin{figure*}[htb]
\centering
\includegraphics[scale=0.31, angle=0.0]{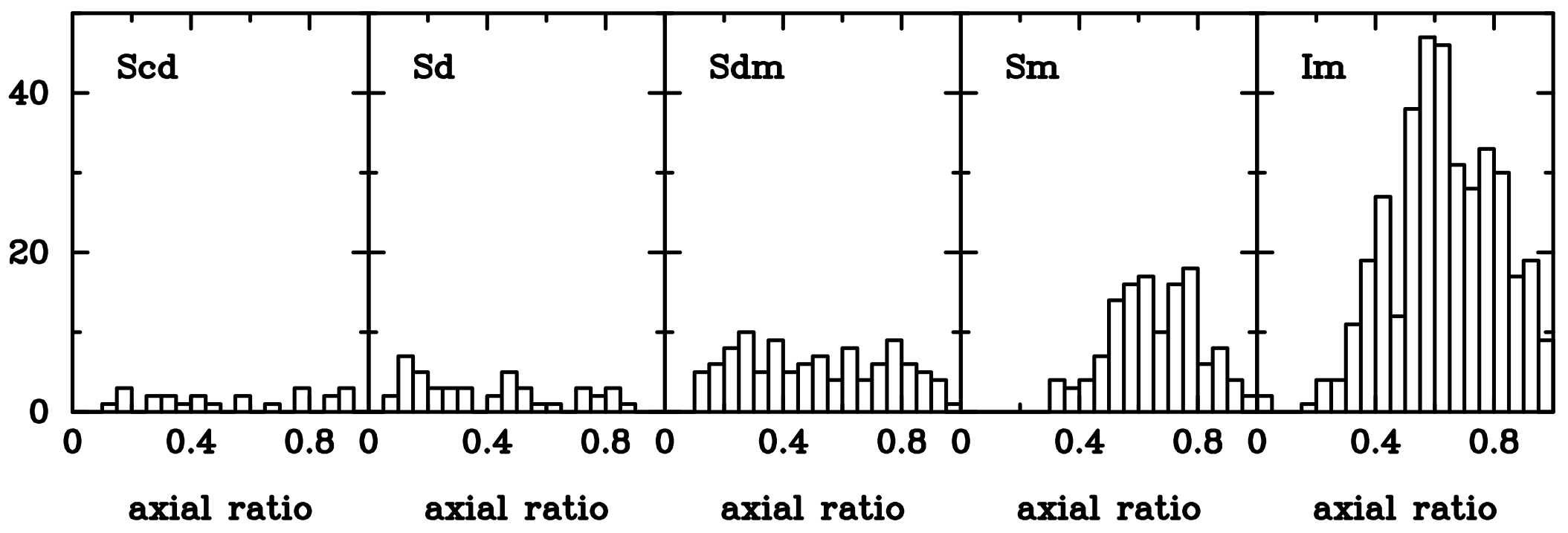}
\caption{Statistics of axial ratios for late-type galaxies (numerical Hubble type $\ge$ 6) in the Local Volume 
Galaxy catalogue (\citealt{Kar13}), split out by morphological type}
\label{fig:lvl-axrat}
\end{figure*}

To examine the problem further, I looked at some of the underlying assumptions in the modelling of the observational data. One is that galaxy disks are thin, with a typical vertical
axial ratio of 0.2. Such a value can be checked statistically, as has been done in the exemplary work of \citet{San70}. For data in the Second Reference Catalog of Bright Galaxies 
(\citealt {Vauc76}), \citet{BdeVauc81} point out that for Hubble types close to the end of the sequence (type 10), the distribution of apparent axial ratios does not indicate flattened disks. This was 
discussed further in \citet{Bos94}, and is illustrated here again in 
Fig.~\ref{fig:lvl-axrat}, now using data from the Local Volume Galaxy
catalogue (\citealt{Kar13}), split by morphological type for the later types. 
While Scd, Sd and Sdm galaxies have histograms that
seem compatible with those for flattened disks viewed from different orientation angles, the data for Sm and in particular Im galaxies show an apparent axial ratio distribution not compatible with this. It is thus likely that some of the dwarf galaxies used in the core-cusp debate are not modelled correctly when it is assumed that they are thin disks: instead, models with a considerable thickness should be explored. The results from the FIGGS sample (\citealt{Beg08, Roy10}) confirm this, and show that the thickness of the H{\sc i} maps peaks around
0.5. \citet{Roy13} also analyze the Local Volume catalogue, and find a similar thickening of the galaxies of later type, but their analysis considers only a few bins. As discussed further in Sect.~\ref{subsec:irr-disp},
thick H{\sc i} disks have been recognized as such by \citet{Puc92}, and even earlier by \citet{Bot86}.
\index{disk thickness}

\begin{figure*}[htb]
\centering
\includegraphics[scale=0.29, angle=0.0]{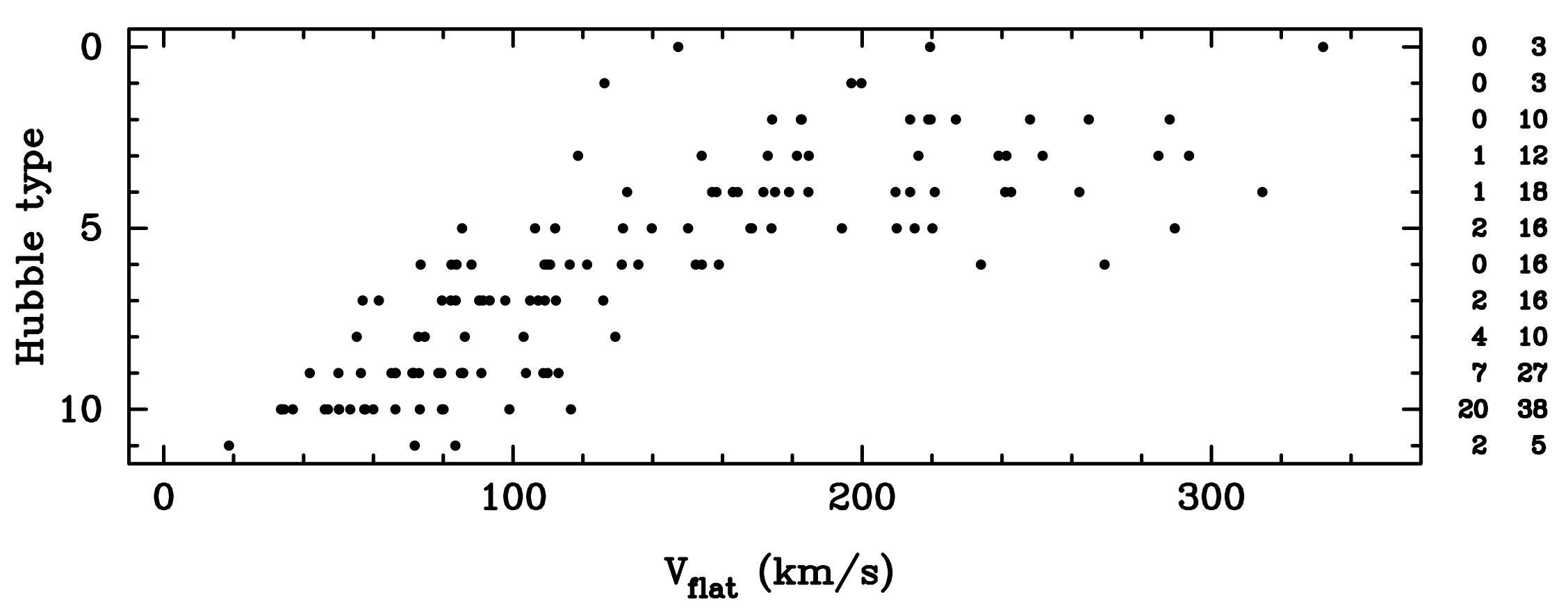}
\caption{Statistics of the $V_{\rm flat}$ value from the SPARC sample (\citealt{Lel16}) split out by 
morphological type. The number of galaxies not reaching a $V_{\rm flat}$ value, and the total number
per type bin, are indicated at the right. Note the overlap in  $V_{\rm flat}$ for types $7 - 10$}
\label{fig:sparc}
\end{figure*}

In Fig.~\ref{fig:sparc} I use data from the recent collection of rotation curves studied by \citet{Lel16} concerning
the mean velocity in the outer parts, $V_{\rm flat}$, if defined. For the thin disk galaxies of type Sd and Sdm (numerical Hubble type 7 and 8)  the range in $V_{\rm flat}$  overlaps the one of the thicker disk galaxies of type Sm, Im and BCD (types 9, 10 and 11, resp.). There is thus more to a galaxy than just its rotation curve, since the  three-dimensional morphology of a galaxy does not follow automatically from the one-dimensional rotation curve. Therefore,
for galaxies with $V_{\rm flat}$ between 70 and 110\,km/s, corresponding to a stellar mass range
of ${\sim} 5\times 10^{8} - 4 \times 10^{9}\,M_{\odot}$,  numerical simulations of galaxy formation 
and evolution ought to reproduce a variety of shapes in the stellar mass distribution, rather than a single
one, and for smaller galaxies a thicker stellar disk should be produced.

I further consider the data on the LITTLE THINGS project discussed by \citet{Oh15}.
In that paper, the H{\sc i} layer is assumed to be infinitesimally thin, and the thickness of the stellar
component is computed from the ratio of disk scale length to disk scale height (h/z$_{\rm 0}$) of 5.0,
which holds for large spiral galaxies seen edge-on, but not necessarily for dwarfs. To minimize problems, I
selected only galaxies for which the difference in position angle and inclination derived from the H{\sc i} data and those from the axial ratio of the optical images, as given in \citet{Hun12}, is less than 25 degrees (as 
$\sqrt{\Delta({\rm PA})^2 + \Delta({\rm Inc})^2}$). 
This leaves 15 galaxies, which are shown in Fig.~\ref{fig:corecusp} (right panel). Selecting 
``better behaving" dwarf galaxies, with little warping of the H{\sc i} disk, thus 
does not alleviate the core-cusp problem. 
A more elaborate analysis is beyond the scope of this review.

\section{Alternative Gravity Theories}
\label{sec:mond}

A recurrent issue with the interpretation of 
H{\sc i} rotation curves is the argument that the dark matter interpretation is not
necessarily correct, starting with a paper by \citet{Mil83} proposing 
MOND (``MOdified Newtonian Dynamics"). \citet{Beg91} made models 
for 10 galaxies, both for Newtonian gravity and for 
MOND, and pointed out that a common critical
acceleration parameter can be found once some leeway is
allowed for galaxy distances. Reviews about MOND and its relative success in reproducing rotation curves
have been produced by \citet{San02} and \citet{Fam12}. However, the applicability of MOND to 
larger scales, such as clusters of galaxies, is problematic, and \citet{Ang09} argues for the presence of
11\,eV neutrinos to cure this. Such particles have not been found yet, but then, the nature of the dark matter
remains unknown. The fate of the Vulcan hypothesis by Le Verrier (cf. \citealt{Font73}) is frequently quoted 
to illustrate the idea behind MOND (modifying the theory of
gravity rather than searching for additional planets), but the history of
dark matter detection resembles now more the discovery story of Pluto, rather than of Neptune.

My own contribution to this has been limited to refereeing some of the papers, and I will neither treat the 
debate here, nor discuss other alternative theories. 
What is of interest are the attempts to disprove MOND, by emphasizing the 
discrepancies with H{\sc i} imaging data. There are manifest problems with some galaxies if their 
Cepheid distance is adopted, 
the clearest case being NGC~3198 (cf. \citealt{Bot02, Gen11, Gen13}). For late-type
dwarf galaxies, the recent results of \citet{Car13} for NGC~3109, and \citet{Ran14} for DDO~154, IC 2574,
NGC~925 and NGC~7793---in addition to NGC~3109 and NGC~3198---clearly bring out discrepancies as well. However, \citet{Ang12} could reduce the 
discrepancy for DDO~154 by considering a thick gaseous disk, up to the point of getting a reasonable
fit for most of the radial extent of this galaxy. Such thick H{\sc i} gas disks are not unreasonable for
DDO~154 and IC 2574, in view of the discussion above, in Sect.~\ref{sec:core-cusp}, and further in
Sect.~\ref{subsec:irr-disp}, but less so for
the Sd galaxies NGC~925 and NGC~7793. 
Finally, the often cited ``dip" in the rotation curve of NGC~1560 
(\citealt{Bro92, Gen10}) poses a problem, since the curve determined from the data for the 
northern half of this galaxy does not have a dip, while the one for the southern part does, due 
to a local absence of gas and stars there; the mean thus has ``half a dip"!? 
\index{disk thickness}
\index{MOND}

MOND does not go away easily, and the recent result from 153 galaxies in the SPARC sample
(12 galaxies too face-on and 10 galaxies too asymmetric were rejected, cf. \citealt{McGa16}), i.e., a
tight relation, with a scatter of $\sim$30\% on a log-log plot, of the acceleration determined from
rotation curves compared to the one based on 3.6\,$\mu$m radial surface brightness profiles assuming a
constant $M/L$-ratio for galactic disks, revived some interest in it. \citet{Lud16} argue that the
small scatter in this mass-discrepancy - acceleration relation is due to different feedback processes moving
data points along this line, rather than deviating strongly from it, so that the $\rm {\Lambda}$CDM 
theory does not have much difficulty explaining it. However, this leaves the core-cusp problem unsolved
(cf. Fig.~\ref{fig:corecusp}, inset of left panel).

The diversity of the 3.6\,$\mu$m radial surface brightness profiles (e.g., Type I, II and III 
profiles---\citealt{MarN12, Mun13, Kim14, Lai14}), and the various physical processes invoked to explain these down- or up-bending profiles, cannot be ignored. Furthermore, in \citet{Fam12}, the
two galaxies presented as examples for good MOND fits, NGC~6946 and NGC~1560, have a different
$K$-band $M/L$-ratio (0.37 vs 0.18), i.e., a factor of two difference, which is just the order of magnitude
difference discussed in Sect.~\ref{sec:darkm-dh}. Moreover, data from the high
angular resolution THINGS and LITTLE THINGS samples are not included in the SPARC sample, since,
according to \citet{Lel16}, their rotation curves are characterized by many small-scale bumps and wiggles 
thought to be due to non-circular components such as streaming motions along spiral arms. This
exclusion is ironic, seen that when MOND is discussed, the capacity of reproducing such bumps is
deemed very important. Anyway, these ``details" seem to operate on a different level of 
complexity than the overall mass-discrepancy - acceleration relation. Perusal of the individual
mass models of each of the SPARC galaxies confirms this: the limiting surface brightness of
the 3.6\,$\mu$m profiles used in those mass models is $23.8 \pm 1.8$\,mag\,arcsec$^{-2}$, i.e., 
faint surface brightness levels are not considered for every galaxy.


\section{Irregular Galaxies}
\label{sec:irr}
\index{irregular galaxies}

\subsection{Very Large HI Envelopes}
\label{subsec:irr-env}

\begin{figure*}[htb]
\centering
\includegraphics[scale=0.424, angle=0.0]{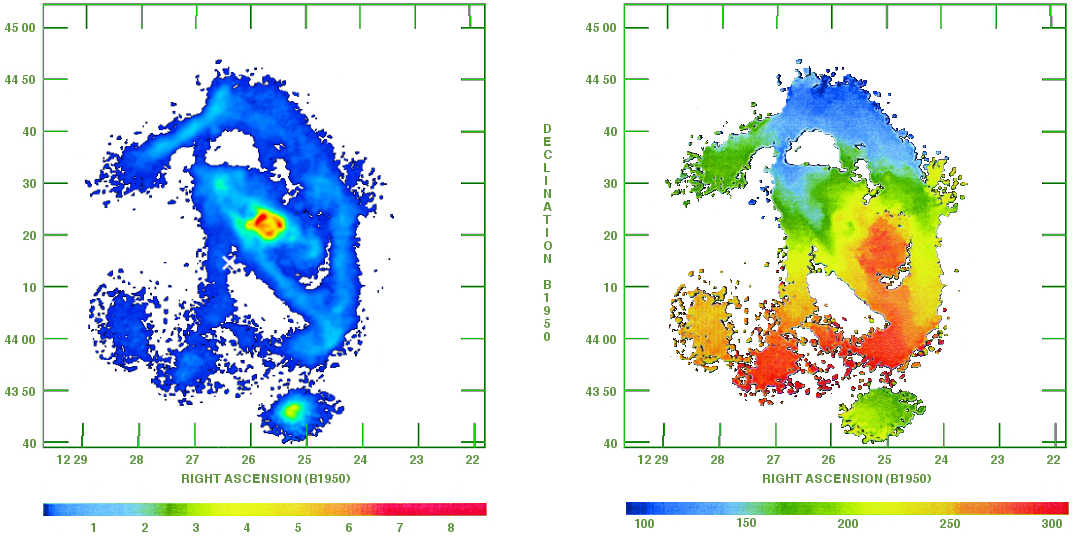}
\includegraphics[scale=0.33, angle=0.0]{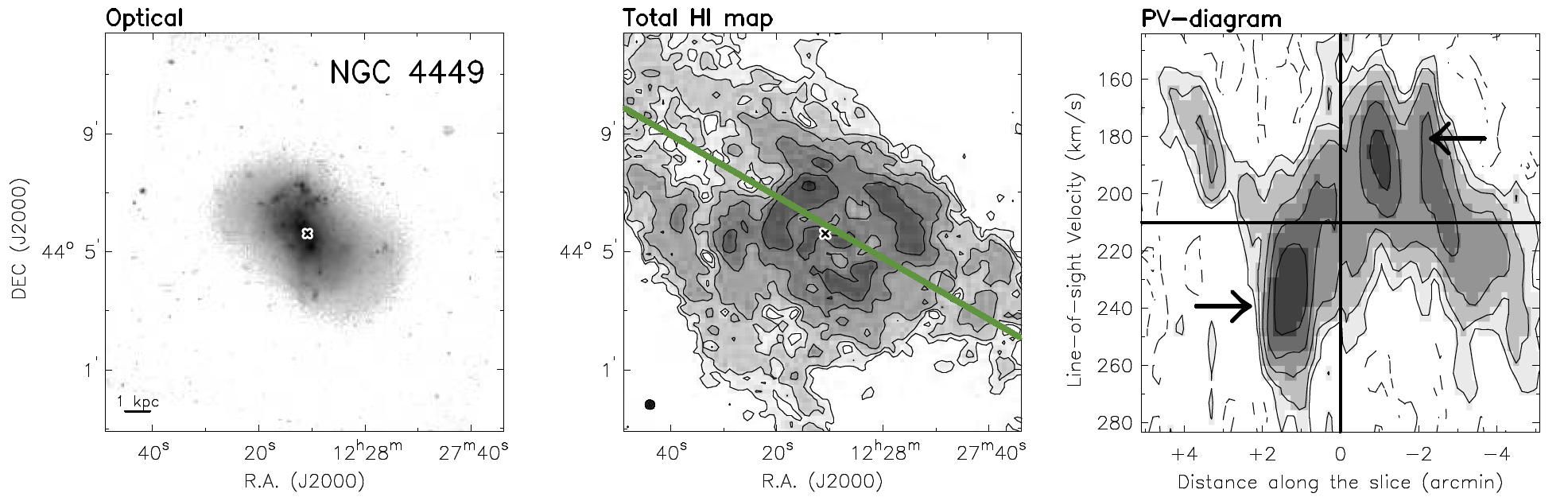}
\caption{{\it Top}: H{\sc i} image ({\it left}) and velocity field ({\it right}) of the Magellanic irregular galaxy NGC~4449 (adapted with permission from
\citealt{Hun98}). The southernmost blob is the companion galaxy DDO~125. The cross is the position
of the stellar stream as given in \citet{Rich12}.
{\it Bottom}: details of the central parts of this galaxy (reproduced with permission from \citealt{Lel14}): a $V$-band optical image ({\it left}), 
the H{\sc i} image ({\it centre}), and a position-velocity diagram ({\it right}) at
a position angle of 60$^{\circ}$ along the green line indicated in the H{\sc i} image showing the inversion of the
line-of-sight velocities in the inner parts}
\index{NGC~4449}
\label{fig:n4449}
\end{figure*}

Irregular galaxies can be intriguing, as is shown by observations of the giant 
magellanic irregular galaxy NGC~4449, whose stellar mass is $1.8\times 10^9\,M_{\odot}$ (\citealt{Mun15}), 
i.e., 89\%  of that of NGC~300.
A large H{\sc i} envelope around this galaxy had already been discovered by \citet{vWBM75}, using the 100\,m
Effelsberg telescope. A $3 \times 3$-point mosaic with the VLA of this galaxy has been made by \citet{Hun98}. The small irregular galaxy DDO~125 is present as the southernmost blob in the images in Fig.~\ref{fig:n4449}. A further faint dwarf, or rather an extended stellar stream, discussed more recently by \citet{Rich12}, \citet{MarD12} and \citet{Tolo16}, is indicated by a cross in the H{\sc i} image: there is no immediate connection between this object and the H{\sc i} features. The outer envelope shows a
regular velocity gradient from north to south, while the body of the main galaxy seen in the visible light
shows a regular velocity
gradient at a position angle of $\sim$60$^\circ$ in the opposite sense. \citet{Lel14} remark that if for NGC~4449 
the rotation velocity of the outer envelope is used this galaxy falls on the baryonic Tully-Fisher relation. This
suggests that it is the outer envelope which traces the dark halo, and that the inner parts, including the visible
galaxy, are not yet settled.

\begin{figure*}[ht]
\centering
\includegraphics[scale=0.375, angle=0.0]{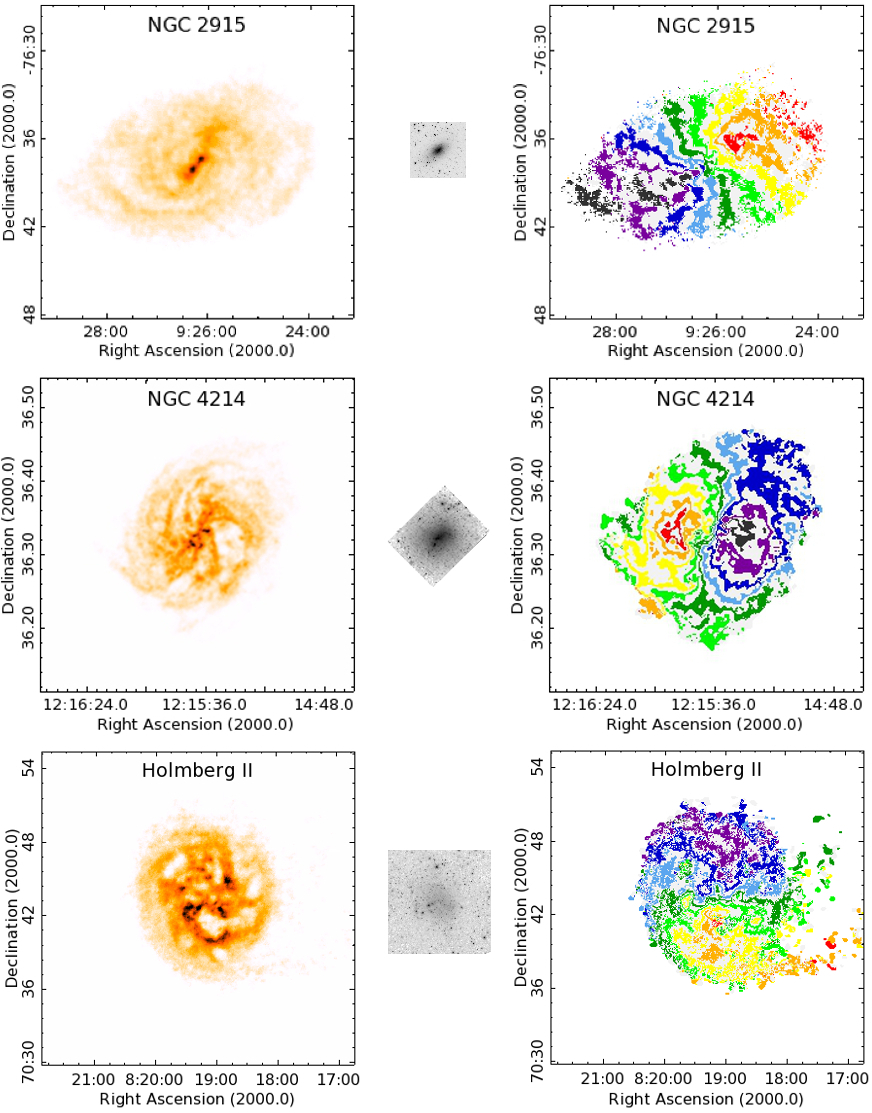}
\caption{{\it Top}: H{\sc i} image ({\it left}) and velocity field ({\it right}) of the blue compact dwarf NGC~2915, made from data kindly supplied by E. Elson, as reported in \citet{Els10}. The 3.6\,$\mu$m image in the middle is on the same 
scale. {\it Middle}:
H{\sc i} image ({\it left}) and velocity field ({\it right}) of NGC~4214, from the THINGS project, and a
3.6\,$\mu$m image. {\it Bottom}: H{\sc i} image ({\it left}) and velocity field ({\it right}) of Holmberg~II, from the ANGST project, and a 3.6\,$\mu$m image. The THINGS, ANGST and 3.6\,$\mu$m data were downloaded from the NED}
\index{NGC~2915}
\label{fig:n2915}
\end{figure*}

Galaxies with such large H{\sc i} envelopes (more than five times larger than the optical radius) are relatively rare,
and only a handful of other cases have been discussed. Blue compact dwarfs can sometimes be surprisingly large in H{\sc i}, as is the case for NGC~2915, which has a relatively
regular H{\sc i} disk (cf. \citealt{Meu96, Els10, Els11a, Els11b}, see Fig.~\ref{fig:n2915}, upper panels), UGC~5288
(\citealt{Zee04}), NGC~3741 (\citealt{Beg05}), ADBS~113845+2008 (\citealt{Can09}),
IZw18 (\citealt{Lel14}), and IIZw40 (\citealt{Bri88}). Note that DDO154 also has a very large H{\sc i} size 
compared to its optical size (\citealt{Car88, dBlo08}). 

Some galaxies look a bit surprising, such as NGC~4214, shown in Fig.~\ref{fig:n2915} (middle panels). If the H{\sc i} is in
a circular disk seen in projection, there is a large misalignment of $\sim$55$^{\circ}$
between the position angle of its  major axis, and the one
derived from the velocity field. \citet{Lel14} fit a tilted ring model to the observations of this galaxy, and
derive a variable inclination, which tends to 0$^{\circ}$ in the outer parts. This seems hard to square 
with the apparent axial ratio of the H{\sc i} distribution.

Also shown in Fig.~\ref{fig:n2915} (lower panels) 
is the galaxy Holmberg II, imaged already by \citet{Puc92}, and later 
in the THINGS and VLA-ANGST projects. From the velocity field, a clear warp can be inferred, 
and the H{\sc i} image also shows a tail towards the west side. Nevertheless, several analyses
have been performed on this galaxy. One of the striking aspects in the H{\sc i} distribution is the presence of 
holes, also discussed for bright galaxies, such as M101 by \citet{Hul88}, and NGC~6946 by
\citet{Boom08}. \citet{Puc92} remark that H{\sc i} holes in
late-type dwarf galaxies are larger than the H{\sc i} holes in large spiral galaxies.
 
\subsection{Velocity Dispersions in Dwarf Irregular Galaxies}
\label{subsec:irr-disp}

A number of recent surveys concern almost uniquely dwarf irregular galaxies. In particular, the 
ANGST survey (\citealt{Ott12}) has observed or re-observed a number a dwarfs with the VLA
at high spectral resolution. An extensive discussion of H{\sc i} gas velocity dispersions based on these 
observations is given in \citet{Stilp13b, Stilp13a}. The dispersions have again been calculated
on the basis of ``super-profiles", derived after derotating the data cube and stacking
the individual profiles. The general shape of the profiles can be described by a double Gaussian,
with a narrow  centre of order $7.2 - 8.5$\,km/s and a wider wing of order $20 - 25$\,km/s. The latter is
presumably due to the influence of star formation, while the former is attributed to turbulence.
\citet{Stilp13b} do not think it is possible to discriminate between the ``cold neutral medium" 
(CNM) and the ``warm neutral medium" (WNM), as done by \citet{Ian12} for the THINGS data, 
since the typical velocity dispersions expected for the typical temperatures associated with 
these do not match. They also argue that the data cubes with robust weighting should be 
used, and that the signal-to-noise threshold matters.
  
Interestingly, thicknesses of the gas layer have been derived, based on a hydrostatic equilibrium
used already in \citet{Ott01} and even before that by \citet{Puc92}.
\citet{Ban11} studied the flaring of the gas layer in four of the THINGS dwarf galaxies,
again using the hydrostatic equilibrium approach. 
For DDO~154, the
disk flares to $\sim$1\,kpc at a radius of $\sim$5\,kpc. This is to be compared with an overall 
thickness derived by \citet{Stilp13a} of 708 $\pm$ 139\,pc based on the method discussed by 
\citet{Ott01}, and a thickness of 650\,pc
calculated by \citet{Ang12} to get a more or less acceptable MOND fit (cf. Sect.~\ref{sec:mond}).
Very recently, \citet{John17} also argue that the H{\sc i} disks in galaxies in the LITTLE THINGS survey 
are thick, rather than thin.
\index{disk thickness}

\section{The Relation Between HI Extent and the Optical Radius}
\label{sec:surveys}

\begin{figure*}[htb]
\centering
\includegraphics[scale=0.329, angle=0.0]{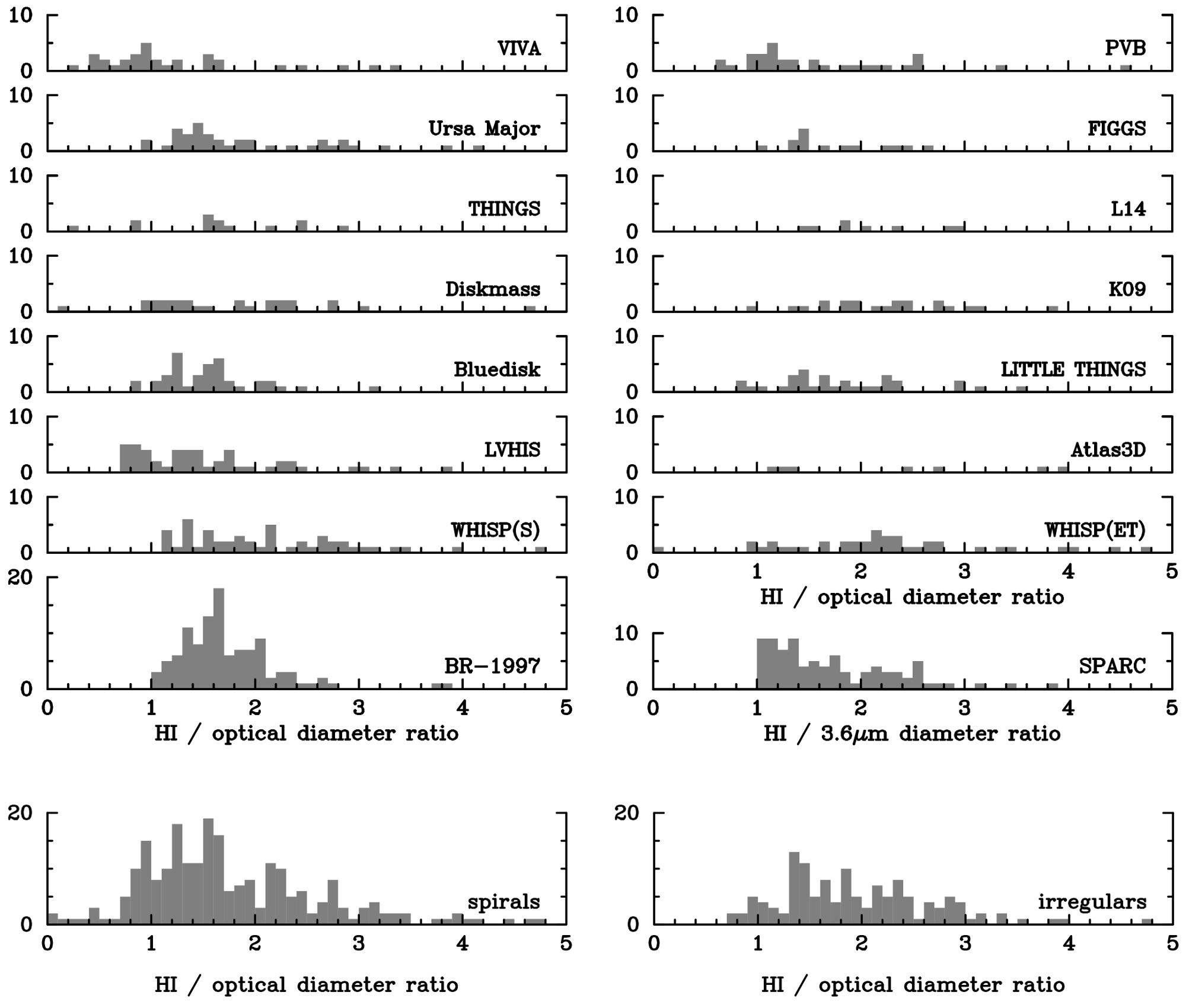}
\caption{Statistics of the ratio of H{\sc i} diameter, evaluated at ${\rm \Sigma}_{\rm HI}$ = 1\,$M_{\odot}$\,pc$^{-2}$, to the optical diameter, following \citet{Wang16}, except for the Bluedisk sample, for which \citet{Wang13} data are used. 
The samples are: BR-1997$-$\citealt{BR97}; WHISP(S)$-$\citealt{Swa02}; LVHIS$-$\citealt{Kor08, WestBK11, WestKB13};
Bluedisk$-$\citealt{Wang13}; Diskmass$-$\citealt{Mar16}; THINGS$-$\citealt{Wal08};
Ursa Major$-$\citealt{Verh01}; VIVA$-$\citealt{Chu09}; WHISP(ET)$-$\citealt{Noo05};
Atlas3D$-$\citealt{Ser12, Ser14}; LITTLE THINGS$-$\citealt{Hun12}; 
K09$-$\citealt{Kov09}; L14$-$\citealt{Lel14}; FIGGS$-$\citealt{Beg08}; PVB$-$\citealt{Pon16}.
Note that several large galaxies are missing from the LVHIS, THINGS, LITTLE THINGS, FIGGS
and VIVA samples, while for the WHISP samples a maximum diameter of 400$^{\prime\prime}$ is imposed,  
on account of missing flux when comparing the interferometric data with single-dish observations or too large in extent compared to the primary beam. For the SPARC sample (\citealt{Lel16}) I determined the 3.6\,$\mu$m radius from the tables associated
with their publication, by interpolating those radial luminosity profiles which reached a depth of $1.0 \,L_{\odot}$pc$^{-2}$, which could be done for 84 of the 174 galaxies in that sample. The {\it bottom} row
refers to the total number of spirals (left) and irregulars (numerical Hubble type $\geq$ 8.5, right) in the samples
studied by \citet{Wang16}}
\index{H{\sc i} to optical diameter ratio}
\label{fig:wang}
\end{figure*}

A specific search for galaxies large in H{\sc i} has been executed by \citet{Bro94} and \citet{BR97}, using 
short observations with the WSRT.  Broeils and Rhee  reported an interesting correlation 
between the H{\sc i} mass and the H{\sc i} size, the latter defined as an isophotal radius at the level of
$1\,M_{\odot}$\,pc$^{-2}$.  More recently, a number of surveys have been executed on various telescopes, typically with sample sizes of order $10 - 60$ galaxies. \citet{Wang16} collected H{\sc i} sizes as 
defined by \citet{BR97} for 437 galaxies---spread over 14 projects---although not for every galaxy an
H{\sc i} size was determined, and there might be a slight overlap in the sense that several galaxies are
in more than one sample. They found again a very tight relationship between the H{\sc i} mass and the H{\sc i} size. 
It is instructive to examine for each survey they 
considered the distribution of the ratio of H{\sc i} to optical 
size, which is shown in Fig.~\ref{fig:wang}. 
The survey done by \citet{BR97} shows a peak at $R_{\rm HI}/R_{\rm opt}\sim1.5$, and they found relatively few
galaxies with extended H{\sc i} disks. The statistics based on the data by \citet{Wang16}
show that the proportion of extended H{\sc i} disks is not too different for spirals compared to irregulars 
(even though statistically there are more irregulars with large $R_{\rm HI}/R_{\rm opt}$), and the
fraction of very extended disks with an H{\sc i} size larger than, e.g., three times the optical size is about 10\%.

Even though \citet{BR97} found
a very good relation between the H{\sc i} mass and the size of the H{\sc i} disk, indicating a roughly constant mean 
H{\sc i} surface
density, there is still no clear indication about 
when we can expect an H{\sc i} disk which is much larger than the
optical disk. This is due to the fact that the amount of H{\sc i} in the extended H{\sc i} disk is usually only a modest fraction
of the total H{\sc i} mass, so that differences are drowned in the ``bigger things are bigger in many quantities" effect.

Not all surveys peak at the same H{\sc i}-to-optical size ratio. This is to 
be expected for the Virgo cluster survey, VIVA (\citealt{Chu09}), since in the centre of that cluster the H{\sc i}
gas in galaxy outskirts is subject to ram-pressure stripping against the hot intergalactic medium (IGM) 
probed by the X-ray
emission. Indeed, several galaxies have been reported undergoing stripping, such as NGC~4522 
(\citealt{Ken04}), NGC~4388, where a long H{\sc i} tail has been observed (\citealt{Oos05}), NGC~4254 and its H{\sc i} tail, including an H{\sc i} dwarf almost
devoid of stars (\citealt{Min07}), and NGC~4569, where \citet{Bose16} report very extended H$\alpha$
emission.

Studies have shown that H{\sc i} sizes of galaxies in a group environment could also be affected by the presence 
of the IGM. Obviously this is the case for compact groups (\citealt{Ver01}), where an evolutionary scenario is suggested depending on the state of the merging. 

For loose groups, the effects are rather more subtle. In a recent extensive study, \citet{Wol16} analysed data for the Ursa Major region, and constructed a complete sample of 1209 galaxies limited in systemic velocity 
($300 \leq V_{\rm LG}\leq 3000$\,km/s), absolute magnitude ($M_{\rm r}\leq -15.3$) and
stellar mass ($M_* \geq 10^8\,M_{\odot}$). They identified six groups with more than three members, 
centred around NGC's 4449, 4258, 4278, 4026, 3938 and 5033, while 74\% of the galaxies in their sample
reside outside those groups. The high resolution interferometer data on the ``Ursa Major cluster" 
(\citealt{Verh01}) and the CVn region (\citealt{Kov09}) are part of this sample, but for other parts of the 
region  single-dish 
data from Jodrell Bank and Arecibo were used. There are H{\sc i}-deficient galaxies associated with the 
densest parts of some of the groups, but not all, as, e.g., NGC~4449 (shown in Fig.~\ref{fig:n4449}), 
which does not seem H{\sc i}-deficient.  
There are also galaxies with excess H{\sc i} mass, usually in regions of low galactic density.

For the Sculptor group, \citet{WestBK11} find that the deep H{\sc i} image of NGC~300 is asymmetric in
the very outer parts, and attribute this to a possible interaction with the IGM in that group. The
process of the formation of warps due to the interaction with the IGM has been 
explored further by \citet{Haan14}.
Deep H{\sc i} imaging data on M83 with the KAT-7 telescope (\citealt{Hea16}) shows sharp edges 
in the outer H{\sc i} there as well, which can be either due to a ram pressure effect with respect to  the surrounding 
IGM, or photoionisation, as discussed below. 

The Void Galaxy Survey (\citealt{Kre11, Kre12}) investigated the properties of H{\sc i} disks in galaxies
in voids, and found that on average the statistics of the size ratio $R_{\rm HI}/R_{\rm opt}$
are roughly similar to those observed by \citet{Swa02} for the late-type WHISP sample. The galaxy 
with the largest ratio, VGS\_12, has an H{\sc i} disk in the polar direction of a small S0-like galaxy 
(cf. \citealt{Sta09}). No star formation is associated with this H{\sc i} gas. However, apart from this, there
is nothing outstanding about the properties of void galaxies with respect to  the field galaxies studied in most 
other surveys.

The outermost H{\sc i} is prone to be ionized by the ultraviolet background radiation,
as  already  suggested by \citet{Sun69}. An early attempt to get limits on 
this has been done for NGC~3198 by \citet{Mal93}, using an unpublished deep VLA H{\sc i} image.
The realisation that there ought to be detectable ionized gas emission around H{\sc i} envelopes, 
which could be used to probe the dark matter at larger radii, has 
led to some further studies of this gas
using Fabry-P\'erot H$\alpha$ observations, but these are technically
very challenging (cf. \citealt{Bla97}). 
 
\citet{Hla11a, Hla11b} report deep H$\alpha$ observations of three Sculptor group galaxies with the wide-field Fabry-P\'erot system on the 36\,cm Marseille Telescope in La Silla. For NGC~247, the H$\alpha$ and H{\sc i} data extend out to $\sim13.5$\,arcmin, barely beyond the Holmberg radius of 12.2\,arcmin. For NGC~300, the 
field of view 
was limited, and the region of the outer, warped H{\sc i} disk not imaged in H$\alpha$. For NGC~253, the H$\alpha$ disk goes out to similar radii (11.5\,arcmin) as the H{\sc i} disk seen with the VLA, but 
faint [N{\sc ii}] emission has been detected out to 19.0 arcmin at the southwest part of the galaxy. 
The kinematic data seem to indicate a declining rotation curve, but the galaxy is heavily perturbed 
there as seen on recent deep optical images, and recent KAT-7 H{\sc i} data show the presence of 
extraplanar cold gas (\citealt{Luc15}). 
Earlier H$\alpha$ observations of the Sculptor group
galaxy NGC~7793 with the same instrument (\citealt{Dic08}) led to the conclusion that the ionizing sources
of the H$\alpha$ disk of this galaxy are likely to be internal, rather than the UV background.  

For more on
the interface between the cold H{\sc i} gas and the hotter circumgalactic medium, see 
the review by Chen (this volume).

\section{Concluding Remarks}

In this review, I have discussed a number of issues related to detailed H{\sc i} imaging of 
the outskirts of nearby spiral and
irregular galaxies done with current interferometers, such as the WSRT, the VLA, the ATCA and the GMRT. 
I have been selective in my topics, and avoided to discuss the issue of star formation in extended H{\sc i} envelopes, since this is covered elsewhere in this volume. Most topics are related to the dark matter 
problem in one way or another. A number of results have been obtained by making case studies
with the emphasis on improved sensitivity, rather than by simply observing more objects from a list.

The Square Kilometer Array project is advancing at great strides, and the associated approach of 
setting a challenging imaging target
requiring the development of new instrumentation to reach it is soon going to bear fruit. Indeed, 
extensive H{\sc i} imaging surveys of relatively nearby galaxies will start next year with SKA pathfinders, such 
as the WSRT Apertif survey, and the SKA precursors ASKAP (the WALLABY survey) and MeerKAT 
(the deep, targeted MHONGOOSE survey, and also the MALS survey). Some of the projects for the
first phase of the SKA telescope,
SKA1, are described in the SKA science book (e.g., \citealt{Bly15, dBlo15}).
All these new surveys will bring fresh data of high quality, which can be brought to bear on the scientific 
problems discussed above and on other, related subjects, and will most likely lead to new insights 
and discoveries. 

\begin{acknowledgement}
I thank Lia Athanassoula for fruitful discussions and a careful reading of the manuscript. I also thank
the editors for inviting me to write this review, and in particular Johan Knapen for his careful
editing. I thank David Malin for sending me the deep image of M83 with the shallower image superimposed, 
and David Mart\'inez-Delgado and Taylor Chonis for providing the image in
the lower right panel of Fig.~\ref{fig:n5055}.
Erwin de Blok supplied the script and Se-Heon Oh and Andrea Macci\`o
supplied data for the construction of Fig.~\ref{fig:corecusp}. Deidre Hunter helped out with 
Fig.~\ref{fig:n4449}. Ed Elson supplied the H{\sc i} data on NGC 2915 from which I could make 
the top panel of Fig.~\ref{fig:n2915}.
Jing Wang clarified an issue with the Bluedisk sample. 

I acknowledge financial support
from the People Programme (Marie Curie Actions) of the European
Union's Seventh Framework Programme FP7/2007-2013/ under REA
grant agreement number PITN-GA-2011-289313 to the DAGAL network,  from the CNES 
(Centre National d'Etudes Spatiales -
France), and from the ``Programme National de
Cosmologie et Galaxies" (PNCG) of CNRS/INSU, France. This research has made use of NASA's Astrophysics Data System. Use was made of the NASA/IPAC Extragalactic Database (NED) which is operated by the Jet Propulsion Laboratory, California Institute of Technology, under contract with the National Aeronautics and Space Administration to produce parts of Fig.~\ref{fig:n5055}, Fig.~\ref{fig:flaring}, Fig.~\ref{fig:gasdisp} and Fig.~\ref{fig:n2915}.
\end{acknowledgement}



\begin{thebibliography}{293}
\providecommand{\natexlab}[1]{#1}
\providecommand{\url}[1]{{#1}}
\providecommand{\urlprefix}{URL }
\expandafter\ifx\csname urlstyle\endcsname\relax
  \providecommand{\doi}[1]{DOI~\discretionary{}{}{}#1}\else
  \providecommand{\doi}{DOI~\discretionary{}{}{}\begingroup
  \urlstyle{rm}\Url}\fi
\providecommand{\eprint}[2][]{\url{#2}}

\bibitem[{{Adams} et~al.(2014){Adams}, {Simon}, {Fabricius}, 
et~al.}]{Ada14}
{Adams}, J.J., {Simon}, J.D., {Fabricius}, M.H., et~al. (2014),  
\apj, 789, 63

\bibitem[{{Allaert} et~al.(2015){Allaert}, {Gentile}, {Baes}, {De Geyter}, 
 et~al.}]{All15}
{Allaert}, F., {Gentile}, G., {Baes}, M., 
 et~al. (2015), \aap, 582, A18

\bibitem[{{Allen} et~al.(1978){Allen}, {Sancisi}, and {Baldwin}}]{All78}
{Allen}, R.J., {Sancisi}, R., {Baldwin}, J.E. (1978), 
\aap, 62, 397--409

\bibitem[{{Angus}(2009)}]{Ang09}
{Angus}, G.W. (2009), 
\mnras, 394, 527--532 

\bibitem[{{Angus} et~al.(2012){Angus}, {van der Heyden}, {Famaey}, 
{Gentile},
  {McGaugh}, and {de Blok}}]{Ang12}
{Angus}, G.W., {van der Heyden}, K.J., {Famaey}, B., {Gentile}, G., {McGaugh}, S.S., {de
  Blok} W.J.G. (2012), 
\mnras, 421, 2598--2609

\bibitem[{{Aniyan} et~al.(2016){Aniyan}, {Freeman}, {Gerhard}, {Arnaboldi}, and
  {Flynn}}]{Ani16}
{Aniyan}, S., {Freeman}, K.C., {Gerhard}, O.E., {Arnaboldi}, M., {Flynn}, C. (2016), 
\mnras,  456, 1484--1494 

\bibitem[{{Aparicio} and {Gallart}(2004)}]{Apa04}
{Aparicio}, A., {Gallart}, C. (2004), 
\aj, 128, 1465--1477

\bibitem[{{Arnaboldi} et~al.(1997){Arnaboldi}, {Oosterloo}, {Combes}, {Freeman},
  and {Koribalski}}]{Arna97}
{Arnaboldi}, M., {Oosterloo}, T., {Combes}, F., {Freeman}, K.C., {Koribalski}, B. (1997),
\aj, 113, 585--598 

\bibitem[{{Athanassoula}(2002)}]{Atha02}
{Athanassoula}, E. (2002), 
\apjl,  569, L83--L86 

\bibitem[{{Athanassoula}(2003)}]{Atha03}
{Athanassoula}, E. (2003), 
\mnras, 341, 1179--1198 

\bibitem[{{Athanassoula}(2008)}]{Atha08}
{Athanassoula}, E. (2008), 
\mnras, 390, L69--L72

\bibitem[{{Athanassoula} et~al.(1987){Athanassoula}, {Bosma}, and
  {Papaioannou}}]{ABP87}
{Athanassoula}, E., {Bosma}, A., {Papaioannou}, S. (1987), 
\aap, 179, 23--40

\bibitem[{{Babcock}(1939)}]{Bab39}
{Babcock}, H.W. (1939), 
Lick Observatory Bulletin, 19, 41--51 

\bibitem[{{Baldwin}(1974)}]{Bal74a}
{Baldwin}, J.E. (1974), 
In: {Shakeshaft}, J.R. (ed.) The Formation and Dynamics of Galaxies, IAU
  Symposium, vol~58, p. 139

\bibitem[{{Baldwin}(1975)}]{Bal74b}
{Baldwin}, J.E. (1975), 
In: {Hayli} A (ed) Dynamics
  of the Solar Systems, IAU Symposium, vol~69, p. 341

\bibitem[{{Banerjee} et~al.(2011){Banerjee}, {Jog}, {Brinks}, and
  {Bagetakos}}]{Ban11}
{Banerjee}, A., {Jog}, C.J., {Brinks}, E., {Bagetakos}, I. (2011), 
\mnras, 415, 687--694

\bibitem[{{Barnab{\`e}} et~al.(2012){Barnab{\`e}}, {Dutton}, {Marshall},
  {Auger}, {Brewer}, {Treu}, {Bolton}, {Koo}, and {Koopmans}}]{Bar12}
{Barnab{\`e}}, M,, {Dutton}, A.A., {Marshall}, P.J., {Auger}, M.W., {Brewer}, B.J., {Treu}, T.,
  {Bolton}, A.S., {Koo}, D.C., {Koopmans}, L.V.E. (2012), 
\mnras, 423, 1073--1088

\bibitem[{{Beck}(2007)}]{Beck07}
{Beck}, R. (2007), 
\aap, 470, 539--556

\bibitem[{{Becquaert} and {Combes}(1997)}]{Bec97}
{Becquaert}, J.F., {Combes}, F. (1997), 
\aap, 325, 41--56 

\bibitem[{{Begeman} et~al.(1991){Begeman}, {Broeils}, and {Sanders}}]{Beg91}
{Begeman}, K.G., {Broeils}, A.H., {Sanders}, R.H. (1991), 
\mnras, 249, 523--537

\bibitem[{{Begum} et~al.(2005){Begum}, {Chengalur}, and {Karachentsev}}]{Beg05}
{Begum}, A., {Chengalur}, J.N., {Karachentsev}, I.D. (2005), 
\aap, 433, L1--L4 

\bibitem[{{Begum} et~al.(2008){Begum}, {Chengalur}, {Karachentsev}, {Sharina},
  and {Kaisin}}]{Beg08}
{Begum}, A., {Chengalur}, J.N., {Karachentsev}, I.D., {Sharina}, M.E., {Kaisin}, S.S. (2008),
\mnras, 386, 1667--1682

\bibitem[{{Bershady} et~al.(2005){Bershady}, {Andersen}, {Verheijen},
  {Westfall}, {Crawford}, and {Swaters}}]{Ber05}
{Bershady}, M.A., {Andersen}, D.R., {Verheijen}, M.A.W., {Westfall}, K.B., {Crawford}, S.M.,
  {Swaters}, R.A. (2005), 
\apjs, 156, 311--344

\bibitem[{{Bershady} et~al.(2010{\natexlab{a}}){Bershady}, {Verheijen},
  {Swaters}, {Andersen}, {Westfall}, and {Martinsson}}]{Ber10a}
{Bershady}, M.A., {Verheijen}, M.A.W., {Swaters}, R.A., {Andersen}, D.R., {Westfall}, K.B.,
  {Martinsson}, T. (2010{\natexlab{a}}), 
\apj, 716, 198--233

\bibitem[{{Bershady} et~al.(2010{\natexlab{b}}){Bershady}, {Verheijen},
  {Westfall}, {Andersen}, {Swaters}, and {Martinsson}}]{Ber10b}
{Bershady}, M.A., {Verheijen}, M.A.W., {Westfall}, K.B., {Andersen}, D.R., {Swaters}, R.A.,
  {Martinsson}, T. (2010{\natexlab{b}}), 
\apj, 716, 234--268 

\bibitem[{{Bershady} et~al.(2011){Bershady}, {Martinsson}, {Verheijen},
  {Westfall}, {Andersen}, and {Swaters}}]{Ber11}
{Bershady}, M.A., {Martinsson}, T.P.K., {Verheijen}, M.A.W., {Westfall}, K.B., {Andersen}, D.R.,
  {Swaters}, R.A. (2011),
\apjl, 739, L47 

\bibitem[{{Bertone} and {Hooper}(2016)}]{Ber16}
{Bertone}, G., {Hooper}, D. (2016), 
ArXiv e-prints, \eprint{1605.04909}

\bibitem[{{Binney} and {de Vaucouleurs}(1981)}]{BdeVauc81}
{Binney}, J., {de Vaucouleurs}, G. (1981), 
\mnras, 194, 679--691

\bibitem[{{Blais-Ouellette} et~al.(2004){Blais-Ouellette}, {Amram}, {Carignan},
  and {Swaters}}]{Blais04}
{Blais-Ouellette}, S., {Amram}, P., {Carignan}, C., {Swaters}, R. (2004), 
\aap, 420, 147--161

\bibitem[{{Bland-Hawthorn} et~al.(1997){Bland-Hawthorn}, {Freeman}, and
  {Quinn}}]{Bla97}
{Bland-Hawthorn}, J., {Freeman}, K.C., {Quinn}, P.J. (1997), 
\apj, 490, 143--155

\bibitem[{{Blyth} et~al.(2015){Blyth}, {van der Hulst}, {Verheijen}, et~al.}]{Bly15}
{Blyth}, S., {van der Hulst}, J.M., {Verheijen}, M.A.W., et~al.
(2015), 
Advancing Astrophysics with the Square Kilometre Array (AASKA14), 128

\bibitem[{{Boomsma} et~al.(2008){Boomsma}, {Oosterloo}, {Fraternali}, {van der
  Hulst}, and {Sancisi}}]{Boom08}
{Boomsma}, R., {Oosterloo}, T.A., {Fraternali}, F., {van der Hulst}, J.M., {Sancisi}, R.
  (2008), 
  \aap, 490, 555--570

\bibitem[{{Boselli} et~al.(2016){Boselli}, {Cuillandre}, {Fossati}, {Boissier},
  {Bomans}, {Consolandi}, {Anselmi}, {Cortese}, {C{\^o}t{\'e}}, {Durrell},
  {Ferrarese}, {Fumagalli}, {Gavazzi}, {Gwyn}, {Hensler}, {Sun}, and
  {Toloba}}]{Bose16}
{Boselli}, A., {Cuillandre}, J.C., {Fossati}, M., et~al.
(2016), 
\aap, 587, A68

\bibitem[{{Bosma}(1978)}]{Bos78}
{Bosma}, A. (1978) {The distribution and kinematics of neutral hydrogen in spiral
  galaxies of various morphological types}. PhD Thesis, Groningen
  University

\bibitem[{{Bosma}(1981{\natexlab{a}})}]{Bos81a}
{Bosma}, A. (1981{\natexlab{a}}), 
\aj, 86, 1791--1824 

\bibitem[{{Bosma}(1981{\natexlab{b}})}]{Bos81b}
{Bosma}, A. (1981{\natexlab{b}}), 
\aj, 86, 1825--1846

\bibitem[{{Bosma}(1994)}]{Bos94}
{Bosma}, A. (1994), {Kinematics and Dynamics of Dwarf Spirals and Irregulars}. In:
  {Meylan} G, {Prugniel} P (eds) European Southern Observatory Conference and
  Workshop Proceedings, European Southern Observatory Conference and Workshop
  Proceedings, vol~49, pp 187--196

\bibitem[{{Bosma}(1999)}]{Bos99}
{Bosma}, A. (1999), {Dark Matter in Disk Galaxies}. In: {Merritt} DR, {Valluri} M,
  {Sellwood} JA (eds) Galaxy Dynamics - A Rutgers Symposium, Astronomical
  Society of the Pacific Conference Series, vol 182

\bibitem[{{Bosma} and {van der Kruit}(1979)}]{Bos79}
{Bosma}, A., {van der Kruit}, P.C. (1979), 
\aap, 79, 281--286

\bibitem[{{Bottema}(1993)}]{Bot93}
{Bottema}, R. (1993),
\aap, 275, 16

\bibitem[{{Bottema}(1997)}]{Bot97}
{Bottema}, R. (1997), 
\aap, 328, 517--525

\bibitem[{{Bottema} et~al.(1986){Bottema}, {Shostak}, and {van der
  Kruit}}]{Bot86}
{Bottema}, R., {Shostak}, G.S., {van der Kruit}, P.C. (1986),
\aap, 167, 34--40

\bibitem[{{Bottema} et~al.(2002){Bottema}, {Pesta{\~n}a}, {Rothberg}, and
  {Sanders}}]{Bot02}
{Bottema}, R., {Pesta{\~n}a}, J.L.G., {Rothberg}, B., {Sanders}, R.H. (2002), 
\aap, 393, 453--460

\bibitem[{{Boulanger} and {Viallefond}(1992)}]{Bou92}
{Boulanger}, F., {Viallefond}, F. (1992), 
\aap, 266, 37--56

\bibitem[{{Bovy} and {Rix}(2013)}]{Bov13}
{Bovy}, J., {Rix}, H.-W. (2013), 
\apj, 779, 115

\bibitem[{{Braun}(1997)}]{Bra97}
{Braun}, R. (1997), 
\apj, 484, 637--655

\bibitem[{{Braun} and {Thilker}(2004)}]{Bra04}
{Braun}, R., {Thilker}, D.A. (2004), 
\aap, 417, 421--435

\bibitem[{{Braun} et~al.(2009){Braun}, {Thilker}, {Walterbos}, and
  {Corbelli}}]{Bra09}
{Braun}, R., {Thilker}, D.A., {Walterbos}, R.A.M., {Corbelli}, E. (2009), 
\apj, 695, 937--953

\bibitem[{{Briggs}(1990)}]{Bri90}
{Briggs}, F.H. (1990),
\apj, 352, 15--29

\bibitem[{{Brinks} and {Klein}(1988)}]{Bri88}
{Brinks}, E., {Klein}, U. (1988), 
\mnras, 231, 63p--67p

\bibitem[{{Broeils}(1992)}]{Bro92}
{Broeils}, A.H. (1992), 
\aap, 256, 19--32

\bibitem[{{Broeils} and {Rhee}(1997)}]{BR97}
{Broeils}, A.H., {Rhee}, M.H. (1997), 
\aap, 324, 877--887

\bibitem[{{Broeils} and {van Woerden}(1994)}]{Bro94}
{Broeils}, A.H., {van Woerden}, H. (1994), 
\aaps, 107, 129--176

\bibitem[{{Burbidge} et~al.(1960){Burbidge}, {Burbidge}, and
  {Prendergast}}]{Bur60}
{Burbidge}, E.M., {Burbidge}, G.R., {Prendergast}, K.H. (1960), 
\apj, 131, 282

\bibitem[{{Burbidge}(1975)}]{Bur75}
{Burbidge}, G. (1975), 
\apjl, 196, L7--L10

\bibitem[{{Burke}(1957)}]{Bur57}
{Burke}, B.F. (1957), 
\aj, 62, 90

\bibitem[{{Cannon} et~al.(2009){Cannon}, {Salzer}, and {Rosenberg}}]{Can09}
{Cannon}, J.M., {Salzer}, J.J., {Rosenberg}, J.L. (2009), 
\apj, 696, 2104--2114

\bibitem[{{Carignan} and {Freeman}(1988)}]{Car88}
{Carignan}, C., {Freeman}, K.C. (1988), 
\apjl, 332, L33--L36

\bibitem[{{Carignan} et~al.(2013){Carignan}, {Frank}, {Hess}, {Lucero},
  {Randriamampandry}, {Goedhart}, and {Passmoor}}]{Car13}
{Carignan}, C., {Frank}, B.S., {Hess}, K.M., {Lucero}, D.M., {Randriamampandry}, T.H.,
  {Goedhart}, S., {Passmoor}, S.S. (2013), 
\aj, 146, 48

\bibitem[{{Chung} et~al.(2009){Chung}, {van Gorkom}, {Kenney}, {Crowl}, and
  {Vollmer}}]{Chu09}
{Chung}, A., {van Gorkom}, J.H., {Kenney}, J.D.P., {Crowl}, H., {Vollmer}, B. (2009), 
\aj, 138, 1741--1816

\bibitem[{{Courteau} and {Dutton}(2015)}]{Cou15}
{Courteau}, S., {Dutton}, A.A. (2015), 
\apjl, 801, L20

\bibitem[{{Dahlem} et~al.(2006){Dahlem}, {Lisenfeld}, and {Rossa}}]{Dah06}
{Dahlem}, M., {Lisenfeld}, U., {Rossa}, J. (2006), 
\aap, 457, 121--131

\bibitem[{{Dalton}(2016)}]{Dal16}
{Dalton}, G. (2016), 
In: {Skillen}, I., {Barcells}, M., {Trager}, S. (eds.) 
Multi-Object Spectroscopy in the Next Decade: Big Questions, Large Surveys, and Wide
  Fields, ASP Conference Series, vol 507, p~97

\bibitem[{{Davies}(1974)}]{Dav74}
{Davies}, R.D. (1974), 
  In: {Shakeshaft}, J.R. (ed.) The Formation and Dynamics of Galaxies, IAU
  Symposium, vol~58, pp 119--128

\bibitem[{{de Blok} et~al.(2015){de Blok}, {Fraternali}, {Heald}, {Adams},
  {Bosma}, and {Koribalski}}]{dBlo15}
{de Blok}, E., {Fraternali}, F., {Heald}, G., {Adams}, B., {Bosma}, A., {Koribalski}, B.
  (2015), 
  Advancing Astrophysics with the Square Kilometre Array (AASKA14), 129

\bibitem[{{de Blok}(2010)}]{dBlo10}
{de Blok}, W.J.G. (2010), 
Advances in Astronomy
  2010:789293

\bibitem[{{de Blok} and {Bosma}(2002)}]{dBlo02}
{de Blok}, W.J.G., {Bosma}, A. (2002), 
\aap, 385, 816--846 

\bibitem[{{de Blok} et~al.(2001){de Blok}, {McGaugh}, {Bosma}, and
  {Rubin}}]{dBlo01}
{de Blok}, W.J.G., {McGaugh}, S.S., {Bosma}, A., {Rubin}, V.C. (2001), 
\apjl, 552, L23--L26

\bibitem[{{de Blok} et~al.(2003){de Blok}, {Bosma}, and {McGaugh}}]{dBlo03}
{de Blok}, W.J.G., {Bosma}, A., {McGaugh}, S.S. (2003), 
\mnras, 340, 657--678 

\bibitem[{{de Blok} et~al.(2008){de Blok}, {Walter}, {Brinks}, {Trachternach},
  {Oh}, and {Kennicutt}}]{dBlo08}
{de Blok}, W.J.G., {Walter}, F., {Brinks}, E., {Trachternach}, C., {Oh}, S.H., {Kennicutt},
  R.C. Jr. (2008), 
\aj, 136, 2648-2719

\bibitem[{{de Blok} et~al.(2014{\natexlab{a}}){de Blok}, {J{\'o}zsa},
  {Patterson}, {Gentile}, {Heald}, {J{\"u}tte}, {Kamphuis}, {Rand}, {Serra},
  and {Walterbos}}]{dBlo14}
{de Blok}, W.J.G., {J{\'o}zsa}, G.I.G., {Patterson}, M., {Gentile}, G., {Heald}, G.H.,
  {J{\"u}tte}, E., {Kamphuis}, P., {Rand}, R.J., {Serra}, P., {Walterbos}, R.A.M.
  (2014{\natexlab{a}}), 
\aap, 566, A80

\bibitem[{{de Blok} et~al.(2014{\natexlab{b}}){de Blok}, {Keating}, {Pisano},
  {Fraternali}, {Walter}, {Oosterloo}, {Brinks}, {Bigiel}, and
  {Leroy}}]{dBloP14}
{de Blok}, W.J.G., {Keating}, .K.M, {Pisano}, D.J., {Fraternali}, F., {Walter}, F.,
  {Oosterloo}, T., {Brinks}, E., {Bigiel}, F., {Leroy}, A. (2014{\natexlab{b}}), 
\aap, 569, A68 

\bibitem[{{de Vaucouleurs}(1958)}]{Vauc58}
{de Vaucouleurs}, G. (1958), 
\apj, 128, 465

\bibitem[{{de Vaucouleurs} et~al.(1976){de Vaucouleurs}, {de Vaucouleurs}, and
  {Corwin}}]{Vauc76}
{de Vaucouleurs}, G., {de Vaucouleurs}, A., {Corwin}, J.R. (1976), 
{Second reference catalogue of bright galaxies}, 
Austin: University of Texas Press

\bibitem[{{Dicaire} et~al.(2008){Dicaire}, {Carignan}, {Amram}, {Marcelin},
  {Hlavacek-Larrondo}, {de Denus-Baillargeon}, {Daigle}, and
  {Hernandez}}]{Dic08}
{Dicaire}, I., {Carignan}, C., {Amram}, P., {Marcelin}, M., {Hlavacek-Larrondo}, J., {de
  Denus-Baillargeon}, M.M., {Daigle}, O., {Hernandez}, O. (2008), 
\aj, 135, 2038--2047

\bibitem[{{Dickey} et~al.(1990){Dickey}, {Hanson}, and {Helou}}]{Dick90}
{Dickey}, J.M., {Hanson}, M.M., {Helou}, G. (1990), 
\apj, 352, 522--531 

\bibitem[{{Dutton} et~al.(2013){Dutton}, {Treu}, {Brewer}, {Marshall}, {Auger},
  {Barnab{\`e}}, {Koo}, {Bolton}, and {Koopmans}}]{Dut13}
{Dutton}, A.A., {Treu}, T., {Brewer}, B.J., {Marshall}, P.J., {Auger}, M.W., {Barnab{\`e}}, M.,
  {Koo}, D.C., {Bolton}, A.S., {Koopmans}, L.V.E. (2013), 
\mnras, 428, 3183--3195

\bibitem[{{Efstathiou} et~al.(1982){Efstathiou}, {Lake}, and
  {Negroponte}}]{ELN82}
{Efstathiou}, G., {Lake}, G., {Negroponte}, J. (1982), 
\mnras, 199, 1069--1088

\bibitem[{{Einasto} et~al.(1974){Einasto}, {Kaasik}, and {Saar}}]{EKS74}
{Einasto}, J., {Kaasik}, A., {Saar}, E. (1974), 
\nat, 250, 309--310

\bibitem[{{Elson} et~al.(2010){Elson}, {de Blok}, and {Kraan-Korteweg}}]{Els10}
{Elson}, E.C., {de Blok}, W.J.G., {Kraan-Korteweg}, R.C. (2010), 
\mnras, 404, 2061--2076

\bibitem[{{Elson} et~al.(2011{\natexlab{a}}){Elson}, {de Blok}, and
  {Kraan-Korteweg}}]{Els11a}
{Elson}, E.C., {de Blok}, W.J.G., {Kraan-Korteweg}, R.C. (2011{\natexlab{a}}), 
\mnras, 411, 200--210

\bibitem[{{Elson} et~al.(2011{\natexlab{b}}){Elson}, {de Blok}, and
  {Kraan-Korteweg}}]{Els11b}
{Elson}, E.C., {de Blok}, W.J.G., {Kraan-Korteweg}, R.C. (2011{\natexlab{b}}),
\mnras, 415, 323--332

\bibitem[{{Emerson} and {Baldwin}(1973)}]{EB73}
{Emerson}, D.T., {Baldwin}, J.E. (1973), 
\mnras, 165, 9P--13P

\bibitem[{{Ewen} and {Purcell}(1951)}]{EP51}
{Ewen}, H.I., {Purcell}, E.M. (1951), 
\nat, 168, 356

\bibitem[{{Faber} and {Gallagher}(1979)}]{Fab79}
{Faber}, S.M., {Gallagher}, J.S. (1979), 
\araa, 17, 135--187

\bibitem[{{Famaey} and {McGaugh}(2012)}]{Fam12}
{Famaey}, B., {McGaugh}, S.S. (2012), 
{Modified Newtonian Dynamics (MOND):
  Observational Phenomenology and Relativistic Extensions},
Living Reviews in Relativity, 15

\bibitem[{{Ferguson} and {Mackey}(2016)}]{Fer16}
{Ferguson}, A.M.N., {Mackey}, A.D. (2016), 
In: {Newberg}, H.J., {Carlin}, J.L. (eds.),
Astrophysics and Space Science
  Library, vol 420, p 191

\bibitem[{{Flores} and {Primack}(1994)}]{Flo94}
{Flores}, R.A., {Primack}, J.R. (1994), 
\apjl, 427, L1--L4 

\bibitem[{{Fontenrose}(1973)}]{Font73}
{Fontenrose}, R. (1973), 
Journal for the History of Astronomy, 4, 145--158

\bibitem[{{Fragkoudi} et~al.(2016){Fragkoudi}, {Athanassoula}, and
  {Bosma}}]{Frag16}
{Fragkoudi}, F., {Athanassoula}, E., {Bosma}, A. (2016), 
ArXiv e-prints, \eprint{1605.05754}

\bibitem[{{Fraternali}(2014)}]{Fra14}
{Fraternali}, F. (2014), 
In: {Feltzing}, S.,
  {Zhao}, G., {Walton}, N.A., {Whitelock}, P. (eds.) Setting the scene for Gaia and
  LAMOST, IAU Symposium, vol 298, pp 228--239

\bibitem[{{Fraternali}(2016)}]{Fra16}
{Fraternali}, F. (2016), {Gas Accretion via Condensation and Fountains}. ArXiv
  e-prints \eprint{1612.00477}

\bibitem[{{Fraternali} and {Binney}(2006)}]{Fra06}
{Fraternali}, F., {Binney}, J.J. (2006), 
\mnras, 366, 449--466

\bibitem[{{Fraternali} and {Binney}(2008)}]{Fra08}
{Fraternali}, F., {Binney}, J.J. (2008), 
\mnras, 386, 935--944 

\bibitem[{{Fraternali} et~al.(2001){Fraternali}, {Oosterloo}, {Sancisi}, and
  {van Moorsel}}]{Fra01}
{Fraternali}, F., {Oosterloo}, T., {Sancisi}, R., {van Moorsel}, G. (2001),
\apjl, 562, L47--L50

\bibitem[{{Fraternali} et~al.(2002){Fraternali}, {van Moorsel}, {Sancisi}, and
  {Oosterloo}}]{Fra02}
{Fraternali}, F., {van Moorsel}, G., {Sancisi}, R., {Oosterloo}, T. (2002), 
\aj, 123, 3124--3140

\bibitem[{{Fraternali} et~al.(2005){Fraternali}, {Oosterloo}, {Sancisi}, and
  {Swaters}}]{Fra05}
{Fraternali}, F., {Oosterloo}, T.A., {Sancisi}, R., {Swaters}, R. (2005), 
In: {Braun}, R. (ed.), Extra-Planar Gas, Astronomical Society of the Pacific Conference Series,
  vol 331, p 239

\bibitem[{{Freeman}(1970)}]{kcf70}
{Freeman}, K.C. (1970), 
\apj, 160, 811

\bibitem[{{Fuchs}(1999)}]{Fuc99}
{Fuchs}, B. (1999), 
In: {Merritt}, D.R., {Valluri}, M., {Sellwood}, J.A. (eds.), Galaxy Dynamics -
  A Rutgers Symposium, Astronomical Society of the Pacific Conference Series,
  vol 182, 365

\bibitem[{{Galaz} et~al.(2015){Galaz}, {Milovic}, {Suc}, {Busta}, {Lizana},
  {Infante}, and {Royo}}]{Gal15}
{Galaz}, G., {Milovic}, C., {Suc}, V., {Busta}, L., {Lizana}, G., {Infante}, L., {Royo}, S.
  (2015), 
\apjl, 815, L29

\bibitem[{{Garc{\'{\i}}a-Ruiz} et~al.(2002){Garc{\'{\i}}a-Ruiz}, {Sancisi}, and
  {Kuijken}}]{Gar02}
{Garc{\'{\i}}a-Ruiz}, I., {Sancisi}, R., {Kuijken}, K. (2002), 
\aap, 394, 769--789

\bibitem[{{Gentile} et~al.(2010){Gentile}, {Baes}, {Famaey}, and {van
  Acoleyen}}]{Gen10}
{Gentile}, G., {Baes}, M., {Famaey}, B., {van Acoleyen}, K. (2010), 
\mnras, 406, 2493--2503

\bibitem[{{Gentile} et~al.(2011){Gentile}, {Famaey}, and {de Blok}}]{Gen11}
{Gentile}, G., {Famaey}, B., {de Blok}, W.J.G. (2011), 
\aap, 527, A76 

\bibitem[{{Gentile} et~al.(2013){Gentile}, {J{\'o}zsa}, {Serra}, {Heald}, {de
  Blok}, {Fraternali}, {Patterson}, {Walterbos}, and {Oosterloo}}]{Gen13}
{Gentile}, G., {J{\'o}zsa}, G.I.G., {Serra}, P., {Heald}, G.H., {de Blok}, W.J.G.,
  {Fraternali}, F., {Patterson}, M.T., {Walterbos}, R.A.M., {Oosterloo}, T. (2013),
\aap, 554, A125

\bibitem[{{Goad} and {Roberts}(1981)}]{Goa81}
{Goad}, J.W., {Roberts}, M.S. (1981), 
\apj, 250, 79--86 

\bibitem[{{Gum} et~al.(1960){Gum}, {Kerr}, and {Westerhout}}]{Gum60}
{Gum}, C.S., {Kerr}, F.J., {Westerhout}, G. (1960), 
\mnras, 121, 132

\bibitem[{{Haan} and {Braun}(2014)}]{Haan14}
{Haan}, S., {Braun}, R. (2014), 
\mnras, 440, L21--L25 

\bibitem[{{Heald}(2015)}]{Hea15}
{Heald}, G. (2015), {The Wsrt Halogas Survey}. In: {Ziegler}, B.L., {Combes}, F.,
  {Dannerbauer}, H., {Verdugo}, M. (eds.) Galaxies in 3D across the Universe, IAU
  Symposium, vol 309, 69--72

\bibitem[{{Heald} et~al.(2016){Heald}, {de Blok}, {Lucero}, {Carignan},
  {Jarrett}, {Elson}, {Oozeer}, {Randriamampandry}, and {van Zee}}]{Hea16}
{Heald}, G., {de Blok}, W.J.G., {Lucero}, D., {Carignan}, C., {Jarrett}, T., {Elson}, E.,
  {Oozeer}, N., {Randriamampandry}, T.H., {van Zee}, L. (2016), 
\mnras, 462, 1238--1255 

\bibitem[{{Hindman}(1967)}]{Hin67}
{Hindman}, J.V. (1967), 
Australian Journal of Physics, 20, 147

\bibitem[{{Hlavacek-Larrondo} et~al.(2011{\natexlab{a}}){Hlavacek-Larrondo},
  {Carignan}, {Daigle}, {de Denus-Baillargeon}, {Marcelin}, {Epinat}, and
  {Hernandez}}]{Hla11a}
{Hlavacek-Larrondo}, J., {Carignan}, C., {Daigle}, O., {de Denus-Baillargeon}, M.M.,
  {Marcelin}, M., {Epinat}, B., {Hernandez}, O. (2011{\natexlab{a}}), 
\mnras, 411, 71--84 

\bibitem[{{Hlavacek-Larrondo} et~al.(2011{\natexlab{b}}){Hlavacek-Larrondo},
  {Marcelin}, {Epinat}, {Carignan}, {de Denus-Baillargeon}, {Daigle}, and
  {Hernandez}}]{Hla11b}
{Hlavacek-Larrondo}, J., {Marcelin}, M., {Epinat}, B., {Carignan}, C., {de
  Denus-Baillargeon}, M.M., {Daigle}, O., {Hernandez}, O. (2011{\natexlab{b}}), 
\mnras, 416, 509--521

\bibitem[{{Hodges-Kluck} and {Bregman}(2013)}]{Hod13}
{Hodges-Kluck}, E.J., {Bregman}, J.N. (2013), 
\apj, 762, 12

\bibitem[{{Hohl}(1971)}]{Hohl71}
{Hohl}, F. (1971), 
\apj, 168, 343

\bibitem[{{Holmberg}(1958)}]{Hol58}
{Holmberg}, E. (1958), 
  Meddelanden fran Lunds Astronomiska Observatorium Serie II, 136, 1

\bibitem[{{Hunter} et~al.(1998){Hunter}, {Wilcots}, {van Woerden}, {Gallagher},
  and {Kohle}}]{Hun98}
{Hunter}, D.A., {Wilcots}, E.M., {van Woerden}, H., {Gallagher}, J.S., {Kohle}, S. (1998),
\apjl, 495, L47--L50

\bibitem[{{Hunter} et~al.(2012){Hunter}, {Ficut-Vicas}, {Ashley}, {Brinks},
  {Cigan}, {Elmegreen}, {Heesen}, {Herrmann}, {Johnson}, {Oh}, {Rupen},
  {Schruba}, {Simpson}, {Walter}, {Westpfahl}, {Young}, and {Zhang}}]{Hun12}
{Hunter}, D.A., {Ficut-Vicas}, D., {Ashley}, T., et~al.
(2012),
\aj, 144, 134

\bibitem[{{Ianjamasimanana} et~al.(2012){Ianjamasimanana}, {de Blok}, {Walter},
  and {Heald}}]{Ian12}
{Ianjamasimanana}, R., {de Blok}, W.J.G., {Walter}, F., {Heald}, G.H. (2012), 
\aj, 144, 96

\bibitem[{{Ianjamasimanana} et~al.(2015){Ianjamasimanana}, {de Blok}, {Walter},
  {Heald}, {Cald{\'u}-Primo}, and {Jarrett}}]{Ian15}
{Ianjamasimanana}, R., {de Blok}, W.J.G., {Walter}, F., {Heald}, G.H., {Cald{\'u}-Primo},
  A., {Jarrett}, T.H. (2015), 
\aj, 150, 47 

\bibitem[{{Johnson} et~al.(2017){Johnson}, {Hunter}, {Kamphuis}, and
  {Wang}}]{John17}
{Johnson}, M.C., {Hunter}, D.A., {Kamphuis}, P., {Wang}, J. (2017), 
\mnras, 465, L49--L53

\bibitem[{{Kahn} and {Woltjer}(1959)}]{KW59}
{Kahn}, F.D., {Woltjer}, L. (1959), 
\apj, 130, 705 

\bibitem[{{Kalberla} and {Dedes}(2008)}]{Kal08}
{Kalberla}, P.M.W., {Dedes}, L. (2008),
\aap, 487, 951--963

\bibitem[{{Kalnajs}(1983)}]{Kal83}
{Kalnajs}, A. (1983), {Mass distribution and dark halos: Discussion}. In:
  {Athanassoula}, E. (ed.) Internal Kinematics and Dynamics of Galaxies, IAU
  Symposium, vol 100, 87--88

\bibitem[{{Kamphuis} and {Sancisi}(1993)}]{Kam93b}
{Kamphuis}, J., {Sancisi}, R. (1993), 
\aap, 273, L31

\bibitem[{{Kamphuis}(1993)}]{Kam93a}
{Kamphuis}, J.J. (1993), {Neutral Hydrogen in nearby Spiral Galaxies: Holes and
  High Velocity Clouds}. PhD Thesis, University of Groningen

\bibitem[{{Karachentsev} et~al.(1999){Karachentsev}, {Karachentseva}, {Kudrya},
  {Sharina}, and {Parnovskij}}]{Kar99}
{Karachentsev}, I.D., {Karachentseva}, V.E., {Kudrya}, Y.N., {Sharina}, M.E., {Parnovskij},
  S.L. (1999), {The revised Flat Galaxy Catalogue.} Bulletin of the Special
  Astrophysics Observatory 47, \eprint{astro-ph/0305566}

\bibitem[{{Karachentsev} et~al.(2013){Karachentsev}, {Makarov}, and
  {Kaisina}}]{Kar13}
{Karachentsev}, I.D., {Makarov}, D.I., {Kaisina}, E.I. (2013), {Updated Nearby Galaxy
  Catalog}. \aj, 145, 101 

\bibitem[{{Kenney} et~al.(2004){Kenney}, {van Gorkom}, and {Vollmer}}]{Ken04}
{Kenney}, J.D.P., {van Gorkom}, J.H., {Vollmer}, B. (2004), 
\aj, 127, 3361--3374

\bibitem[{{Kent}(1986)}]{Kent86}
{Kent}, S.M. (1986), 
\aj, 91, 1301--1327

\bibitem[{{Kent}(1987)}]{Kent87}
{Kent}, S.M. (1987), 
\aj, 93, 816--832 

\bibitem[{{Kent}(1988)}]{Kent88}
{Kent}, S.M. (1988), 
\aj, 96, 514--527

\bibitem[{{Kerr}(1957)}]{Kerr57}
{Kerr}, F.J. (1957), 
\aj, 62, 93--93

\bibitem[{{Khoperskov} et~al.(2014){Khoperskov}, {Moiseev}, {Khoperskov}, and
  {Saburova}}]{Khop14}
{Khoperskov}, S.A., {Moiseev}, A.V., {Khoperskov}, A.V., {Saburova}, A.S. (2014), 
\mnras, 441, 2650--2662

\bibitem[{{Kim} et~al.(2014){Kim}, {Gadotti}, {Sheth}, {Athanassoula}, {Bosma},
  {Lee}, {Madore}, {Elmegreen}, {Knapen}, {Zaritsky}, {Ho}, {Comer{\'o}n},
  {Holwerda}, {Hinz}, {Mu{\~n}oz-Mateos}, {Cisternas}, {Erroz-Ferrer}, {Buta},
  {Laurikainen}, {Salo}, {Laine}, {Men{\'e}ndez-Delmestre}, {Regan}, {de
  Swardt}, {Gil de Paz}, {Seibert}, and {Mizusawa}}]{Kim14}
{Kim} T, {Gadotti} DA, {Sheth} K, et~al.
(2014), 
\apj, 782, 64

\bibitem[{{Koribalski}(2008)}]{Kor08}
{Koribalski}, B.S. (2008), 
Astrophysics and
  Space Science Proceedings, 5, 41 

\bibitem[{{Koribalski} and {L{\'o}pez-S{\'a}nchez}(2009)}]{Kor09}
{Koribalski}, B.S., {L{\'o}pez-S{\'a}nchez}, {\'A}.R. (2009),
\mnras, 400, 1749--1767

\bibitem[{{Kova{\v c}} et~al.(2009){Kova{\v c}}, {Oosterloo}, and {van der
  Hulst}}]{Kov09}
{Kova{\v c}}, K., {Oosterloo}, T.A., {van der Hulst}, J.M. (2009), 
\mnras, 400 ,743--765

\bibitem[{{Kranz} et~al.(2003){Kranz}, {Slyz}, and {Rix}}]{KSR03}
{Kranz}, T., {Slyz}, A., {Rix}, H.-W. (2003), 
\apj, 586, 143--151 

\bibitem[{{Kreckel} et~al.(2011){Kreckel}, {Platen}, {Arag{\'o}n-Calvo}, {van
  Gorkom}, {van de Weygaert}, {van der Hulst}, {Kova{\v c}}, {Yip}, and
  {Peebles}}]{Kre11}
{Kreckel}, K., {Platen}, E., {Arag{\'o}n-Calvo}, M.A., {van Gorkom}, J.H., {van de
  Weygaert}, R., {van der Hulst}, J.M., {Kova{\v c}}, K., {Yip}, C.W., {Peebles}, P.J.E.
  (2011), 
\aj, 141, 4

\bibitem[{{Kreckel} et~al.(2012){Kreckel}, {Platen}, {Arag{\'o}n-Calvo}, {van
  Gorkom}, {van de Weygaert}, {van der Hulst}, and {Beygu}}]{Kre12}
{Kreckel}, K., {Platen}, E., {Arag{\'o}n-Calvo}, M.A., {van Gorkom}, J.H., {van de
  Weygaert}, R., {van der Hulst}, J.M., {Beygu}, B. (2012), 
\aj, 144, 16

\bibitem[{{Kregel} et~al.(2005){Kregel}, {van der Kruit}, and {Freeman}}]{Kre05}
{Kregel}, M., {van der Kruit}, P.C., {Freeman}, K.C. (2005), 
\mnras, 358, 503--520

\bibitem[{{Kuzio de Naray} et~al.(2006){Kuzio de Naray}, {McGaugh}, {de Blok},
  and {Bosma}}]{Kuz06}
{Kuzio de Naray}, R., {McGaugh}, S.S., {de Blok}, W.J.G., {Bosma}, A. (2006),
\apjs, 165, 461--479

\bibitem[{{Kuzio de Naray} et~al.(2008){Kuzio de Naray}, {McGaugh}, and {de
  Blok}}]{Kuz08}
{Kuzio de Naray}, R., {McGaugh}, S.S., {de Blok}, W.J.G. (2008), 
  \apj, 676, 920-943

\bibitem[{{Kwee} et~al.(1954){Kwee}, {Muller}, and {Westerhout}}]{KMW54}
{Kwee}, K.K., {Muller}, C.A., {Westerhout}, G. (1954), 
\bain, 12, 211

\bibitem[{{Laine} et~al.(2014){Laine}, {Laurikainen}, {Salo}, {Comer{\'o}n},
  {Buta}, {Zaritsky}, {Athanassoula}, {Bosma}, {Mu{\~n}oz-Mateos}, {Gadotti},
  {Hinz}, {Erroz-Ferrer}, {Gil de Paz}, {Kim}, {Men{\'e}ndez-Delmestre},
  {Mizusawa}, {Regan}, {Seibert}, and {Sheth}}]{Lai14}
{Laine} J, {Laurikainen} E, {Salo} H, et~al.
(2014), 
\mnras, 441, 1992--2012 

\bibitem[{{Lelli} et~al.(2010){Lelli}, {Fraternali}, and {Sancisi}}]{Lel10}
{Lelli}, F., {Fraternali}, F., {Sancisi}, R. (2010), 
\aap, 516, A11

\bibitem[{{Lelli} et~al.(2014){Lelli}, {Verheijen}, and {Fraternali}}]{Lel14}
{Lelli}, F., {Verheijen}, M., {Fraternali}, F. (2014) 
\aap, 566, A71

\bibitem[{{Lelli} et~al.(2016){Lelli}, {McGaugh}, and {Schombert}}]{Lel16}
{Lelli}, F., {McGaugh}, S.S., {Schombert}, J.M. (2016), 
ArXiv e-prints \eprint{1606.09251}

\bibitem[{{Lewis}(1968)}]{Lew68}
{Lewis}, B.M. (1968), 
Proceedings of
  the Astronomical Society of Australia, 1, 104

\bibitem[{{Lindblad} et~al.(1996){Lindblad}, {Lindblad}, and
  {Athanassoula}}]{LLA96}
{Lindblad}, P.A.B., {Lindblad}, P.O., {Athanassoula}, E. (1996),
\aap, 313, 65--90

\bibitem[{{Lucero} et~al.(2015){Lucero}, {Carignan}, {Elson},
  {Randriamampandry}, {Jarrett}, {Oosterloo}, and {Heald}}]{Luc15}
{Lucero}, D.M., {Carignan}, C., {Elson}, E.C., {Randriamampandry}, T.H., {Jarrett}, T.H.,
  {Oosterloo}, T.A., {Heald}, G.H. (2015), 
\mnras, 450, 3935--3951

\bibitem[{{Ludlow} et~al.(2016){Ludlow}, {Benitez-Llambay}, {Schaller},
  {Theuns}, {Frenk}, {Bower}, {Schaye}, {Crain}, {Navarro}, {Fattahi}, and
  {Oman}}]{Lud16}
{Ludlow}, A.D., {Benitez-Llambay}, A., {Schaller}, M., {Theuns}, T., {Frenk}, C.S., {Bower},
  R., {Schaye}, J., {Crain}, R.A., {Navarro}, J.F., {Fattahi}, A., {Oman}, K.A. (2016), 
ArXiv e-prints \eprint{1610.07663}

\bibitem[{{Malin} and {Hadley}(1997)}]{Mal97}
{Malin}, D., {Hadley}, B. (1997), 
\pasa, 14, 52--58

\bibitem[{{Maloney}(1993)}]{Mal93}
{Maloney}, P. (1993), 
\apj, 414, 41--56 

\bibitem[{{Mart{\'{\i}}n-Navarro} et~al.(2012){Mart{\'{\i}}n-Navarro}, {Bakos},
  {Trujillo}, {Knapen}, {Athanassoula}, {Bosma}, {Comer{\'o}n}, {Elmegreen},
  {Erroz-Ferrer}, {Gadotti}, {Gil de Paz}, {Hinz}, {Ho}, {Holwerda}, {Kim},
  {Laine}, {Laurikainen}, {Men{\'e}ndez-Delmestre}, {Mizusawa},
  {Mu{\~n}oz-Mateos}, {Regan}, {Salo}, {Seibert}, and {Sheth}}]{MarN12}
{Mart{\'{\i}}n-Navarro}, I., {Bakos}, J., {Trujillo}, I., et~al. 
(2012), 
 \mnras, 427, 1102--1134 

\bibitem[{{Mart{\'{\i}}nez-Delgado} et~al.(2008){Mart{\'{\i}}nez-Delgado},
  {Pe{\~n}arrubia}, {Gabany}, {Trujillo}, {Majewski}, and {Pohlen}}]{Mar08}
{Mart{\'{\i}}nez-Delgado}, D., {Pe{\~n}arrubia}, J., {Gabany}, R.J., {Trujillo}, I.,
  {Majewski}, S.R., {Pohlen}, M. (2008), 
\apj, 689, 184-193

\bibitem[{{Mart{\'{\i}}nez-Delgado} et~al.(2010){Mart{\'{\i}}nez-Delgado},
  {Gabany}, {Crawford}, {Zibetti}, {Majewski}, {Rix}, {Fliri},
  {Carballo-Bello}, {Bardalez-Gagliuffi}, {Pe{\~n}arrubia}, {Chonis}, {Madore},
  {Trujillo}, {Schirmer}, and {McDavid}}]{Mar10}
{Mart{\'{\i}}nez-Delgado}, D., {Gabany}, R.J., {Crawford}, K., et~al.
(2010), 
\aj, 140, 962--967

\bibitem[{{Mart{\'{\i}}nez-Delgado} et~al.(2012){Mart{\'{\i}}nez-Delgado},
  {Romanowsky}, {Gabany}, {Annibali}, {Arnold}, {Fliri}, {Zibetti}, {van der
  Marel}, {Rix}, {Chonis}, {Carballo-Bello}, {Aloisi}, {Macci{\`o}},
  {Gallego-Laborda}, {Brodie}, and {Merrifield}}]{MarD12}
{Mart{\'{\i}}nez-Delgado} D, {Romanowsky} AJ, {Gabany} RJ, et~al. 
(2012), 
\apjl, 748, L24 

\bibitem[{{Martinsson} et~al.(2013){Martinsson}, {Verheijen}, {Westfall},
  {Bershady}, {Andersen}, and {Swaters}}]{Mar13}
{Martinsson}, T.P.K., {Verheijen}, M.A.W., {Westfall}, K.B., {Bershady}, M.A., {Andersen}, D.R.,
  {Swaters}, R.A. (2013), 
\aap, 557, A131

\bibitem[{{Martinsson} et~al.(2016){Martinsson}, {Verheijen}, {Bershady},
  {Westfall}, {Andersen}, and {Swaters}}]{Mar16}
{Martinsson}, T.P.K., {Verheijen}, M.A.W., {Bershady}, M.A., {Westfall}, K.B., {Andersen}, D.R.,
  {Swaters}, R.A. (2016), 
\aap, 585, A99

\bibitem[{{Materne} and {Tammann}(1976)}]{MT76}
{Materne}, J., {Tammann}, G.A. (1976), {On the stability of groups of galaxies and
  the question of hidden matter.} In: {Kharadze}, E.K. (ed.) Stars and Galaxies
  from Observational Points of View, 455--462

\bibitem[{{Matthews} and {Uson}(2008)}]{Mat08}
{Matthews}, L.D., {Uson}, J.M. (2008), 
\aj, 135, 291-318

\bibitem[{{Matthews} and {Wood}(2003)}]{Mat03}
{Matthews}, L.D., {Wood}, K. (2003), 
\apj, 593, 721--732

\bibitem[{{McGaugh} et~al.(2016){McGaugh}, {Lelli}, and {Schombert}}]{McGa16}
{McGaugh}, S., {Lelli}, F., {Schombert}, J. (2016), 
ArXiv e-prints \eprint{1609.05917}

\bibitem[{{McGee} and {Milton}(1966)}]{McGee66}
{McGee}, R.X., {Milton}, J.A. (1966), 
  Australian Journal of Physics, 19, 343

\bibitem[{{Meurer} et~al.(1996){Meurer}, {Carignan}, {Beaulieu}, and
  {Freeman}}]{Meu96}
{Meurer}, G.R., {Carignan}, C., {Beaulieu}, S.F., {Freeman}, K.C. (1996), 
\aj, 111, 1551

\bibitem[{{Milgrom}(1983)}]{Mil83}
{Milgrom}, M. (1983), 
\apj, 270, 365--370

\bibitem[{{Minchin} et~al.(2007){Minchin}, {Davies}, {Disney}, {Grossi},
  {Sabatini}, {Boyce}, {Garcia}, {Impey}, {Jordan}, {Lang}, {Marble},
  {Roberts}, and {van Driel}}]{Min07}
{Minchin} R, {Davies} J, {Disney} M, et~al.
(2007), 
\apj, 670, 1056--1064

\bibitem[{{Mogotsi} et~al.(2016){Mogotsi}, {de Blok}, {Cald{\'u}-Primo},
  {Walter}, {Ianjamasimanana}, and {Leroy}}]{Mog16}
{Mogotsi}, K.M., {de Blok}, W.J.G., {Cald{\'u}-Primo}, A., {Walter}, F., {Ianjamasimanana},
  R., {Leroy}, A.K. (2016), 
\aj, 151, 15 

\bibitem[{{Moore}(1994)}]{Moo94}
{Moore}, B. (1994), 
\nat, 370, 629--631 

\bibitem[{{Mouhcine} et~al.(2010){Mouhcine}, {Ibata}, and {Rejkuba}}]{Mou10}
{Mouhcine}, M., {Ibata}, R., {Rejkuba}, M. (2010), 
\apjl, 714, L12--L15 

\bibitem[{{Mu{\~n}oz-Mateos} et~al.(2013){Mu{\~n}oz-Mateos}, {Sheth}, {Gil de
  Paz}, {Meidt}, {Athanassoula}, {Bosma}, {Comer{\'o}n}, {Elmegreen},
  {Elmegreen}, {Erroz-Ferrer}, {Gadotti}, {Hinz}, {Ho}, {Holwerda}, {Jarrett},
  {Kim}, {Knapen}, {Laine}, {Laurikainen}, {Madore}, {Menendez-Delmestre},
  {Mizusawa}, {Regan}, {Salo}, {Schinnerer}, {Seibert}, {Skibba}, and
  {Zaritsky}}]{Mun13}
{Mu{\~n}oz-Mateos}, J.C., {Sheth}, K., {Gil de Paz}, A., et~al.
(2013), 
\apj, 771, 59

\bibitem[{{Mu{\~n}oz-Mateos} et~al.(2015){Mu{\~n}oz-Mateos}, {Sheth}, {Regan},
  {Kim}, {Laine}, {Erroz-Ferrer}, {Gil de Paz}, {Comeron}, {Hinz},
  {Laurikainen}, {Salo}, {Athanassoula}, {Bosma}, {Bouquin}, {Schinnerer},
  {Ho}, {Zaritsky}, {Gadotti}, {Madore}, {Holwerda}, {Men{\'e}ndez-Delmestre},
  {Knapen}, {Meidt}, {Querejeta}, {Mizusawa}, {Seibert}, {Laine}, and
  {Courtois}}]{Mun15}
{Mu{\~n}oz-Mateos}, J.C., {Sheth}, K., {Regan}, M., et~al.
(2015), 
\apjs, 219, 3 

\bibitem[{{Muller} and {Oort}(1951)}]{MO51}
{Muller}, C.A., {Oort}, J.H. (1951),
\nat, 168, 357--358 

\bibitem[{{Navarro}(1998)}]{Nav98}
{Navarro}, J.F. (1998), 
ArXiv Astrophysics e-prints \eprint{astro-ph/9807084}

\bibitem[{{Navarro} et~al.(1996){Navarro}, {Frenk}, and {White}}]{NFW96}
{Navarro}, J.F., {Frenk}, C.S., {White}, S.D.M. (1996), 
\apj, 462, 563 

\bibitem[{{Navarro} et~al.(1997){Navarro}, {Frenk}, and {White}}]{NFW97}
{Navarro}, J.F., {Frenk}, C.S., {White}, S.D.M. (1997), 
\apj, 490, 493--508 

\bibitem[{{Newton}(1980)}]{New80}
{Newton}, K. (1980), 
\mnras, 191,169--184

\bibitem[{{Newton} and {Emerson}(1977)}]{New77}
{Newton}, K., {Emerson}, D.T. (1977), 
\mnras, 181, 573--590 

\bibitem[{{Noordermeer} et~al.(2005){Noordermeer}, {van der Hulst}, {Sancisi},
  {Swaters}, and {van Albada}}]{Noo05}
{Noordermeer}, E., {van der Hulst}, J.M., {Sancisi}, R., {Swaters}, R.A., {van Albada}, T.S.
  (2005), 
\aap, 442, 137--157

\bibitem[{{O'Brien} et~al.(2010{\natexlab{a}}){O'Brien}, {Freeman}, and {van der
  Kruit}}]{OBri10c}
{O'Brien}, J.C., {Freeman}, K.C., {van der Kruit}, P.C. (2010{\natexlab{a}}), 
\aap, 515, A62 

\bibitem[{{O'Brien} et~al.(2010{\natexlab{b}}){O'Brien}, {Freeman}, and {van der
  Kruit}}]{OBri10b}
{O'Brien}, J.C., {Freeman}, K.C., {van der Kruit}, P.C. (2010{\natexlab{b}}), 
\aap, 515, A61 

\bibitem[{{O'Brien} et~al.(2010{\natexlab{c}}){O'Brien}, {Freeman}, and {van der
  Kruit}}]{OBri10d}
{O'Brien}, J.C., {Freeman}, K.C., {van der Kruit}, P.C. (2010{\natexlab{c}}), 
\aap, 515, A63

\bibitem[{{O'Brien} et~al.(2010{\natexlab{d}}){O'Brien}, {Freeman}, {van der
  Kruit}, and {Bosma}}]{OBri10a}
{O'Brien}, J.C., {Freeman}, K.C., {van der Kruit}, P.C., {Bosma}, A. (2010{\natexlab{d}}),
\aap, 515, A60

\bibitem[{{Oh} et~al.(2008){Oh}, {de Blok}, {Walter}, {Brinks}, and
  {Kennicutt}}]{Oh08}
{Oh}, S.-H., {de Blok}, W.J.G., {Walter}, F., {Brinks}, E., {Kennicutt}, R.C. Jr. (2008),
\aj, 136, 2761--2781

\bibitem[{{Oh} et~al.(2011){Oh}, {Brook}, {Governato}, {Brinks}, {Mayer}, {de
  Blok}, {Brooks}, and {Walter}}]{Oh11}
{Oh}, S.-H., {Brook}, C., {Governato}, F., {Brinks}, E., {Mayer}, L., {de Blok}, W.J.G.,
  {Brooks}, A., {Walter}, F. (2011), 
\aj, 142, 24

\bibitem[{{Oh} et~al.(2015){Oh}, {Hunter}, {Brinks}, {Elmegreen}, {Schruba},
  {Walter}, {Rupen}, {Young}, {Simpson}, {Johnson}, {Herrmann}, {Ficut-Vicas},
  {Cigan}, {Heesen}, {Ashley}, and {Zhang}}]{Oh15}
{Oh}, S.-H., {Hunter}, D.A., {Brinks}, E., et~al. 
(2015),
\aj, 149, 180

\bibitem[{{Olling}(1996{\natexlab{a}})}]{Oll96a}
{Olling}, R.P. (1996{\natexlab{a}}), 
\aj, 112, 457

\bibitem[{{Olling}(1996{\natexlab{b}})}]{Oll96b}
{Olling}, R.P. (1996{\natexlab{b}}), 
{The Highly Flattened Dark Matter Halo of NGC 4244}. 
\aj, 112, 481

\bibitem[{{Olling} and {Merrifield}(2000)}]{Oll00}
{Olling}, R.P., {Merrifield}, M.R. (2000) 
\mnras, 311, 361--369

\bibitem[{{Oman} et~al.(2015){Oman}, {Navarro}, {Fattahi}, {Frenk}, {Sawala},
  {White}, {Bower}, {Crain}, {Furlong}, {Schaller}, {Schaye}, and
  {Theuns}}]{Oman15}
{Oman}, K.A., {Navarro}, J.F., {Fattahi}, A., et~al. 
(2015), 
\mnras, 452, 3650--3665

\bibitem[{{Oman} et~al.(2016){Oman}, {Navarro}, {Sales}, {Fattahi}, {Frenk},
  {Sawala}, {Schaller}, and {White}}]{Oman16}
{Oman}, K.A., {Navarro}, J.F., {Sales}, L.V., {Fattahi}, A., {Frenk}, C.S., {Sawala}, T.,
  {Schaller}, M., {White}, S.D.M. (2016), 
  \mnras, 460, 3610--3623

\bibitem[{{Oosterloo} and {van Gorkom}(2005)}]{Oos05}
{Oosterloo}, T., {van Gorkom}, J. (2005), 
\aap, 437, L19--L22

\bibitem[{{Oosterloo} et~al.(2007){Oosterloo}, {Fraternali}, and
  {Sancisi}}]{Oos07}
{Oosterloo}, T., {Fraternali}, F., {Sancisi}, R. (2007), 
\aj, 134, 1019

\bibitem[{{Ostriker} and {Peebles}(1973)}]{OP73}
{Ostriker}, J.P., {Peebles}, P.J.E. (1973), 
\apj, 186, 467--480

\bibitem[{{Ostriker} et~al.(1974){Ostriker}, {Peebles}, and {Yahil}}]{OPY74}
{Ostriker}, J.P., {Peebles}, P.J.E., {Yahil}, A. (1974), 
\apjl, 193, L1--L4

\bibitem[{{Ott} et~al.(2001){Ott}, {Walter}, {Brinks}, {Van Dyk}, {Dirsch}, and
  {Klein}}]{Ott01}
{Ott}, J., {Walter}, F., {Brinks}, E., {Van Dyk}, S.D., {Dirsch}, B., {Klein}, U. (2001),
\aj, 122, 3070--3091

\bibitem[{{Ott} et~al.(2012){Ott}, {Stilp}, {Warren}, {Skillman}, {Dalcanton},
  {Walter}, {de Blok}, {Koribalski}, and {West}}]{Ott12}
{Ott}, J., {Stilp}, A.M., {Warren}, S.R., {Skillman}, E.D., {Dalcanton}, J.J., {Walter}, F.,
  {de Blok}, W.J.G., {Koribalski}, B., {West}, A.A. (2012), 
\aj, 144, 123

\bibitem[{{Peters} et~al.(2013){Peters}, {van der Kruit}, {Allen}, and
  {Freeman}}]{Pet13}
{Peters}, S.P.C., {van der Kruit}, P.C., {Allen}, R.J., {Freeman}, K.C. (2013), 
ArXiv e-prints,  \eprint{1303.2463}

\bibitem[{{Peters} et~al.(2016{\natexlab{a}}){Peters}, {de Geyter}, {van der
  Kruit}, and {Freeman}}]{Pet16c}
{Peters}, S.P.C., {de Geyter}, G., {van der Kruit}, P.C., {Freeman}, K.C.
  (2016{\natexlab{a}}), 
ArXiv e-prints \eprint{1608.05563}

\bibitem[{{Peters} et~al.(2016{\natexlab{b}}){Peters}, {van der Kruit}, {Allen},
  and {Freeman}}]{Pet16a}
{Peters}, S.P.C., {van der Kruit}, P.C., {Allen}, R.J., {Freeman}, K.C. (2016{\natexlab{b}}),
ArXiv e-prints \eprint{1608.05557}

\bibitem[{{Peters} et~al.(2016{\natexlab{c}}){Peters}, {van der Kruit}, {Allen},
  and {Freeman}}]{Pet16b}
{Peters}, S.P.C., {van der Kruit}, P.C., {Allen}, R.J., {Freeman}, K.C. (2016{\natexlab{c}}),
ArXiv e-prints \eprint{1608.05559}

\bibitem[{{Peters} et~al.(2016{\natexlab{d}}){Peters}, {van der Kruit}, {Allen},
  and {Freeman}}]{Pet16d}
{Peters}, S.P.C., {van der Kruit}, P.C., {Allen}, R.J., {Freeman}, K.C. (2016{\natexlab{d}}),
ArXiv e-prints \eprint{1605.08209}

\bibitem[{{Ponomareva} et~al.(2016){Ponomareva}, {Verheijen}, and
  {Bosma}}]{Pon16}
{Ponomareva}, A.A., {Verheijen}, M.A.W., {Bosma}, A. (2016), 
\mnras, 463, 4052--4067

\bibitem[{{Puche} et~al.(1992){Puche}, {Westpfahl}, {Brinks}, and {Roy}}]{Puc92}
{Puche}, D., {Westpfahl}, D., {Brinks}, E., {Roy}, J.R. (1992), 
\aj, 103, 1841--1858

\bibitem[{{Radburn-Smith} et~al.(2014){Radburn-Smith}, {de Jong}, {Streich},
  {Bell}, {Dalcanton}, {Dolphin}, {Stilp}, {Monachesi}, {Holwerda}, and
  {Bailin}}]{Rad14}
{Radburn-Smith}, D.J., {de Jong}, R.S., {Streich}, D., et~al. 
(2014),
\apj, 780, 105 

\bibitem[{{Rand} et~al.(1990){Rand}, {Kulkarni}, and {Hester}}]{Rand90}
{Rand}, R.J., {Kulkarni}, S.R., {Hester}, J.J. (1990), 
\apjl, 352, L1--L4

\bibitem[{{Randriamampandry} and {Carignan}(2014)}]{Ran14}
{Randriamampandry}, T.H., {Carignan}, C. (2014), 
\mnras, 439, 2132--2145 

\bibitem[{{Read} et~al.(2016){Read}, {Iorio}, {Agertz}, and
  {Fraternali}}]{Read16}
{Read}, J.I., {Iorio}, G., {Agertz}, O., {Fraternali}, F. (2016), 
\mnras, 462, 3628--3645 

\bibitem[{{Rich} et~al.(2012){Rich}, {Collins}, {Black}, {Longstaff}, {Koch},
  {Benson}, and {Reitzel}}]{Rich12}
{Rich}, R.M., {Collins}, M.L.M., {Black}, C.M., {Longstaff}, F.A., {Koch}, A., {Benson}, A.,
  {Reitzel}, D.B. (2012), 
\nat, 482, 192--194 

\bibitem[{{Roberts}(1966)}]{Rob66}
{Roberts}, M.S. (1966), 
\apj, 144, 639 

\bibitem[{{Roberts}(1972)}]{Rob72}
{Roberts}, M.S. (1972), 
In: {Evans}, D.S., {Wills}, D., {Wills}, B.J. (eds.) 
External Galaxies and Quasi-Stellar Objects, IAU Symposium, vol~44, p~12

\bibitem[{{Roberts}(1975)}]{Rob74}
{Roberts}, M.S. (1975), 
In: {Hayli}, A. (ed.)
  Dynamics of Stellar Systems, IAU Symposium, vol~69, p 331

\bibitem[{{Roberts}(1988)}]{Rob88}
{Roberts}, M.S. (1988), {How Much of the Universe Do We See?} In: Bicentennial
  commemoration of R.G. Boscovich : Milano, September 15-18,  Bossi, M, Tucci, P. (eds.),
  237-247.

\bibitem[{{Roberts} and {Rots}(1973)}]{Rob73}
{Roberts}, M.S., {Rots}, A.H. (1973), 
\aap, 26, 483--485

\bibitem[{{Roberts} and {Whitehurst}(1975)}]{Rob75}
{Roberts}, M.S., {Whitehurst}, R.N. (1975), 
\apj, 201, 327--346 

\bibitem[{{Rogstad}(1971)}]{Rog71b}
{Rogstad}, D.H. (1971), 
{Aperture synthesis study of neutral hydrogen in the galaxy M101: II. Discussion.} 
\aap, 13, 108--115

\bibitem[{{Rogstad} and {Shostak}(1971)}]{Rog71a}
{Rogstad}, D.H., {Shostak}, G.S. (1971), 
\aap, 13, 99--107

\bibitem[{{Rogstad} and {Shostak}(1972)}]{Rog72}
{Rogstad}, D.H., {Shostak}, G.S. (1972) 
\apj, 176, 315

\bibitem[{{Rogstad} et~al.(1974){Rogstad}, {Lockhart}, and {Wright}}]{Rog74}
{Rogstad}, D.H., {Lockhart}, I.A., {Wright}, M.C.H. (1974), 
\apj, 193, 309--319

\bibitem[{{Rogstad} et~al.(1976){Rogstad}, {Wright}, and {Lockhart}}]{Rog76}
{Rogstad}, D.H., {Wright}, M.C.H., {Lockhart}, I.A. (1976), 
\apj 204, 703--711 

\bibitem[{{Rogstad} et~al.(1979){Rogstad}, {Chu}, and {Crutcher}}]{Rog79}
{Rogstad}, D.H., {Chu}, K., {Crutcher}, R.M. (1979), 
\apj, 229, 509--513 

\bibitem[{{Roychowdhury} et~al.(2010){Roychowdhury}, {Chengalur}, {Begum}, and
  {Karachentsev}}]{Roy10}
{Roychowdhury}, S., {Chengalur}, J.N., {Begum}, A., {Karachentsev}, I.D. (2010),
\mnras 404:L60--L63,

\bibitem[{{Roychowdhury} et~al.(2013){Roychowdhury}, {Chengalur},
  {Karachentsev}, and {Kaisina}}]{Roy13}
{Roychowdhury}, S., {Chengalur}, J.N., {Karachentsev}, I.D., {Kaisina}, E.I. (2013), 
\mnras, 436, L104--L108

\bibitem[{{Rubin} and {Ford}(1970)}]{Rub70}
{Rubin}, V.C., {Ford}, W.K. Jr. (1970), 
\apj, 159, 379 

\bibitem[{{Rubin} et~al.(1978){Rubin}, {Thonnard}, and {Ford}}]{Rub78}
{Rubin}, V.C., {Thonnard}, N., {Ford}, W.K. Jr. (1978), 
\apjl, 225, L107--L111 

\bibitem[{{Rubin} et~al.(1980){Rubin}, {Ford}, and {.~Thonnard}}]{Rub80}
{Rubin}, V.C., {Ford}, W.K. Jr., {~Thonnard}. N. (1980), 
\apj, 238, 471--487

\bibitem[{{Rubin} et~al.(1982){Rubin}, {Ford}, {Thonnard}, and
  {Burstein}}]{Rub82}
{Rubin}, V.C., {Ford}, W.K. Jr., {Thonnard}, N., {Burstein}, D. (1982),
\apj, 261, 439--456 

\bibitem[{{Rubin} et~al.(1985){Rubin}, {Burstein}, {Ford}, and
  {Thonnard}}]{Rub85}
{Rubin}, V.C., {Burstein}, D., {Ford}, W.K. Jr., {Thonnard}, N. (1985), 
\apj, 289, 81--98 

\bibitem[{{Rupen}(1991)}]{Rup91}
{Rupen}, M.P. (1991), 
\aj, 102, 48--106 

\bibitem[{{Sackett} et~al.(1994{\natexlab{a}}){Sackett}, {Morrisoni}, {Harding},
  and {Boroson}}]{Sac94a}
{Sackett}, P.D., {Morrison}, H.L., {Harding}, P., {Boroson}, T.A. (1994{\natexlab{a}}),
\nat, 370, 441--443 

\bibitem[{{Sackett} et~al.(1994{\natexlab{b}}){Sackett}, {Rix}, {Jarvis}, and
  {Freeman}}]{Sac94b}
{Sackett}, P.D., {Rix}, H.-W., {Jarvis}, B.J., {Freeman}, K.C. (1994{\natexlab{b}}), 
\apj, 436, 629--641

\bibitem[{{Saha} et~al.(2009){Saha}, {de Jong}, and {Holwerda}}]{Saha09}
{Saha}, K., {de Jong}, R., {Holwerda}, B. (2009), 
\mnras, 396, 409--422

\bibitem[{{Salpeter}(1978)}]{Sal78}
{Salpeter}, E.E. (1978), 
In: {Berkhuijsen}, E.M., {Wielebinski}, R. (eds.) Structure and Properties of Nearby Galaxies, 
IAU Symposium, vol~77, pp 23--26

\bibitem[{{Sancisi}(1976)}]{San76}
{Sancisi}, R. (1976), 
\aap, 53, 159

\bibitem[{{Sancisi} and {Allen}(1979)}]{San79}
{Sancisi}, R., {Allen}, R.J. (1979), 
\aap, 74, 73--84

\bibitem[{{Sancisi} et~al.(2008){Sancisi}, {Fraternali}, {Oosterloo}, and {van
  der Hulst}}]{San08}
{Sancisi}, R., {Fraternali}, F., {Oosterloo}, T., {van der Hulst}, T. (2008), 
\aapr, 15, 189--223 

\bibitem[{{Sandage}(1961)}]{San61}
{Sandage}, A. (1961), {The Hubble atlas of galaxies}, Washington: Carnegie
  Institution

\bibitem[{{Sandage} et~al.(1970){Sandage}, {Freeman}, and {Stokes}}]{San70}
{Sandage}, A., {Freeman}, K.C., {Stokes}, N.R. (1970), 
\apj, 160, 831 

\bibitem[{{Sanders} and {McGaugh}(2002)}]{San02}
{Sanders}, R.H., {McGaugh}, S.S. (2002), 
\araa, 40, 263--317,

\bibitem[{{Schmidt}(1957)}]{Schmi57}
{Schmidt}, M. (1957), 
\bain, 14, 17

\bibitem[{{Schwarzschild}(1954)}]{Schwa54}
{Schwarzschild}, M. (1954), 
\aj, 59, 273

\bibitem[{{Schweizer} et~al.(1983){Schweizer}, {Whitmore}, and
  {Rubin}}]{Schwe83}
{Schweizer}, F., {Whitmore}, B.C., {Rubin}, V.C. (1983), 
\aj, 88, 909--925

\bibitem[{{Serra} et~al.(2012){Serra}, {Oosterloo}, {Morganti}, {Alatalo},
  {Blitz}, {Bois}, {Bournaud}, {Bureau}, {Cappellari}, {Crocker}, {Davies},
  {Davis}, {de Zeeuw}, {Duc}, {Emsellem}, {Khochfar}, {Krajnovi{\'c}},
  {Kuntschner}, {Lablanche}, {McDermid}, {Naab}, {Sarzi}, {Scott}, {Trager},
  {Weijmans}, and {Young}}]{Ser12}
{Serra} P, {Oosterloo} T, {Morganti} R, et~al.
(2012), 
\mnras, 422, 1835--1862 

\bibitem[{{Serra} et~al.(2014){Serra}, {Oser}, {Krajnovi{\'c}}, {Naab},
  {Oosterloo}, {Morganti}, {Cappellari}, {Emsellem}, {Young}, {Blitz}, {Davis},
  {Duc}, {Hirschmann}, {Weijmans}, {Alatalo}, {Bayet}, {Bois}, {Bournaud},
  {Bureau}, {Crocker}, {Davies}, {de Zeeuw}, {Khochfar}, {Kuntschner},
  {Lablanche}, {McDermid}, {Sarzi}, and {Scott}}]{Ser14}
{Serra} P, {Oser} L, {Krajnovi{\'c}} D, et~al.
(2014),
\mnras, 444, 3388--3407

\bibitem[{{Shobbrook} and {Robinson}(1967)}]{Shob67}
{Shobbrook}, R.R., {Robinson}, B.J. (1967), 
Australian Journal of Physics, 20, 131

\bibitem[{{Shostak} and {van der Kruit}(1984)}]{Shos84}
{Shostak}, G.S., {van der Kruit}, P.C. (1984), 
\aap, 132, 20--32

\bibitem[{{Simon} et~al.(2003){Simon}, {Bolatto}, {Leroy}, and {Blitz}}]{Sim03}
{Simon}, J.D., {Bolatto}, A.D., {Leroy}, A., {Blitz}, L. (2003), 
\apj, 596, 957--981 

\bibitem[{{Simon} et~al.(2005){Simon}, {Bolatto}, {Leroy}, {Blitz}, and
  {Gates}}]{Sim05}
{Simon}, J.D., {Bolatto}, A.D., {Leroy}, A., {Blitz}, L., {Gates}, E.L. (2005),
\apj, 621, 757--776

\bibitem[{{Stanonik} et~al.(2009){Stanonik}, {Platen}, {Arag{\'o}n-Calvo}, {van
  Gorkom}, {van de Weygaert}, {van der Hulst}, and {Peebles}}]{Sta09}
{Stanonik}, K., {Platen}, E., {Arag{\'o}n-Calvo}, M.A., {van Gorkom}, J.H., {van de
  Weygaert}, R., {van der Hulst}, J.M., {Peebles}, P.J.E. (2009), 
\apjl, 696, L6--L9

\bibitem[{{Stilp} et~al.(2013{\natexlab{a}}){Stilp}, {Dalcanton}, {Skillman},
  {Warren}, {Ott}, and {Koribalski}}]{Stilp13b}
{Stilp}, A.M., {Dalcanton}, J.J., {Skillman}, E., {Warren}, S.R., {Ott}, J., {Koribalski}, B.
  (2013{\natexlab{a}}), 
\apj,  773, 88 

\bibitem[{{Stilp} et~al.(2013{\natexlab{b}}){Stilp}, {Dalcanton}, {Warren},
  {Skillman}, {Ott}, and {Koribalski}}]{Stilp13a}
{Stilp}, A.M., {Dalcanton}, J.J., {Warren}, S.R., {Skillman}, E., {Ott}, J., {Koribalski}, B.
  (2013{\natexlab{b}}), 
\apj, 765, 136

\bibitem[{{Streich} et~al.(2016){Streich}, {de Jong}, {Bailin}, {Bell},
  {Holwerda}, {Minchev}, {Monachesi}, and {Radburn-Smith}}]{Str16}
{Streich}, D., {de Jong}, R.S., {Bailin}, J., {Bell}, E.F., {Holwerda}, B.W., {Minchev}, I.,
  {Monachesi}, A., {Radburn-Smith}, D.J. (2016), 
 \aap, 585, A97 

\bibitem[{{Sunyaev}(1969)}]{Sun69}
{Sunyaev}, R.A. (1969), 
\aplett, 3, 33

\bibitem[{{Swaters} et~al.(1997){Swaters}, {Sancisi}, and {van der
  Hulst}}]{Swa97}
{Swaters}, R.A., {Sancisi}, R., {van der Hulst}, J.M. (1997), 
 \apj, 491, 140--145

\bibitem[{{Swaters} et~al.(2002){Swaters}, {van Albada}, {van der Hulst}, and
  {Sancisi}}]{Swa02}
{Swaters}, R.A., {van Albada}, T.S., {van der Hulst}, J.M., {Sancisi}, R. (2002), 
\aap, 390 ,829--861

\bibitem[{{Swaters} et~al.(2003){Swaters}, {Madore}, {van den Bosch}, and
  {Balcells}}]{Swa03}
{Swaters}, R.A., {Madore}, B.F., {van den Bosch}, F.C., {Balcells}, M. (2003), 
\apj, 583, 732--751 

\bibitem[{{Tamburro} et~al.(2009){Tamburro}, {Rix}, {Leroy}, {Mac Low},
  {Walter}, {Kennicutt}, {Brinks}, and {de Blok}}]{Tam09}
{Tamburro}, D., {Rix}, H.-W., {Leroy}, A.K., {Mac Low}, M.M., {Walter}, F., {Kennicutt}, R.C.,
  {Brinks}, E., {de Blok}, W.J.G. (2009), 
\aj, 137, 4424--4435 

\bibitem[{{Thilker} et~al.(2007){Thilker}, {Bianchi}, {Meurer}, {Gil de Paz},
  {Boissier}, {Madore}, {Boselli}, {Ferguson}, {Mu{\~n}oz-Mateos}, {Madsen},
  {Hameed}, {Overzier}, {Forster}, {Friedman}, {Martin}, {Morrissey}, {Neff},
  {Schiminovich}, {Seibert}, {Small}, {Wyder}, {Donas}, {Heckman}, {Lee},
  {Milliard}, {Rich}, {Szalay}, {Welsh}, and {Yi}}]{Thi07}
{Thilker}, D.A., {Bianchi}, L., {Meurer}, G., et~al.
(2007), 
\apjs, 173, 538--571 

\bibitem[{{Tollet} et~al.(2016){Tollet}, {Macci{\`o}}, {Dutton}, {Stinson},
  {Wang}, {Penzo}, {Gutcke}, {Buck}, {Kang}, {Brook}, {Di Cintio}, {Keller},
  and {Wadsley}}]{Tol16}
{Tollet}, E., {Macci{\`o}}, A.V., {Dutton}, A.A., et~al.
(2016), 
\mnras, 456, 3542--3552

\bibitem[{{Toloba} et~al.(2016){Toloba}, {Guhathakurta}, {Romanowsky}, {Brodie},
  {Mart{\'{\i}}nez-Delgado}, {Arnold}, {Ramachandran}, and
  {Theakanath}}]{Tolo16}
{Toloba}, E., {Guhathakurta}, P., {Romanowsky}, A.J., {Brodie}, J.P.,
  {Mart{\'{\i}}nez-Delgado}, D., {Arnold}, J.A., {Ramachandran}, N., {Theakanath}, K.
  (2016), 
\apj, 824, 35

\bibitem[{{Toomre}(1981)}]{Toom81}
{Toomre}, A. (1981), 
In: {Fall}, S.M., {Lynden-Bell}, D.
(eds.) Structure and Evolution of Normal Galaxies, pp 111--136

\bibitem[{{Trott} et~al.(2010){Trott}, {Treu}, {Koopmans}, and
  {Webster}}]{Tro10}
{Trott}, C.M., {Treu}, T., {Koopmans}, L.V.E., {Webster}, R.L. (2010), 
\mnras, 401, 1540--1551

\bibitem[{{Uson} and {Matthews}(2003)}]{Uson03}
{Uson}, J.M., {Matthews}, L.D. (2003), 
\aj, 125, 2455--2472 

\bibitem[{{Van Albada} and {Sancisi}(1986)}]{vAS86}
{Van Albada}, T.S., {Sancisi}, R. (1986),
 Philosophical Transactions of the Royal Society of London Series A,
320, 447--464 

\bibitem[{{van de Hulst} et~al.(1957){van de Hulst}, {Raimond}, and {van
  Woerden}}]{Hulst57}
{van de Hulst}, H.C., {Raimond}, E., {van Woerden}, H. (1957),
\bain, 14, 1

\bibitem[{{van der Hulst} and {Sancisi}(1988)}]{Hul88}
{van der Hulst}, T., {Sancisi}, R. (1988), 
\aj, 95, 1354--1359 

\bibitem[{{van der Kruit}(1979)}]{PCK79}
{van der Kruit}, P.C. (1979), 
\aaps, 38, 15--38

\bibitem[{{van der Kruit}(1981)}]{pck81}
{van der Kruit}, P.C. (1981), 
\aap, 99, 298--304

\bibitem[{{van der Kruit}(2007)}]{pck07}
{van der Kruit}, P.C. (2007), 
\aap, 466, 883--893

\bibitem[{{van der Kruit} and {Shostak}(1982)}]{pck82}
{van der Kruit}, P.C., {Shostak}, G.S. (1982), 
\aap, 105, 351--358

\bibitem[{{van der Kruit} and {Shostak}(1984)}]{pck84}
{van der Kruit}, P.C., {Shostak}, G.S. (1984), 
\aap, 134, 258--267

\bibitem[{{van Woerden} et~al.(1975){van Woerden}, {Bosma}, and
  {Mebold}}]{vWBM75}
{van Woerden}, H., {Bosma}, A., {Mebold}, U. (1975), 
In: {Weliachew}, L.
  (ed.) La Dynamique des galaxies spirales, vol 241, 483

\bibitem[{{van Zee}(2004)}]{Zee04}
{van Zee}, L. (2004), 
Bulletin of the American Astronomical Society, vol~36, 1495

\bibitem[{{Verdes-Montenegro} et~al.(2001){Verdes-Montenegro}, {Yun},
  {Williams}, {Huchtmeier}, {Del Olmo}, and {Perea}}]{Ver01}
{Verdes-Montenegro}, L., {Yun}, M.S., {Williams}, B.A., {Huchtmeier}, W.K., {Del Olmo}, A.,
  {Perea}, J. (2001), 
\aap, 377, 812--826 

\bibitem[{{Verdes-Montenegro} et~al.(2002){Verdes-Montenegro}, {Bosma}, and
  {Athanassoula}}]{Ver02}
{Verdes-Montenegro}, L., {Bosma}, A., {Athanassoula}, E. (2002),
\aap, 389, 825--835 

\bibitem[{{Verheijen} and {Sancisi}(2001)}]{Verh01}
{Verheijen}, M.A.W., {Sancisi}, R. (2001), 
\aap, 370, 765--867 

\bibitem[{{Volders}(1959)}]{Vol59}
{Volders}, L.M.J.S. (1959), 
\bain, 14, 323

\bibitem[{{Walter} et~al.(2008){Walter}, {Brinks}, {de Blok}, {Bigiel},
  {Kennicutt}, {Thornley}, and {Leroy}}]{Wal08}
{Walter}, F., {Brinks}, E., {de Blok}, W.J.G., {Bigiel}, F., {Kennicutt}, R.C. Jr.,
  {Thornley}, M.D., {Leroy}, A. (2008), 
\aj, 136, 2563-2647

\bibitem[{{Wang} et~al.(2013){Wang}, {Kauffmann}, {J{\'o}zsa}, {Serra}, {van der
  Hulst}, {Bigiel}, {Brinchmann}, {Verheijen}, {Oosterloo}, {Wang}, {Li}, {den
  Heijer}, and {Kerp}}]{Wang13}
{Wang}, J., {Kauffmann}, G., {J{\'o}zsa}, G.I.G., et~al.
(2013), 
\mnras, 433, 270--294

\bibitem[{{Wang} et~al.(2016){Wang}, {Koribalski}, {Serra}, {van der Hulst},
  {Roychowdhury}, {Kamphuis}, and {Chengalur}}]{Wang16}
{Wang}, J., {Koribalski}, B.S., {Serra}, P., {van der Hulst}, T., {Roychowdhury}, S.,
  {Kamphuis}, P., {Chengalur}, J.N. (2016), 
\mnras, 460, 2143--2151 

\bibitem[{{Weiner}(2004)}]{Wei04}
{Weiner}, B.J. (2004), 
  In: {Ryder}, S., {Pisano}, D., {Walker}, M., {Freeman}, K. (eds.) Dark Matter in
  Galaxies, IAU Symposium, vol 220, 265 

\bibitem[{{Weiner} et~al.(2001){Weiner}, {Sellwood}, and {Williams}}]{Wei01}
{Weiner}, B.J., {Sellwood}, J.A., {Williams}, T.B. (2001),
\apj, 546, 931--951 

\bibitem[{{Westfall} et~al.(2011){Westfall}, {Bershady}, and
  {Verheijen}}]{Wes11}
{Westfall}, K.B., {Bershady}, M.A., {Verheijen}, M.A.W. (2011), 
\apjs, 193, 21

\bibitem[{{Westmeier} et~al.(2011){Westmeier}, {Braun}, and
  {Koribalski}}]{WestBK11}
{Westmeier}, T., {Braun}, R., {Koribalski}, B.S. (2011), 
\mnras, 410, 2217--2236

\bibitem[{{Westmeier} et~al.(2013){Westmeier}, {Koribalski}, and
  {Braun}}]{WestKB13}
{Westmeier}, T., {Koribalski}, B.S., {Braun}, R. (2013), 
\mnras, 434, 3511--3525 

\bibitem[{{White} and {Rees}(1978)}]{WR78}
{White}, S.D.M., {Rees}, M.J. (1978),
\mnras, 183, 341--358

\bibitem[{{Whitmore} et~al.(1987){Whitmore}, {McElroy}, and
  {Schweizer}}]{Whitm87}
{Whitmore}, B.C., {McElroy}, D.B., {Schweizer}, F. (1987), 
\apj, 314, 439--456 

\bibitem[{{Wilkinson}(1991)}]{Wilk91}
{Wilkinson}, P.N. (1991), 
In: {Cornwell}, T.J., {Perley}, R.A.
  (eds.) IAU Colloq. 131: Radio Interferometry. Theory, Techniques, and
  Applications, ASP Conference Series, vol~19, 428--432

\bibitem[{{Wolfe} et~al.(2013){Wolfe}, {Pisano}, {Lockman}, {McGaugh}, and
  {Shaya}}]{Wol13}
{Wolfe}, S.A., {Pisano}, D.J., {Lockman}, F.J., {McGaugh}, S.S., {Shaya}, E.J. (2013),
\nat, 497, 224--226 

\bibitem[{{Wolfinger} et~al.(2016){Wolfinger}, {Kilborn}, {Ryan-Weber}, and
  {Koribalski}}]{Wol16}
{Wolfinger}, K., {Kilborn}, V.A., {Ryan-Weber}, E.V., {Koribalski}, B.S. (2016), 
\pasa, 33, e038

\bibitem[{{Z{\'a}nmar S{\'a}nchez} et~al.(2008){Z{\'a}nmar S{\'a}nchez},
  {Sellwood}, {Weiner}, and {Williams}}]{Zan08}
{Z{\'a}nmar S{\'a}nchez}, R., {Sellwood}, J.A., {Weiner}, B.J., {Williams}, T.B. (2008),
\apj, 674, 797-813

\bibitem[{{Zheng} et~al.(1999){Zheng}, {Shang}, {Su}, {Burstein}, {Chen},
  {Deng}, {Byun}, {Chen}, {Chen}, {Deng}, {Fan}, {Fang}, {Hester}, {Jiang},
  {Li}, {Lin}, {Sun}, {Tsay}, {Windhorst}, {Wu}, {Xia}, {Xu}, {Xue}, {Yan},
  {Zheng}, {Zhou}, {Zhu}, {Zou}, and {Lu}}]{Zhe99}
{Zheng}, Z., {Shang}, Z., {Su}, H., et~al. 
(1999),
\aj, 117, 2757--2780

\end{thebibliography}



\end{document}